\documentclass{jfm}
\usepackage{natbib}
\usepackage{upmath}
\usepackage[british]{babel}
\usepackage{amsmath,bm}
\usepackage{amssymb}
\usepackage{amsbsy}
\usepackage{amscd}
\usepackage{amstext}
\usepackage{tabularx}
\usepackage{float}
\usepackage{makeidx}
\usepackage{amsmath}
\usepackage{subfigure}
\usepackage{afterpage}
\usepackage[T1]{fontenc}
\usepackage[latin1]{inputenc}
\usepackage{multirow}
\usepackage{color}

\usepackage{graphicx} 
\usepackage{graphics,epsfig}
\usepackage{amsfonts}
\usepackage{psfrag}

\ifCUPmtlplainloaded \else
  \checkfont{eurm10}
  \iffontfound
    \IfFileExists{upmath.sty}
      {\typeout{^^JFound AMS Euler Roman fonts on the system,
                   using the 'upmath' package.^^J}%
       \usepackage{upmath}}
      {\typeout{^^JFound AMS Euler Roman fonts on the system, but you
                   dont seem to have the}%
       \typeout{'upmath' package installed. JFM.cls can take advantage
                 of these fonts,^^Jif you use 'upmath' package.^^J}%
      }
  \else
  \fi
\fi

\ifCUPmtlplainloaded \else
  \checkfont{msam10}
  \iffontfound
    \IfFileExists{amssymb.sty}
      {\typeout{^^JFound AMS Symbol fonts on the system, using the
                'amssymb' package.^^J}%
       \usepackage{amssymb}%
         
       \let\ge=\geqslant  
      }{}
  \fi
\fi

\ifCUPmtlplainloaded \else
  \IfFileExists{amsbsy.sty}
    {\typeout{^^JFound the 'amsbsy' package on the system, using it.^^J}%
     \usepackage{amsbsy}}
    {\providecommand\boldsymbol[1]{\mbox{\boldmath $##1$}}}
\fi

\newsavebox{\astrutbox}
\sbox{\astrutbox}{\rule[-5pt]{0pt}{20pt}}

\newcommand\bu{\boldsymbol{u}}
\newcommand\bc{\boldsymbol{c}}
\newcommand\bgamma{\boldsymbol{\gamma}}

\newcommand\ad[2]{\left< #1, \ #2 \right>}

\newcommand\Dongdong[1]{\color{black} #1\color{black}}

\title[Subcritical and supercritical bifurcations in axisymmetric viscoelastic pipe flows]
{Subcritical and supercritical bifurcations in axisymmetric viscoelastic pipe flows}

\author[D. Wan, G. Sun, M. Zhang]%
{
Dongdong Wan, Guangrui Sun \and Mengqi Zhang
}

\affiliation{
Department of Mechanical Engineering, National University of Singapore, 9 Engineering Drive 1, 117575 Singapore 
  \\[\affilskip]
}

\date{xx}

\graphicspath{{figures/}}
\newlength\savewidth

\begin{document}
\maketitle

Axisymmetric viscoelastic pipe flow of Oldroyd-B fluids has been recently found to be linearly unstable by Garg \textit{et al.} \textit{Phys. Rev. Lett.}, 121.024502 (2018). From a nonlinear point of view, this means that the flow can transition to turbulence supercritically, in contrast to the subcritical Newtonian pipe flows. Experimental evidences of subcritical and supercritical bifurcations of viscoelastic pipe flows have been reported, but these nonlinear phenomena have not been examined theoretically. In this work, we study the weakly nonlinear stability of this flow by performing a multiple-scale expansion of the disturbance around linear critical conditions. \Dongdong{The perturbed parameter is Reynolds number with the others being unperturbed. } A third-order Ginzburg-Landau equation is derived with its coefficient indicating the bifurcation type of the flow. After exploring a large parameter space, we found that polymer concentration plays an important role: at high polymer concentrations (or small \Dongdong{solvent-to-solution } viscosity ratio $\beta\lessapprox0.785$), the nonlinearity stabilises the flow, indicating that the flow will bifurcate supercritically, while at low polymer concentrations ($\beta\gtrapprox 0.785$), the flow bifurcation is subcritical. The results agree qualitatively with experimental observations where critical $\beta\approx0.855$. The pipe flow of UCM fluids can be linearly unstable and its bifurcation type is also supercritical. \Dongdong{At a fixed value of $\beta$, the Landau coefficient scales with the inverse of Weissenberg number ($Wi$) when $Wi$ is sufficiently large. } The present analysis provides a theoretical understanding of the recent studies on the supercritical and subcritical routes to the elasto-inertial turbulence in viscoelastic pipe flows.

\begin{abstract}

\end{abstract}

\section{Introduction}

Upon a minute addition of polymers, turbulent drag reduction can occur in viscoelastic polymeric flows. This feature has attracted much attention of researchers for a long time. As one of the challenging problems, transition to turbulence in viscoelastic pipe flow has not been fully understood so far. \Dongdong{Newtonian pipe flows are linearly stable at very high Reynolds numbers \citep{Davey1969,Meseguer2003Linearized} (in fact, the current consensus holds that this type of flow is unconditionally linearly stable although there is no rigorous mathematical proof of this)}; thus, at Reynolds numbers explored so far, the laminar Newtonian pipe flow can only transition to turbulence via a subcritical route (meaning that nonlinearity destabilises the flow because of lacking a linear instability mechanism). It has been recently found that the balance of inertial, viscous and elastic effects can lead to a linear instability of centre mode in viscoelastic pipe flows (of Oldroyd-B fluids), providing a possible supercritical transition route to turbulence in this flow \citep{Garg2018Viscoelastic}. Much attention has been diverted to this novel instability; nevertheless, the subsequent nonlinear development of the centre-mode instability has not been well explored so far and the bifurcation type of this flow has not been investigated extensively in a large parameter space, both of which will be the topics of the present study. In the following subsections, we will first review the relevant works in the literature on viscoelastic shear flows and then discuss the position of the current work.

\subsection{Linear stability/instability in viscoelastic shear flows}

In the linear stability analysis of viscoelastic parallel shear flows, the plane Couette flow of upper convected Maxwell fluids (UCM, which is a limiting case of the Oldroyd-B model without solvent viscosity) was first studied, pioneered by the milestone work of \cite{Gorodtsov1967Linear}. Many subsequent works confirmed the linear stability of this flow in the whole parameter space of Reynolds number $Re$ and Weissenberg number $Wi$ \citep{Lee1986Stability,Renardy1992Rigorous,Chokshi2009Stability,Chaudhary2019Elasto-inertial}. In contrast, the plane Poiseuille flow of UCM fluids was found to present a more complex linear stability/instability diagram. The first investigation into the linear stability of this flow was carried out by \cite{Porteous1972Linear}. Their analysis at high $Re>1400$ showed that polymer elasticity (characterized by the elasticity number $E=Wi/Re$) destabilizes the flow, as the critical Reynolds number $Re_c$ decreases monotonically with increasing $E$ up to 0.0025. This effect is more profound for sufficiently high $E$ at which they found two new unstable modes, qualitatively different from the elastically modified Tollmien-Schlichting (TS) mode. After a more extensive exploration up to $E=0.0030$, the minimum $Re_c$ was found to exist at about $E=0.0025$ \citep{Sureshkumar1995Linear}. At low $Re$, however, the plane Poiseuille flow of highly elastic UCM fluids was found to be linearly stable \citep{Ho1977Stability,Lee1986Stability}, implying the necessity of a nonlinear instability to account for the ``melt fracture'' phenomenon observed in polymer extrusion process. The importance of such nonlinearity was indeed confirmed in a weakly nonlinear stability analysis by \cite{Meulenbroek2004Weakly}, to be discussed below. In terms of the symmetry properties of eigenfunctions, all these studies found merely antisymmetric unstable modes, while a more recent comprehensive linear stability analysis performed by \cite{Chaudhary2019Elasto-inertial} revealed up to seven unstable modes with some of them being symmetric after an extensive search in the parameter space of the plane Poiseuille flow of UCM fluids. They interpreted these unstable modes as part of infinite hierarchy of elasto-inertial instabilities resulting from the competing effects between polymer elasticity and fluid inertia. 

Oldroyd-B model, compared to the UCM model, more realistically models the effect of solvent viscosity, quantified by the ratio $\beta$ of solvent viscosity to total viscosity. By definition, the Oldroyd-B model reduces to the UCM model at $\beta=0$ and Newtonian fluid at $\beta=1$. In the same study mentioned above, \cite{Sureshkumar1995Linear} also investigated the linear stability of plane Poiseuille flow of Oldroyd-B fluids. They reported a pronounced stabilizing effect of non-zero solvent viscosity (i.e. $\beta>0$) at $E=0.001$, with the $Re_c$ increasing monotonically from the UCM limit to the Newtonian limit. At a fixed viscosity ratio ($\beta=0.5$ in their study), their $Re_{c}$-$E$ plot also presents a minimum similar to what they found for UCM fluids. The mechanism underlying such minimum was revealed years later and it involves two competing contributions to the transport of perturbation vorticity at highly elastic regimes: one from the shear stress, stabilizing the flow, and the other from the normal stress, destabilizing the flow \citep{Sadanandan2002Viscoelastic}. \cite{Zhang2013Linear} proposed a dimensionless parameter characterizing the non-monotonic effect of elasticity on the linear stability, that is, the ratio of the polymer relaxation time to the characteristic instability time scale, applicable to both short-time and long-time horizons. In addition to these wall-mode instabilities reviewed above, a new centre-mode instability in plane Poiseuille flow of Oldroyd-B fluids has been reported recently by \cite{Khalid2021Centre}. This novel mode only exists for dilute polymer concentrations (with the viscosity ratio $\beta>0.5$) and highly elastic fluids (with the elasticity number $E>0.01$). Moreover, the $Re_c$ they calculated is in a qualitative agreement with that estimated in the experiment conducted by \cite{Srinivas2017Effect}. 

Unlike the linear stability analyses of viscoelastic plane Poiseuille flows, very few studies have focused on viscoelastic pipe flows. \cite{Hansen1973Stability} addressed the linear stability problem of the pipe flow with simplified UCM and Oldroyd-B fluid models. The author found that polymer additive has a stabilizing effect on disturbances when the polymer relaxation time is small, while a destabilizing effect appears if this relaxation time exceeds a critical value. However, this analysis was restricted to high-phase-velocity axisymmetric disturbances and the conclusions were built upon an oversimplification of UCM and Oldroyd-B models by ignoring the convected derivatives in the polymer constitutive equations. Although no linear instability was explicitly found in this study, the author held that the ``destabilizing effect'' identified might support the viewpoint that early turbulence observed in the viscoelastic pipe flow may possibly result from hydrodynamic instability \citep{Forame1972Observations}. Such a speculation has proved to be reasonable in the work by \cite{Garg2018Viscoelastic} based on the complete UCM and Oldroyd-B models, as summarised below.

Using both spectral collocation method and shooting method to solve the linear eigenvalue problem of the viscoelastic pipe flow, \cite{Garg2018Viscoelastic} identified an unstable mode travelling at a phase speed close to the maximum base flow, in contrast to the wall-mode. The authors interpreted this unstable centre mode as an instability in velocity field reinforced by the polymeric force localized near the pipe centreline. The Oldroyd-B model with $\beta \to 0$ reaches the UCM limit, at which they found that the linear instability is absent, different from the plane Poiseuille flow of UCM fluids where linear instability exists \Dongdong{\citep{Porteous1972Linear,Sureshkumar1995Linear,Chaudhary2019Elasto-inertial}}. The authors concluded that the novel linear instability is a subtle balance among fluid inertia, viscosity and polymer elasticity in viscoelastic pipe flows. Recently, the same research group extended their work to the exploration of instability in a larger parameter space and compared their predictions with direct numerical simulation (DNS) results and experimental measurements \citep{Chaudhary2021Linear}. Some of their new observations include that the eigenfunction of unstable modes is not localized near pipe centreline as $\beta$ approaches one with $E$ on the order of 0.1; and that the laminar flow remains stable at quite low $Re$ regardless the values of $E$ and $\beta$. However, one of their comparisons showed that there is one order of magnitude difference between the elasticity number $E$ at which they predicted instability and the $E$ measured in experiments near the onset of turbulence \citep{Samanta2013Elasto-inertial}. They attributed this discrepancy to the characterization procedure in measuring the polymer relaxation time in experiments. Nevertheless, the comparison of the predicted $Re_c$ with the transitional Reynolds number $Re_t$ observed in an experiment carried out by \cite{Chandra2018Onset} showed general consistency. 
Upon considering the shear thinning effect by using the FENE-P model (finitely extensible nonlinear elastic model with Peterlin closure) in their scaling analysis, \cite{Chaudhary2021Linear} were able to derive $Re_c \sim (E(1-\beta))^{-5/8}$, which is consistent with the scaling $Re_t \sim (E(1-\beta))^{-1/2}$ observed in the experiment. However, in the FENE-P model if the polymer maximum extensibility is small enough (when this parameter approaches infinity, the FENE-P model reduces to the Oldroyd-B model), the linear instability ceases to exist \citep{Zhang2021}. The existence of such linear instability in viscoelastic pipe flow indicates a possibility of supercritical bifurcation route. The bifurcation type can be more systematically studied in a weakly nonlinear analysis framework (as will be reviewed next) to further understand the nonlinear development. 

\subsection{Weakly nonlinear stability/instability in Newtonian and viscoelastic flows}

When the amplitude of the disturbance increases to a certain degree, the linear framework becomes inapplicable and the effect of nonlinearity starts to manifest itself. In principle, it is difficult to analyse the nonlinearity, especially when it is strong, but if the weakly nonlinear phase is of interest, the flow can be studied in an analytical manner by applying a multiple-scale expansion method. 

The weakly nonlinear stability theory has initially been developed in the context of Newtonian plane shear flows \citep{Landau1944,Stuart1960Non-linear,Reynolds1967Finite,Herbert1983a,Fujimura1989Equivalence}, reaching the conclusion that the transition in plane Poiseuille flow is subcritical, i.e., the nonlinearity will destabilise the flow just beyond the linear criticality. The weakly nonlinear stability theory was conventionally applied around the linear criticality in order to guarantee the convergence of the expansion scheme. Thus, historically, some controversies were caused when this theory was applied to the Newtonian pipe flow, which is linearly stable at all $Re$ investigated \citep{Davey1969,Meseguer2003Linearized}. Finite-amplitude equilibrium solutions were reported by \cite{Davey1971Finite} who adopted an equilibrium amplitude method (based on the false problem method developed by \cite{Reynolds1967Finite}) to analyse the axisymmetric Newtonian pipe Poiseuille flow. On the contrary, \cite{Itoh1977Nonlinear} found no such equilibrium solutions using Stuart's method. The issue was then  partially solved by \cite{Davey1978} who pointed out the major weakness of both expansion methods for the problems without neutral curves. Later, \cite{Patera1981Finite-amplitude} conducted DNS and confirmed that there is no finite-amplitude equilibrium in the axisymmetric Newtonian pipe flow. From this perspective, the newly found linear instability in \cite{Garg2018Viscoelastic} can facilitate the application of the expansion method in viscoelastic pipe flows in general, because linear critical conditions now exist in this flow and can guarantee the convergence of the expansion scheme.

In addition to the Newtonian fluids, weakly nonlinear stability analyses have also proven to be useful in revealing the bifurcation for more complex fluids. For example, \cite{Bouteraa2015WeaklyA} and \cite{Bouteraa2015WeaklyB} adopted an amplitude expansion method to investigate the bifurcation nature of Rayleigh-B\'{e}nard convention in shear-thinning fluids between two horizontal plates. In the viscoelastic flows, the corresponding weakly nonlinear stability analysis has been mainly performed by Morozov \& van Saarloos and their co-workers. Their efforts aimed at understanding the melt fracture instability occurring in polymer extrusion through a die. In this case, the flow is dominated by polymer elasticity and also linearly stable. In the zero-$Re$ limit, \cite{Meulenbroek2003Intrinsic} studied the weakly nonlinear stability of pipe Poiseuille flow of UCM fluids and reported a subcritical instability for $Wi>5$. \Dongdong{This subcritical mechanism has been experimentally demonstrated by \cite{Bonn2011Large} who observed large velocity fluctuations in this flow, supporting the existence of elastic turbulence at very low Reynolds numbers. } An extension to plane Poiseuille flow showed that the subcritical instability exists for $Wi>4$ \citep{Meulenbroek2004Weakly}. Both these subcritical instabilities are believed to be intrinsic routes to the melt fracture phenomenon (see \cite{Bertola2003Experimental} for the corresponding experimental evidences). More recently, \cite{Morozov2019Subcritical} further extended their analysis in the zero-$Re$ limit to the plane Poiseuille flow of Oldroyd-B fluids and again the instability was found to be subcritical. In addition to these weakly nonlinear analyses of Poiseuille flows,  subcriticality was also found to exist in plane Couette flow of UCM fluids at very small Reynolds numbers \citep{Morozov2005}. A more comprehensive discussion on these subcritical mechanisms in viscoelastic parallel shear flows can be found in their introductory essay \citep{Morozov2007}. Overall, the works summarised above focused on the inertialess limit, relevant to the flow problems of their interest. The present work, instead, aims at performing weakly nonlinear stability analyses of the viscoelastic pipe flows in the elasto-inertial regime, where drag reduction happens.

\subsection{Recent discussions on the bifurcation type in viscoelastic flows}

Recently, in the research community of viscoelastic flows, elasto-inertial turbulence (EIT) has received much attention owing to its unique flow features relevant to the MDR (maximum drag reduction, an asymptotic statistical state first discussed by \cite{Virk1975}), and has been investigated experimentally and numerically by many researchers \citep{Samanta2013Elasto-inertial,Dubief2013Mechanism,Sid2018Two-dimensional,Lopez2019Dynamics,Shekar2020Self,Shekar2019Critical-Layer,Page2020Exact,Choueiri2021Experimental,Shekar2021Tollmien}, among many others. \Dongdong{In the DNS study by \cite{Page2020Exact}, the exact coherent structures were calculated in subcritical 2D viscoelastic channel flows of FENE-P fluids and the authors demonstrated that the flow possesses a subcritical transition mechanism in terms of both Reynolds number and Weissenberg number}. As for the bifurcation type of pipe flows, since the transition to turbulence in Newtonian pipe flow is known to be subcritical, a naive speculation would be to argue that the viscoelastic pipe flow will also transition to turbulence only via subcritical routes. However, the finding of the centre-mode instability enriches the picture \citep{Garg2018Viscoelastic,Chaudhary2021Linear}, implying and confirming the possibility of a supercritical transition route as envisioned by \cite{Graham2014Drag}. 

\cite{Samanta2013Elasto-inertial} first studied experimentally the EIT phenomenon in a viscoelastic pipe flow at a high polymer concentration of 500 ppm polyacrylamide solution, and found that the flow became turbulent at the same $Re$ ($\approx 800$) regardless of the perturbation level. Even though this non-hysteresis behaviour can be considered to support a supercritical transition (see also the discussion in \cite{Garg2018Viscoelastic}), the authors noted that a subcritical transition mechanism could not be ruled out, because it is difficult to reduce the disturbance in experiments to a quite low level for the possible subcriticality to manifest itself in this flow (indicating that the viscoelastic pipe flows may be very sensitive to disturbance). When the concentration is lower, \cite{Samanta2013Elasto-inertial} found a clear hysteresis loop for 100 ppm solutions and the flow is Newtonian-like, signifying a subcritical transition. The occurrence of such hysteresis seems to be dependent on the polymer concentration (related to viscosity ratio $\beta$), which implies that there may exist a boundary between supercritical and subcritical transitions according to this parameter. In a recent experiment of pipe flows of 600 ppm polymer solutions \citep{Choueiri2021Experimental}, the measured pressure fluctuation amplitude $p_m$ grows continuously with increasing $Re$ near the instability onset, following a scaling of $p_m \propto \sqrt{Re-Re_c}$. Even though this continuous change is also suggestive of a supercritical transition, the authors warned the reader that other scaling relations may exist, considering experimental uncertainties. On the other hand, they revealed a disordered chevron pattern flow at $Re\sim 5$ while the lowest $Re_c$ predicted by the linear stability theory is of $O(80)$. This observation, along with the weakly chaotic fluid motion in the chevron pattern, is believed to be consistent with a subcritical scenario. However, due to experimental uncertainties, it is difficult to confirm the existence of this subcriticality. More coordinated comparisons between experiments and theoretical analyses need to be conducted in order to dispel the doubts on parameter choices and experimental uncertainties. 

The current situation calls for a systematic investigation of the flow bifurcation in viscoelastic pipe flows from the perspective of governing equations. The possible existence of both subcriticality and supercriticality in a large parameter space in the reviewed experiments and numerical simulations implies a bifurcation boundary in viscoelastic pipe flows. This view seems to be supported also by the experimental observations in \cite{Chandra2020Early} where they suggested a possible crossover of the transition type from subcritical at low polymer concentrations to supercritical at high polymer concentrations in microtubes. We are motivated by these recent studies to conduct weakly nonlinear analyses of viscoelastic pipe flows to distinguish the two types of transition and identify the boundary between them. 

\subsection{The position of the current work}

As reviewed above, there is currently no theoretical work studying the bifurcation mechanism near linear critical conditions in viscoelastic pipe flows; both supercritical and subcritical transitions have been observed in experiments where disturbances are likely of finite amplitude. It is difficult to distinguish between the genuine supercritical bifurcation and \Dongdong{the subcritical bifurcation that is very sensitive to system-level disturbances (this kind of subcritical bifurcation may appear to be supercritical bifurcation in experiments as the laboratory background noise makes it difficult to differentiate the two, see Appendix \ref{Appendix_subsuper} for an illustration)}. Such difficulty necessitates a systematic theoretical investigation. In the current work, we will perform a weakly nonlinear analysis of axisymmetric viscoelastic pipe flows (of Oldroyd-B fluids) based on multiple-scale expansion around the linear critical point (damped mode was investigated in \cite{Meulenbroek2003Intrinsic}). The significance of the current work lies in (1) supplementing the works by Morozov \& van Saarloos on the weakly nonlinear stability analysis of viscoelastic plane shear flows; (2) extending the works by \cite{Garg2018Viscoelastic} and \cite{Chaudhary2021Linear} to the study of the weakly nonlinear phase of the linearly unstable centre mode; and (3) providing a theoretical investigation of the bifurcation types in viscoelastic pipe flows in a large parameter space to understand and reconcile the experimental observations and numerical results in the studies reviewed above.

The rest of the paper is organized as follows. Section \ref{problemformulation} introduces the governing equation in two different formulations, the non-dimensional control parameters, the multiple-scale expansion method in the weakly nonlinear framework and the resulting Ginzburg-Landau equation. In section \ref{numericalmethod}, the numerical method used to evaluate the Landau coefficients is briefly presented. We show the results in section \ref{results}, including a validation step, neutral curves in linear stability analysis, effects of nonlinearity in weakly nonlinear stability analysis, bifurcation type of the flow and a scaling law of the Landau coefficient. We conclude the paper in section \ref{Conclusions} with some discussions on the results. In the five appendices, we provide more information on the bifurcation types, the linear/nonlinear operators in the weakly nonlinear stability theory, validation by a DNS method and more results on the UCM pipe flows.

\section{Problem formulation}\label{problemformulation}

\subsection{Governing equations and parameters}

We consider viscoelastic fluids in a circular pipe with density $\rho^{*}$, dynamic viscosity of the solvent $\mu_{s}^{*}$ and the additional dynamic viscosity $\mu_{p}^{*}$ due to polymers where ${}^*$ is used to indicate dimensional quantities. The governing equations are the incompressible Navier-Stokes equation and the constitutive equation modelling the polymer dynamics. The characteristic scales used to nondimensionalise the governing equations include: the pipe radius $R^{*}$ as the length scale, the centreline velocity of the laminar flow $U_{c}^{*}$ as the velocity scale, $\rho^* U_{c}^{*2}$ as the reference pressure and $\mu_{p}^{*} U_{c}^{*}/R^{*}$ for the polymeric shear stress. Then using a hat to denote a nondimensionalised variable, we can write the nondimensional continuity and Navier-Stokes equations as
\Dongdong{
\begin{equation}\label{eq:continuity-NS}
\boldsymbol{\nabla} \cdot \hat{\bu} = 0, \ \ \ \ \ \ \ \ \
\frac{\partial \hat{\bu}}{\partial t} + \hat{\bu} \cdot \boldsymbol{\nabla}  \hat{\bu} = - \boldsymbol{\nabla} \hat{p} + \frac{\beta}{Re} \nabla^{2} \hat{\bu} + \frac{1-\beta}{Re} \boldsymbol{\nabla} \cdot \hat{\boldsymbol{\tau}}_{p},
\end{equation}}
where $\hat{\bu} = (\hat{u}_{r}, \hat{u}_{\theta}, \hat{u}_{z})$ is the velocity vector with the subscripts $r$, $\theta$ and $z$ representing the radial, azimuthal and axial directions, respectively; $\hat{p}$ is the pressure; $\hat{\boldsymbol{\tau}}_{p}$ is the polymeric stress tensor, to be discussed below. The Reynolds number is defined as $Re = \rho^{*} U_{c}^{*} R^{*} / \mu^{*}$, and the viscosity ratio as $\beta = \mu_{s}^{*} / \mu^{*}$ with $\mu^{*} = \mu_{s}^{*} + \mu_{p}^{*}$ being the total dynamic viscosity.

To model the polymeric stress $\hat{\boldsymbol{\tau}}_{p}$, the Oldroyd-B model  \citep{Bird1987Dynamics} is adopted in the present study. In this model, polymer chains are treated as non-interacting dumbbells with two beads connected with Hookean springs, described by an end-to-end vector $\boldsymbol{q}^{*}$. Coarse-graining modelling introduces the conformation tensor $\bc^{*}=\langle \boldsymbol{q}^{*}\boldsymbol{q}^{*}\rangle$ to characterize the configuration of the polymers where $\langle\cdot\rangle$ denotes ensemble average. The polymeric stress can then be expressed as $\boldsymbol{\tau}_{p}^{*} = \left( \bc^{*} - \bc^{*eq} \right)\mu_{p}^{*} H^{*}/(\lambda^{*} k_{B}^{*} T^{*})$ where $\lambda^{*}$ is the polymer relaxation time, $k_{B}^{*}$ the Boltzmann constant, $T^{*}$ the absolute temperature and $H^{*}$ the spring constant. Normalizing $\bc^{*}$ by $k_{B}^{*}T^{*}/H^{*}$ results in the nondimensional expression $\hat{\boldsymbol{\tau}}_{p} = (\hat{\bc}-\boldsymbol{I})/{Wi}$ where $\boldsymbol{I}$ denotes the identity matrix, corresponding to the equilibrium state $\bc^{*eq}$, and the Weissenberg number is defined as $Wi = \lambda^{*} U_{c}^{*} /R^{*}$, quantifying the polymer relaxation time to the flow turn-over time. Consequently, the evolution equation of conformation tensor $\hat{\bc}$ which has six components $(\hat{c}_{rr}, \hat{c}_{r\theta}, \hat{c}_{rz}, \hat{c}_{\theta\theta}, \hat{c}_{\theta z}, \hat{c}_{zz})$ in nondimensional form reads
\Dongdong{
\begin{equation}\label{eq:conformation-tensor}
\frac{\partial \hat{\bc}}{\partial t} + \hat{\bu} \cdot \boldsymbol{\nabla} \hat{\bc} - \hat{\bc} \cdot (\boldsymbol{\nabla} \hat{\bu}) - (\boldsymbol{\nabla} \hat{\bu})^{T} \cdot \hat{\bc} = - \hat{\boldsymbol{\tau}}_{p}.
\end{equation}}
Despite the simplicity of the Oldroyd-B model, it has been reported to be able to reliably reproduce the purely elastic instabilities observed in experiments of viscometric flows \citep{Shaqfeh1996Purely} and discover the centre-mode linear instability in the stability analysis of viscoelastic pipe flows \citep{Garg2018Viscoelastic}. This model has been used in many previous works on the stability analyses of viscoelastic flows \citep{Sureshkumar1995Linear,Morozov2007,Zhang2013Linear,Morozov2019Subcritical,Khalid2021Centre}.

The governing equations \eqref{eq:continuity-NS} and \eqref{eq:conformation-tensor} admit a steady solution and it will hereafter be referred to as the laminar base flow. To perform the stability analysis, the state variables are decomposed as \Dongdong{$\hat{\bu} = \boldsymbol{U} + \bu$, $\hat{p} = P + p$ and $\hat{\bc} = \boldsymbol{C} + \bc$ } where the uppercase variables correspond to profiles of the laminar base flow and those lowercase variables without hat symbols denote the perturbations (note that the Reynolds decomposition of $\hat{\bc}$ is formally presented here, but a geometric decomposition will be introduced later for $\hat{\bc}$ in a different formulation of the same problem). By substituting this decomposition into equations \eqref{eq:continuity-NS} and \eqref{eq:conformation-tensor} and then subtracting equations for the laminar base flow, we can obtain the following nonlinear evolution equations for perturbations
\begin{subequations}\label{eq:disturbance-continuity-NS-conf}
\begin{equation}
\boldsymbol{\nabla}  \cdot \bu = 0, \ \ \ \ \ \ \ \frac{\partial \bu}{\partial t} + \bu \cdot \boldsymbol{\nabla} \boldsymbol{U} + \boldsymbol{U} \cdot \boldsymbol{\nabla} \bu  + \boldsymbol{N}_{\bu} = - \boldsymbol{\nabla} p + \frac{\beta}{Re} \nabla^{2} \bu + \frac{1-\beta}{ReWi} \boldsymbol{\nabla} \cdot \bc,
\end{equation}
\begin{equation}
\frac{\partial \bc}{\partial t} + \bu \cdot \boldsymbol{\nabla} \boldsymbol{C} - \bc \cdot (\boldsymbol{\nabla} \boldsymbol{U}) - (\boldsymbol{\nabla} \bu)^{T} \cdot \boldsymbol{C} + \boldsymbol{U} \cdot \boldsymbol{\nabla} \bc - \boldsymbol{C} \cdot (\boldsymbol{\nabla} \bu) - (\boldsymbol{\nabla} \boldsymbol{U})^{T} \cdot \bc  + \boldsymbol{N}_{\bc} = - \frac{\bc}{Wi},
\end{equation}
\end{subequations}
where the nonlinear terms are $\boldsymbol{N}_{\bu} =  \bu \cdot \boldsymbol{\nabla} \bu$, $\boldsymbol{N}_{\bc} =  \bu \cdot \boldsymbol{\nabla} \bc - \bc \cdot (\boldsymbol{\nabla} \bu) - (\boldsymbol{\nabla}\bu)^{T} \cdot \bc$, and the laminar base flow profiles are
\begin{subequations}
\begin{align}
&U_r=U_{\theta}=0, \ \ \ U_z = 1-r^2, \\ &C_{rr}=C_{\theta\theta}=1, \ \ C_{r\theta}=C_{\theta z} = 0, \ \  C_{rz}=WiU_z', \ \ C_{zz}=1+2Wi^2 U_z'^2,
\end{align}
\end{subequations}
where the symbol prime $'$ denotes the derivative with respect to $r$.

In this work, we will focus on extending the centre-mode instability as discovered by \cite{Garg2018Viscoelastic}. \Dongdong{Till now, only the axisymmetric mode in viscoelastic pipe flows have been found to be linearly unstable in the literature \citep{Garg2018Viscoelastic,Chaudhary2021Linear,Zhang2021}; therefore, it is legitimate to focus on the weakly nonlinear development of the axisymmetric mode in this work. } For the axisymmetric viscoelastic pipe flow, the component-wise governing equations can be obtained from equation \eqref{eq:disturbance-continuity-NS-conf} as
\begin{subequations}\label{eq:up-c}
\begin{equation}
\frac{\partial u_{r}}{\partial r} + \frac{u_{r}}{r} +\frac{\partial u_{z}}{\partial z} = 0,
\end{equation}
\begin{equation}\label{eq:ur-p}
\frac{\partial u_{r}}{\partial t} + U_{z} \frac{\partial u_{r}}{\partial z} + N_{u_{r}} = - \frac{\partial p}{\partial r} + \frac{\beta}{Re} \left( \nabla^{2}u_{r} - \frac{u_{r}}{r^{2}} \right) + \frac{1-\beta}{ReWi} \left( \frac{\partial c_{rr}}{\partial r} + \frac{c_{rr}}{r} - \frac{c_{\theta \theta}}{r} + \frac{\partial c_{rz}}{\partial z} \right),
\end{equation}
\begin{equation}\label{eq:uz-p}
\frac{\partial u_{z}}{\partial t} + U_{z} \frac{\partial u_{z}}{\partial z} + u_{r} U_{z}' + N_{u_{z}} = - \frac{\partial p}{\partial z} + \frac{\beta}{Re} \left( \nabla^{2}u_{z} \right) + \frac{1-\beta}{ReWi} \left( \frac{\partial c_{rz}}{\partial r} + \frac{c_{rz}}{r} + \frac{\partial c_{zz}}{\partial z} \right),
\end{equation}
\begin{equation}\label{eq:crr-u}
\frac{\partial c_{rr}}{\partial t} + u_{r} C_{rr}' + U_{z} \frac{\partial c_{rr}}{\partial z} - 2 C_{rr}\frac{\partial u_{r}}{\partial r} - 2 C_{rz}\frac{\partial u_{r}}{\partial z} + N_{c_{rr}} = - \frac{c_{rr}}{Wi},
\end{equation}
\begin{equation}\label{eq:crz-u}
\frac{\partial c_{rz}}{\partial t} + u_{r} C_{rz}' + U_{z} \frac{\partial c_{rz}}{\partial z} - c_{rr} U_{z}' - C_{rr}\frac{\partial u_{z}}{\partial r} - C_{rz}\frac{\partial u_{z}}{\partial z} - C_{rz}\frac{\partial u_{r}}{\partial r} - C_{zz}\frac{\partial u_{r}}{\partial z} + N_{c_{rz}} = - \frac{c_{rz}}{Wi},
\end{equation}
\begin{equation}\label{eq:ctt-u}
\frac{\partial c_{\theta \theta}}{\partial t} + u_{r} C_{\theta \theta}' + U_{z} \frac{\partial c_{\theta \theta}}{\partial z} - 2 C_{\theta \theta}\frac{u_{r}}{r} + N_{c_{\theta \theta}} = - \frac{c_{\theta \theta}}{Wi},
\end{equation}
\begin{equation}\label{eq:czz-u}
\frac{\partial c_{zz}}{\partial t} + u_{r} C_{zz}' + U_{z} \frac{\partial c_{zz}}{\partial z} - 2 c_{rz} U_{z}' - 2 C_{rz}\frac{\partial u_{z}}{\partial r} - 2 C_{zz}\frac{\partial u_{z}}{\partial z} + N_{c_{zz}} = - \frac{c_{zz}}{Wi},
\end{equation}
\end{subequations}
where $\nabla^{2}=\frac{\partial^2}{\partial r^2} + \frac{1}{r}\frac{ \partial}{\partial r} + \frac{\partial^2}{\partial z^2}$ and the nonlinear terms are given as
\begin{subeqnarray}
N_{u_{r}} &=& u_{r} \frac{\partial u_{r}}{\partial r} + u_{z} \frac{\partial u_{r}}{\partial z}, \ \ \ \
N_{u_{z}} = u_{r} \frac{\partial u_{z}}{\partial r} + u_{z} \frac{\partial u_{z}}{\partial z}, \\
N_{c_{rr}} &=& u_{r} \frac{\partial c_{rr}}{\partial r} + u_{z} \frac{\partial c_{rr}}{\partial z} - 2 c_{rr} \frac{\partial u_{r}}{\partial r} - 2 c_{rz} \frac{\partial u_{r}}{\partial z}, \\
N_{c_{rz}} &=& u_{r} \frac{\partial c_{rz}}{\partial r} + u_{z} \frac{\partial c_{rz}}{\partial z} - c_{rr} \frac{\partial u_{z}}{\partial r} - c_{rz} \frac{\partial u_{z}}{\partial z} - c_{rz} \frac{\partial u_{r}}{\partial r} - c_{zz} \frac{\partial u_{r}}{\partial z}, \\
N_{c_{\theta \theta}} &=& u_{r} \frac{\partial c_{\theta \theta}}{\partial r} + u_{z} \frac{\partial c_{\theta \theta}}{\partial z} - \frac{2 u_{r}}{r} c_{\theta \theta},\\
N_{c_{zz}} &=& u_{r} \frac{\partial c_{zz}}{\partial r} + u_{z} \frac{\partial c_{zz}}{\partial z} - 2 c_{rz} \frac{\partial u_{z}}{\partial r} - 2 c_{zz} \frac{\partial u_{z}}{\partial z}.
\end{subeqnarray}
No-slip boundary conditions at the pipe wall for the velocity components are $u_{r}(1)=0$, $u_{z}(1)=0$, and we do not need to specify boundary conditions for the conformation tensor components at the wall. At the pipe axis, the conditions for both $\boldsymbol{u}$ and $\boldsymbol{c}$ are specified by the parity conditions as will be presented in Section \ref{numericalmethod}. 

The above formulation in equation \eqref{eq:up-c} having explicit pressure terms is referred to as the $up$-$c$ formulation hereafter. The traditional Reynolds decomposition has been applied in this formulation to the conformation tensor. However, this decomposition may jeopardise the positive definiteness of the conformation tensor $\hat{\bc}$, leading to non-physical results. In the current work, we will also work with another formulation, i.e., the geometric decomposition proposed by \cite{Hameduddin2018Geometric,Hameduddin2019Perturbative} and \cite{Hameduddin2019b} to decompose $\hat{\bc}$. Another advantage of this formulation is that the elastic energy can be defined with a more physical significance (see the discussions on the inner product below). The perturbation $\bc$ is expressed as $\bc = \boldsymbol{F} \cdot \boldsymbol{g} \cdot \boldsymbol{F}^{T}$ where $\boldsymbol{F}$ is a deformation gradient tensor and is given by the Cholesky decomposition of the conformation tensor of the base flow as $\boldsymbol{F} \cdot \boldsymbol{F}^{T}=\boldsymbol{C}$; the perturbation tensor $\boldsymbol{g}$ can be interpreted as the polymer deformation \textit{relative to} the mean configuration. By substituting the Cholesky decomposition into $\bc = \boldsymbol{F} \cdot \boldsymbol{g} \cdot \boldsymbol{F}^{T}$, the explicit expression of $\bc=\boldsymbol{Q}\boldsymbol{g}$ can be obtained as follows (see also \cite{Zhang2021})
\begin{equation}\label{eq:cg_transformation}
\begin{pmatrix}
c_{rr} \\ c_{rz} \\ c_{\theta\theta} \\ c_{zz}
\end{pmatrix} = 
\begin{pmatrix}
C_{rr} & 0 & 0 & 0\\
C_{rz} & S & 0 & 0 \\
0 & 0 & C_{\theta\theta} & 0 \\
C_{rz}^{2}/C_{rr} & 2SC_{rz}/C_{rr} & 0 & S^2/C_{rr}
\end{pmatrix}
\begin{pmatrix}
g_{rr} \\ g_{rz} \\ g_{\theta\theta} \\ g_{zz}
\end{pmatrix},
\end{equation}
where the notation $S=\sqrt{C_{rr}C_{zz}-C_{rz}^2}$ is used. Besides, in the Navier-Stokes equations, we eliminate the pressure term using the continuity condition (so that the inner product of the variable array will bear the significance of the disturbance energy directly). To this end, we use the streamfunction $\phi$, which is related to the velocity components as $u_{r} = - \frac{1}{r} \frac{\partial \phi}{\partial z}, \quad u_{z} = \frac{1}{r} \frac{\partial \phi}{\partial r}$. Instead of using $\phi$ directly for pipe flows, we will work with $\psi = \phi/r$ for the convenience of implementing the boundary conditions at $r = 0$ \citep{Orlandi2012Fluid}. The governing equation of $\psi$ reads
\begin{align}
&\frac{\partial}{\partial t} \left( -\nabla^{2} \psi + \frac{\psi}{r^{2}} \right) = U_{z} \frac{\partial \nabla^{2} \psi}{\partial z} - \left( U_{z}'' - \frac{1}{r} U_{z}' + \frac{1}{r^2} U_{z} \right) \frac{\partial \psi}{\partial z} \label{eqnpsi}  \\
&- \frac{\beta}{Re} \left( \frac{\partial^{4}}{\partial r^{4}} + \frac{2}{r} \frac{\partial^{3}}{\partial r^{3}} - \frac{3}{r^{2}} \frac{\partial^{2}}{\partial r^{2}} + 2\frac{\partial^{4}}{\partial r^{2} \partial z^{2}} + \frac{3}{r^{3}} \frac{\partial}{\partial r} + \frac{2}{r} \frac{\partial^{3}}{\partial r \partial z^{2}} - \frac{3}{r^{4}} - \frac{2}{r^{2}} \frac{\partial^{2}}{\partial z^{2}} + \frac{\partial^{4}}{\partial z^{4}} \right) \psi \notag \\
&- \frac{1 - \beta}{ReWi} \left( F(g_{rr}) + F(g_{rz}) + \frac{C_{\theta\theta}}{r} \frac{\partial g_{\theta\theta}}{\partial z} +  \frac{S^2}{C_{rr}} \frac{\partial^2 g_{zz}}{\partial r \partial z} + \left(\frac{S^2}{C_{rr}}\right)' \frac{\partial g_{zz}}{\partial z} \right) + N_{\psi}, \notag
\end{align}\label{eq:psi-c}
where the nonlinear term is $N_{\psi} = \frac{\partial N_{u_{z}}}{\partial r} - \frac{\partial N_{u_{r}}}{\partial z}$; \Dongdong{explicit expressions of the short notations $F(g_{rr})$ and $F(g_{rz})$, along with the governing equations of $(g_{rr},g_{rz},g_{\theta\theta},g_{zz})$, can be found in Appendix \ref{Appendix_psig}}; the boundary conditions for $\psi$ at the pipe wall are $\psi(1)=0$, $\frac{\partial \psi}{\partial r}|(1)=0$. This new equation system involving $\psi$ and $\boldsymbol{g}$ is denoted as the $\psi$-$g$ formulation.

A general compact form of the equation systems in the two formulations is
\begin{equation}\label{eq:compact-form}
\boldsymbol{M} \frac{\partial \boldsymbol{\gamma}}{\partial t} = \boldsymbol{L} \boldsymbol{\gamma} + \boldsymbol{N}, \ \  \text{or} \ \ \left( \boldsymbol{M} \frac{\partial }{\partial t} - \boldsymbol{L} \right) \boldsymbol{\gamma} = \boldsymbol{N},
\end{equation}
where $\boldsymbol{\gamma}=(u_{r},u_{z},p,c_{rr},c_{rz},c_{\theta\theta},c_{zz})^T$ in the $up$-$c$ formulation, and $\boldsymbol{\gamma}=(\psi,g_{rr},g_{rz},g_{\theta\theta},g_{zz})^T$ in the $\psi$-$g$ formulation; $\boldsymbol{M}$ is the weight matrix, $\boldsymbol{L}$ the linear operator, $\boldsymbol{N}$ the nonlinear operator. The explicit expressions of these operators for the $up$-$c$ formulation are given in Appendix \ref{Appendix_original_operators} and those for the $\psi$-$g$ formulation can be derived in a similar way.

\subsection{Multiple-scale expansion for the weakly nonlinear stability analysis}

There are different ways to conduct the multiple-scale expansion of disturbances and equations. The specific expansion scheme below follows our previous work \cite{Zhang2016Weakly} (which followed and adapted the methods of \cite{Stewartson1971Non-linear} and \cite{Fujimura1989Equivalence}) and expands the time $t$, the spatial coordinate in the streamwise direction $z$, the disturbance $\bgamma$ and the governing parameter $Re$ as function of a small dimensionless parameter $\epsilon$
\begin{subeqnarray}\label{eq:expansion-tzgammaRe}
\frac{\partial}{\partial t} &=& \frac{\partial}{\partial t_{0}} + \epsilon \frac{\partial}{\partial t_{1}} + \epsilon^{2} \frac{\partial}{\partial t_{2}} + O(\epsilon^{3}), \ \ \ \ \frac{\partial}{\partial z} = \frac{\partial}{\partial z_{0}} + \epsilon \frac{\partial}{\partial z_{1}} + O(\epsilon^{2}), \\
\bgamma &=& \epsilon \bgamma_{1} + \epsilon^2 \bgamma_{2} + \epsilon^3 \bgamma_{3} +  O(\epsilon^{4}), \ \ \ \ \ \ \  Re = Re_{c} + \epsilon^{2} + O(\epsilon^{4}).
\end{subeqnarray}
There are three parameters $Re$, $Wi$ and $\beta$, while only $Re$ is perturbed around its linear critical value $Re_{c}$ at a given combination of $\beta$ and $Wi$. The expansion form of $Re$ is due to the parabolic shape of neutral curves around $Re_{c}$, originating from \cite{Stuart1960Non-linear} in a weakly nonlinear analysis of Newtonian plane Poiseuille flows. Specifically, the small quantity $\epsilon$ can be considered as a measure of the distance between $Re$ and the linear critical condition $Re_c$. \Dongdong{It should be mentioned that the definition of the expansion parameter $\epsilon$ is not unique and rescaling of the present $\epsilon = \sqrt{Re-Re_c}$ does not affect the Landau coefficient $a_3$ (derived in equation \eqref{eq:GL-t2 scale} to follow) as long as the computation is restricted to the linear critical conditions. An example illustrating this point can be found in table 1 in \cite{Zhang2016Weakly}, where the Landau coefficient obtained from the weakly nonlinear analysis of Newtonian plane Poiseuille flow agrees well with that in \cite{Fujimura1989Equivalence} where a different expansion scheme of $Re$ was used}. Since there are temporal and spatial derivatives $\partial /\partial t$ and $\partial/\partial z$ in operators as introduced in equation \eqref{eq:compact-form}, these operators also need to be expanded as series of $\epsilon$
\begin{subeqnarray}\label{eq:expansion-MLN}
\boldsymbol{M} &=& \boldsymbol{M}_{0} + \epsilon \boldsymbol{M}_{1} + \epsilon^{2} \boldsymbol{M}_{2} + O(\epsilon^{3}),  \ \ \ \ 
\boldsymbol{L} = \boldsymbol{L}_{0} + \epsilon \boldsymbol{L}_{1} + \epsilon^{2} \boldsymbol{L}_{2} + O(\epsilon^{3}), \\
\boldsymbol{N} &=& \epsilon^{2} \boldsymbol{N}_{2} + \epsilon^{3} \boldsymbol{N}_{3} + O(\epsilon^{4}).
\end{subeqnarray}
The explicit expression of these subscale operators are provided in Appendix \ref{Appendix_expansion_operators} for the $up$-$c$ formulation.

Next, we substitute equations \eqref{eq:expansion-tzgammaRe} and \eqref{eq:expansion-MLN} into equation \eqref{eq:compact-form}, and collect the terms at the same order of $\epsilon$. This step gives the equation at order $\epsilon$
\begin{equation}\label{eq:epsilon1}
\left( \boldsymbol{M}_{0} \frac{\partial}{\partial t_{0}} - \boldsymbol{L}_{0} \right) \bgamma_{1} = 0,
\end{equation}
the equation at order $\epsilon^{2}$
\begin{equation}\label{eq:epsilon2}
\left( \boldsymbol{M}_{0} \frac{\partial}{\partial t_{0}} - \boldsymbol{L}_{0} \right) \bgamma_{2} = \left( \boldsymbol{L}_{1} - \boldsymbol{M}_{1} \frac{\partial}{\partial t_{0}} - \boldsymbol{M}_{0} \frac{\partial}{\partial t_{1}} \right) \bgamma_{1} + \boldsymbol{N}_{2},
\end{equation}
and the equation at order $\epsilon^{3}$
\begin{align}\label{eq:epsilon3}
\left( \boldsymbol{M}_{0} \frac{\partial}{\partial t_{0}} - \boldsymbol{L}_{0} \right) \bgamma_{3} = &\left( \boldsymbol{L}_{2} - \boldsymbol{M}_{2} \frac{\partial}{\partial t_{0}} - \boldsymbol{M}_{1} \frac{\partial}{\partial t_{1}} - \boldsymbol{M}_{0} \frac{\partial}{\partial t_{2}} \right) \bgamma_{1} \notag \\
&+ \left( \boldsymbol{L}_{1} - \boldsymbol{M}_{1} \frac{\partial}{\partial t_{0}} - \boldsymbol{M}_{0} \frac{\partial}{\partial t_{1}} \right) \bgamma_{2} + \boldsymbol{N}_{3}.
\end{align}

At order $\epsilon$, the equation \eqref{eq:epsilon1} is exactly the linearised equation in the framework of linear stability analysis. Its solution can be assumed to take the wave-like form of
\begin{equation}\label{eq:gamma1}
\bgamma_{1}(z_{0},z_{1},r,t_{0},t_{1},t_{2};Re) = A(z_{1},t_{1},t_{2})\tilde{\bgamma}_1(r)e^{i \alpha z_{0} + \mu t_{0}} + c.c.
\end{equation}
and by substituting \eqref{eq:gamma1} into equation \eqref{eq:epsilon2}, then the second-order solution can be expressed in the form of
\begin{equation}\label{eq:gamma2}
\bgamma_{2} = A^*A\tilde{\bgamma}_{20}(r)  + \frac{ \partial A}{ \partial z_1}\tilde{\bgamma}_{21}(r) e^{i \alpha z_{0} + \mu t_{0}} + c.c. + A^2 \tilde{\bgamma}_{22}(r) e^{2i \alpha z_{0} + 2\mu t_{0}} + c.c.
\end{equation}
where $c.c.$ stands for the complex conjugate of the preceding term; $i=\sqrt{-1}$ is the imaginary unit; $\alpha$ is the wavenumber; $\mu=-i\omega$  and $\tilde{\bgamma}_1$ are the complex eigenvalue and eigenfunction respectively of the eigenvalue problem $\mu \tilde{\boldsymbol{M}}_{0} \tilde{\bgamma}_1= \tilde{\boldsymbol{L}}_{0} \tilde{\bgamma}_1$ resulting from equation \eqref{eq:epsilon1} (the symbol tilde $\ \tilde{} \ $ denotes matrices/variables in the spectral space); $A$ is the complex amplitude of the disturbance \Dongdong{(the physical meaning of $A$ depends on the normalization of $\tilde{\bgamma}_1$ as will be discussed in the end of this section)}, and its value cannot be determined from the linear equation as the linear problem can be arbitrarily scaled. Instead, the evolution equation of $A$ (i.e. Ginzburg-Landau equation) can be obtained by applying solvability conditions on equations \eqref{eq:epsilon2} and \eqref{eq:epsilon3} at orders $\epsilon^{2}$ and $\epsilon^{3}$ respectively. The aim is to remove the secular terms from the inhomogeneous terms \citep{Bender1999Advanced} and this can be done via adjoint variables. 

To formulate the adjoint problem corresponding to the linear eigenvalue problem \eqref{eq:epsilon1}, we introduce an inner product defined as
\begin{equation}\label{eq:inner-product-definition}
\langle \boldsymbol{a},\boldsymbol{b} \rangle = \frac{1}{2 L_p} \int_0^{L_p} \int_{0}^{1} \boldsymbol{a} \cdot \boldsymbol{W} \boldsymbol{b} \, rdr \, dz_{0} = \frac{\Dongdong{A_{\boldsymbol{a}}A_{\boldsymbol{b}}^*}}{2}  \int_{0}^{1} \tilde{\boldsymbol{a}} \cdot \boldsymbol{W} {\tilde{\boldsymbol{b}}}^{*} \, rdr + c.c. = \Dongdong{A_{\boldsymbol{a}}A_{\boldsymbol{b}}^*}{\langle \tilde{\boldsymbol{a}}, \tilde{\boldsymbol{b}} \rangle}_{s} + c.c.
\end{equation}
where $L_p=2\pi/\alpha$ is the wavelength; $\boldsymbol{a}$ and $\boldsymbol{b}$ are real-valued vectors in the same form as the linear solution in equation \eqref{eq:gamma1} \Dongdong{($A_{\boldsymbol{a}}$ and $A_{\boldsymbol{b}}$ are the corresponding complex amplitudes)}; $\boldsymbol{W}$ is a real coefficient matrix; ${\langle \tilde{\boldsymbol{a}}, \tilde{\boldsymbol{b}} \rangle}_{s}$ can be seen as the inner product defined in the spectral space. Under this definition, the inner product $\langle \boldsymbol{\gamma}_{1},\boldsymbol{M}_{0} \boldsymbol{\gamma}_{1} \rangle$ in the $\psi$-$g$ formulation represents the total energy (kinetic energy plus elastic energy) of the first order disturbance with the coefficient matrix being specified as $\boldsymbol{W} = diag(I,wI,2wI,wI,wI)$, where $w = (1-\beta)/Re/Wi$. We illustrate this point as follows.

For the kinetic energy part in $\langle \boldsymbol{\gamma}_{1},\boldsymbol{M}_{0} \boldsymbol{\gamma}_{1} \rangle$, we single out $\langle \psi_{1}, (-\nabla^2+\frac{1}{r^2}) \psi_{1} \rangle$ (where the element in $\boldsymbol{M}_{0}$ for $\psi$ is $-\nabla^2+\frac{1}{r^2}$, see equation \eqref{eqnpsi}) and one is able to get
\begin{align} \label{kineticenergy}
E_{k} & = \frac{1}{2 L_p} \int_0^{L_p} \int_0^1 \left(u_{r,1}^2 + u_{z,1}^2 \right) \, rdr \, dz_{0} = \ad{\psi_{1}}{\left(-\nabla^2+\frac{1}{r^2}\right) \psi_{1}},
\end{align}
which calculates the disturbance kinetic energy. The elastic energy part in 
$\langle \boldsymbol{\gamma}_1,\boldsymbol{M}_{0}\boldsymbol{\gamma}_1 \rangle$ can be directly written as
\begin{equation}\label{elastic_inner}
E_e =\frac{1}{2 L_p} \int_0^{L_p} \int_{0}^{1} \left( \frac{1-\beta}{Re Wi}(g_{rr,1}^{2}+2g_{rz,1}^{2}+g_{\theta\theta, 1}^{2}+g_{zz,1}^2) \right) \, rdr \, dz_{0},
\end{equation}
where we have used the same concept of geodesic distance as in the calculation of elastic energy proposed by \cite{Hameduddin2018Geometric,Hameduddin2019Perturbative}. Thus, the inner product defined with the coefficient matrix $\boldsymbol{W}$ as above bears the physical meaning of total energy. Nevertheless, we find that the results in Section \ref{results} (especially the value of the Landau coefficient $a_3$ in equation \eqref{eq:Landau-coefficients}) do not depend on the specific choice of the coefficient matrix. Therefore, in the $up$-$c$ formulation, for simplicity, we use an identity matrix as the coefficient matrix, resulting in
\begin{equation}\label{upc_inner}
\langle \boldsymbol{\gamma}_1,\boldsymbol{M}_{0}\boldsymbol{\gamma}_1 \rangle =\frac{1}{2 L_p} \int_0^{L_p} \int_{0}^{1} ( u_{r,1}^{2}+u_{z,1}^{2}+c_{rr,1}^{2}+c_{rz,1}^{2}+c_{\theta\theta,1}^{2}+c_{zz,1}^{2} ) \, rdr \, dz_{0}.
\end{equation}
 
With the inner product defined, through integration by parts, we can obtain the adjoint equation \citep{Luchini2014Adjoint}
\begin{equation}\label{eq:adjoint}
\left\langle \left( \boldsymbol{M}_{0} \frac{\partial}{\partial t_{0}} - \boldsymbol{L}_{0} \right) \bgamma_{1}, \bgamma_{1}^{\dagger} \right\rangle = \left\langle \bgamma_{1}, \left( \boldsymbol{M}_{0}^{\dagger} \frac{\partial}{\partial t_{0}} - \boldsymbol{L}_{0}^{\dagger} \right) \bgamma_{1}^{\dagger} \right\rangle, \ \ \ \rightarrow \ \ \ \left( \boldsymbol{M}_{0}^{\dagger} \frac{\partial}{\partial t_{0}} - \boldsymbol{L}_{0}^{\dagger} \right) \bgamma_{1}^{\dagger} = 0,
\end{equation}
where the symbol dagger $\dagger$ denotes adjoint variables and operators; explicit expressions of the adjoint operators $\boldsymbol{M}_{0}^{\dagger}$ and $\boldsymbol{L}_{0}^{\dagger}$ can be found in Appendix \ref{Appendix_adjoint_operators}. The boundary conditions for the adjoint problem are the same as those for the direct problem: at the pipe wall, the adjoint velocity components $u_{r,1}^{\dagger}(1)=0$, $u_{z,1}^{\dagger}(1)=0$, and there is no need to specify boundary conditions for the adjoint conformation tensor components; at the pipe axis, parity conditions are applied to $\boldsymbol{u}_1^{\dagger}$ and $\boldsymbol{c}_1^{\dagger}$. As mentioned earlier, in the inhomogeneous equations \eqref{eq:epsilon2} and \eqref{eq:epsilon3}, the secular terms---terms related to the wave component $e^{i \alpha z_{0} + \mu t_{0}}$ and its complex conjugate---on the right hand side may cause the solution to blow up and thus the solvability conditions need to be enforced, dictating the projection of the inhomogeneous terms of equations \eqref{eq:epsilon2} and \eqref{eq:epsilon3} on the adjoint waves $\bgamma_{1}^{\dagger}$ to be zero, i.e.,
\begin{subequations}\label{eq:solvability}
\begin{equation}\label{eq:solvability_1}
\left\langle \left( \boldsymbol{L}_{1} - \boldsymbol{M}_{1} \frac{\partial}{\partial t_{0}} - \boldsymbol{M}_{0} \frac{\partial}{\partial t_{1}} \right)  \bgamma_{1}  + \boldsymbol{N}_2, \bgamma_{1}^{\dagger} \right\rangle = 0, \\
\end{equation}
\begin{equation}\label{eq:solvability_2}
\left\langle \left( \boldsymbol{L}_{2} - \boldsymbol{M}_{2} \frac{\partial}{\partial t_{0}} - \boldsymbol{M}_{1} \frac{\partial}{\partial t_{1}} - \boldsymbol{M}_{0} \frac{\partial}{\partial t_{2}} \right) \bgamma_{1} + \left( \boldsymbol{L}_{1} - \boldsymbol{M}_{1} \frac{\partial}{\partial t_{0}} - \boldsymbol{M}_{0} \frac{\partial}{\partial t_{1}} \right) \bgamma_{2} + \boldsymbol{N}_3, \bgamma_{1}^{\dagger} \right\rangle = 0.
\end{equation}
\end{subequations}
 
By singling out the terms related to the wave component $e^{i \alpha z_{0} + \mu t_{0}}$ and its complex conjugate, and expressing the above projection in the spectral space, the following equation with group velocity $c_{g}$ can be derived from equation \eqref{eq:solvability_1}
\begin{equation}\label{eq:group_velocity}
\frac{\partial A}{\partial t_{1}} + c_{g}\frac{\partial A}{\partial z_{1}} = 0 \ \ \ \ \text{with} \ \ \ \ c_{g} = -\frac{ {\langle (\tilde{\boldsymbol{L}}_{1}^{\circ} - \mu_{c}\tilde{\boldsymbol{M}}_{1}^{\circ} )\tilde{\bgamma}_{1}, \tilde{\bgamma}_{1}^{\dagger} \rangle}_{s}}{{\langle \tilde{\boldsymbol{M}}_{0}\tilde{\bgamma}_1, \tilde{\bgamma}_{1}^{\dagger} \rangle}_{s}},
\end{equation}
where $\mu_{c}=\mu_{c,r} + i \mu_{c,i}$ is the eigenvalue at the critical condition $(Re_c, \alpha_{c})$, i.e., $\mu_{c,r}$ (bearing the meaning of growth rate) should be zero theoretically or a very small value numerically. The superscript $^\circ$ means that the $\partial/\partial z_{1}$-related derivatives and the $A$-related coefficients have been moved out of the corresponding operators (see Appendix \ref{Appendix_expansion_operators} for details in the $up$-$c$ formulation).  A Ginzburg-Landau equation (GLE) governing the evolution of the complex amplitude $A$ can be derived from equation \eqref{eq:solvability_2}
\begin{equation}\label{eq:GL-t2 scale}
\frac{\partial A}{\partial t_{2}} = a_{1}A + a_{2}\frac{\partial^{2} A}{\partial z_{1}^{2}} + a_{3}{|A|}^{2}A,
\end{equation}
where the three coefficients can be expressed as
\begin{subeqnarray}\label{eq:Landau-coefficients}
a_{1} &=& \frac{{\langle \tilde{\boldsymbol{L}}_{Re}\tilde{\bgamma}_1, \tilde{\bgamma}_{1}^{\dagger} \rangle}_{s}}{{\langle \tilde{\boldsymbol{M}}_{0}\tilde{\bgamma}_1, \tilde{\bgamma}_{1}^{\dagger} \rangle}_{s}}, \\
a_{2} &=& \frac{{\langle (\tilde{\boldsymbol{L}}_{2}^{\circ} - \mu_{c}\tilde{\boldsymbol{M}}_{2}^{\circ} - c_{g}\tilde{\boldsymbol{M}}_{1}^{\circ})\tilde{\bgamma}_{1} + (\tilde{\boldsymbol{L}}_{1}^{\circ} - \mu_{c}\tilde{\boldsymbol{M}}_{1}^{\circ} - c_{g}\tilde{\boldsymbol{M}}_{0}^{\circ})\tilde{\bgamma}_{21}, \tilde{\bgamma}_{1}^{\dagger} \rangle}_{s}}{{\langle \tilde{\boldsymbol{M}}_{0}\tilde{\bgamma}_1, \tilde{\bgamma}_{1}^{\dagger} \rangle}_{s}}, \\
a_{3} &=& \frac{{\langle \tilde{\boldsymbol{N}}_{3}^{\circ}, \tilde{\bgamma}_{1}^{\dagger} \rangle}_{s}}{{\langle \tilde{\boldsymbol{M}}_{0}\tilde{\bgamma}_1, \tilde{\bgamma}_{1}^{\dagger} \rangle}_{s}}.
\end{subeqnarray}
Among the above coefficients, the complex Landau coefficient $a_{3}=a_{3,r}+i a_{3,i}$ is of particular interest in this study as the sign of its real part can \Dongdong{reveal the bifurcation type of the flow studied (i.e., the bifurcation is supercritical if $a_{3,r}<0$) and subcritical if $a_{3,r}>0$) as long as the amplitudes of the higher-order coefficients do not increase significantly fast with right signs.  
The present GLE is truncated to the third order (up to $a_3$). The expansion (even to higher orders) is known to be well convergent \textit{a priori} as long as the calculation is performed near the linear critical condition \citep{Herbert1980,Sen1983}, which will be followed here. Besides, in the present study, we have also used DNS to  determine the value of $a_3$ and the DNS results validate the bifurcation type (see Appendix \ref{Appendix_DNS_validation}), suggesting that the present truncation of GLE up to third order is reasonable. We mention in passing that even though we focus solely on the significance of $a_3$ in this paper, the application of the GLE to our flow can be general (see \cite{Cross1993}). } Another issue to note is that, to avoid ambiguity, $\tilde{\bgamma}_1$ should be normalized so that the Landau coefficient $a_3$ can be uniquely determined. In the current work, we normalize $\tilde{\bgamma}_1$ so that \Dongdong{
 \begin{equation}\label{eq:total_energy1}
 \sqrt{\frac{1}{2} \int_{0}^{1} \left( \left(|\tilde{u}_{r,1}|^2 + |\tilde{u}_{z,1}|^2 \right)  +  \frac{1-\beta}{Re Wi}  \left( |\tilde{g}_{rr,1}|^{2}+2|\tilde{g}_{rz,1}|^{2}+|\tilde{g}_{\theta\theta,1}|^{2}+|\tilde{g}_{zz,1}|^2\right) \right) \, rdr} = 1.
 \end{equation}
With this normalization, the physical meaning of $|A|$ in equation \eqref{eq:gamma1} is the square root of the total energy (kinetic energy plus elastic energy) in both formulations (conversion is needed where different variables are used).}

\section{Numerical method}\label{numericalmethod}

A spectral collocation method has been adopted to solve the linear equation \eqref{eq:epsilon1} at order $\epsilon$, equation \eqref{eq:epsilon2} at order $\epsilon^{2}$ for the second-order solution $\bgamma_{2}$ and the adjoint equation \eqref{eq:adjoint}, as well as to evaluate the Landau coefficients in equation \eqref{eq:Landau-coefficients}. To enforce the no-slip boundary condition for velocity at the pipe wall, we remove the corresponding rows and columns of the node $r=1$ in the matrices. We apply no boundary conditions for the conformation tensor $\boldsymbol{c}$ or $\boldsymbol{g}$ at the pipe wall. As for the singularity problem at $r=0$, we avoid placing a node at $r=0$ \citep{Mohseni2000} and the derivative matrices are constructed with an even-odd property following \cite{Trefethen2000Spectral}. In our axisymmetric pipe flows, in the $up$-$c$ formulation, $u_{r}$, $c_{rz}$ are odd functions and $u_z$, $p$, $c_{rr}$, $c_{\theta\theta}$, $c_{zz}$ are even; in the $\psi$-$g$ formulation, $\psi$, $g_{rz}$ are odd and $g_{rr}$, $g_{\theta\theta}$, $g_{zz}$ are even. The same parity conditions are applied to the corresponding adjoint variables. The present weakly nonlinear code is adapted from the code used by \cite{Zhang2016Weakly}, which has been validated by comparing with the results in the literature. The adaptation is mainly for implementing the constitutive governing equations for polymers, while the main framework of the code remains unchanged.

\section{Results and Discussions} \label{results}
\subsection{Validation}\label{validation}

\begin{figure}
	\centering
	\includegraphics[width=0.52\textwidth,trim= 10 0 30 0,clip]{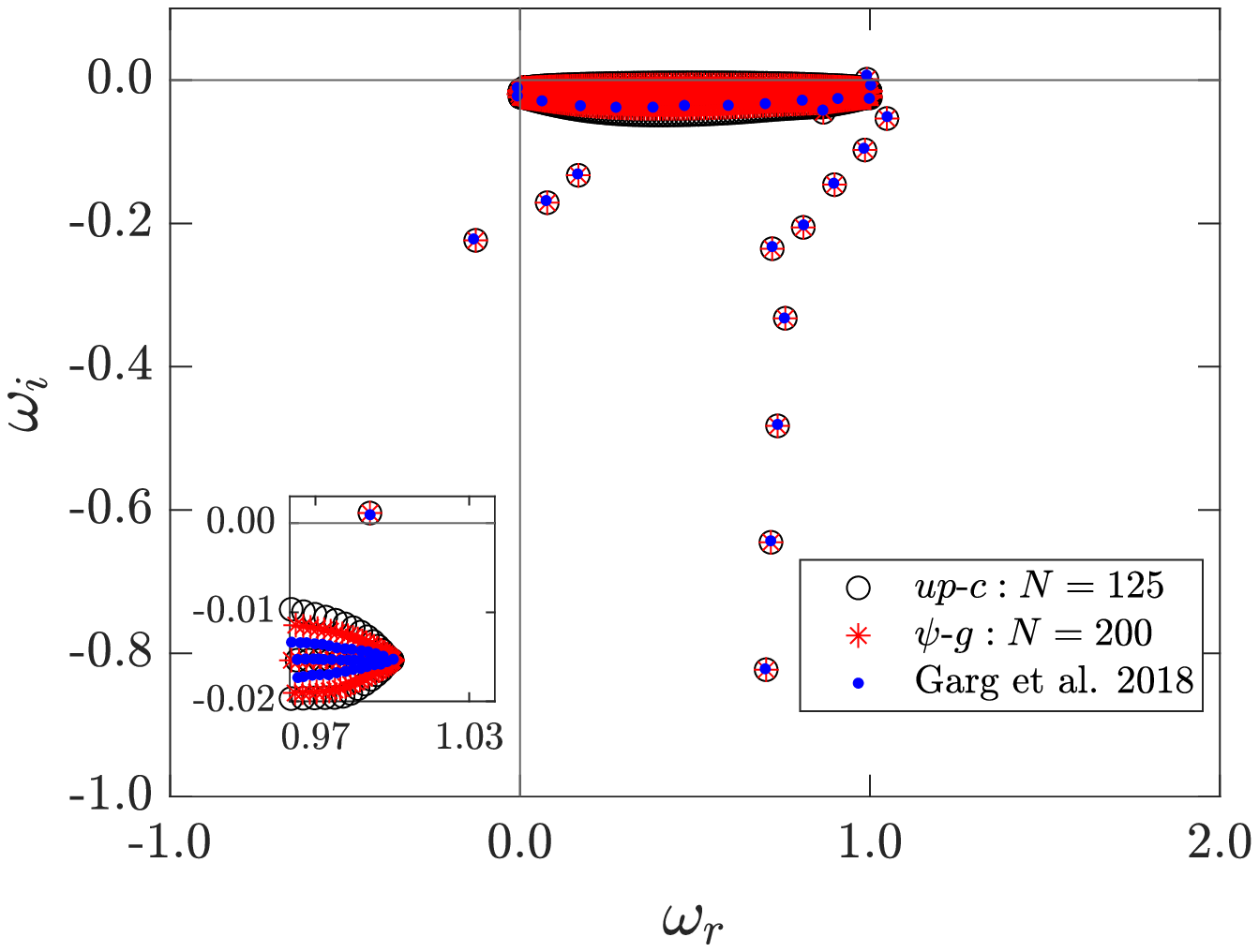}
	\put(-204,145){$(a)$}
	\includegraphics[width=0.47\textwidth,trim= 20 0 60 0,clip]{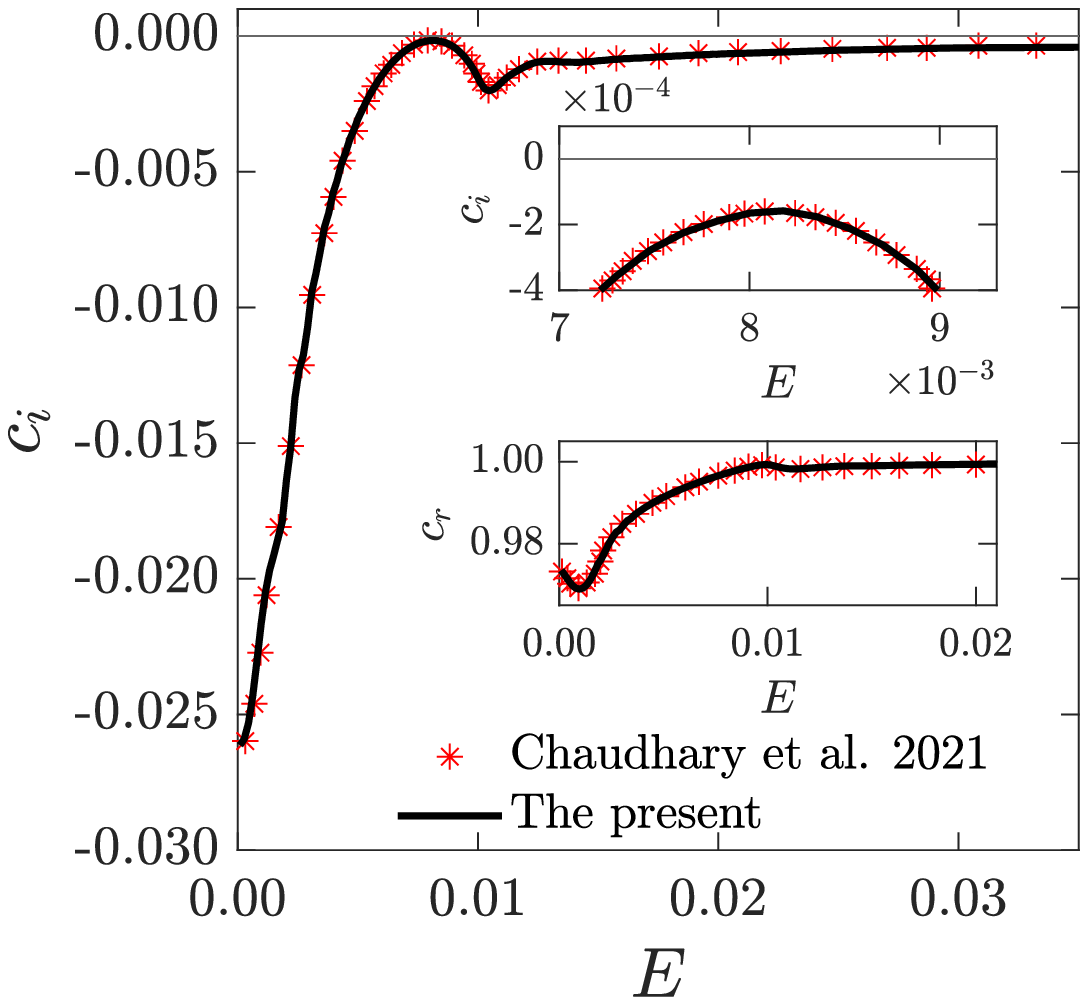}
	\put(-180,145){$(b)$}
	\caption{$(a)$ Comparison of the eigenspectra of viscoelastic pipe flow of Oldroyd-B fluids at $Re=800$, $\alpha=1$, $Wi=65$, $\beta=0.65$; the inset is an enlarged view near the unstable mode. $(b)$ Effect of $E$ on the eigenvalue of the least stable centre mode ($c=\omega/\alpha$ is used to facilitate the comparison and $c_i$ is the imaginary part and $c_r$ the real part) for UCM fluids at $Re=6000$ and $\alpha=2$.}
	\label{Fig:Garg_Chaudhary}
\end{figure}

Linear instability of viscoelastic pipe flow of Oldroyd-B fluids was firstly reported by \cite{Garg2018Viscoelastic}. A typical parameter setting at which they found a single unstable mode is $Re=800$, $\alpha=1$, $Wi=65$ and $\beta=0.65$. We solved the linear eigenvalue problem with the same parameters and validated our code by comparing our results with theirs. Firstly, the convergence of the eigenvalue of the unstable mode has been examined by varying the total number of interior nodes $N$ \Dongdong{within $0<r<1$ } in the spectral collocation method. The results are listed in table \ref{Tab:convergence-Garg} for both the $up$-$c$ formulation and the $\psi$-$g$ formulation. It is clear that, in the $up$-$c$ formulation, $N=125$ gives well-converged results (up to 9-11 decimal numbers), while in the $\psi$-$g$ formulation, it is more difficult to converge to the same accuracy as $up$-$c$ formulation. Secondly, we compare the present eigenspectra of Oldroyd-B pipe flow with that obtained by \cite{Garg2018Viscoelastic} in figure \ref{Fig:Garg_Chaudhary}$(a)$. An overall good agreement can be observed regarding the locations of the continuous spectra and the discrete modes (the data labelled with blue dot were manually digitised from figure 1 in  \cite{Garg2018Viscoelastic}). Moreover, we validate our result of linear stability analysis of UCM pipe flow by comparing with that reported in \cite{Chaudhary2021Linear}. As shown in figure \ref{Fig:Garg_Chaudhary}$(b)$, the variations of the growth rate $c_i$ and phase speed $c_r$ of the least stable mode with elasticity number $E$ at $Re=6000, \alpha=2$ agree very well with theirs.
\begin{table}
	\begin{center}
		\begin{tabular}{c c c}
			$N$ & $\omega$ ($up$-$c$ formulation) & $\omega$ ($\psi$-$g$ formulation) \\
			75 & $0.99126877439+0.00113308966i$ & $0.99126780409+0.00113572338i$\\
			100 & $0.99126860219+0.00113325078i$ & $0.99126893542+0.00113322959i$ \\
			125 & $0.99126859594+0.00113325246i$ & $0.99126917410+0.00113275031i$ \\
			150 & $0.99126859586+0.00113325243i$ &  $0.99126909401+0.00113278275i$ \\
			200 & $0.99126859574+0.00113325247i$ & $0.99126882844+0.00113303075i$ \\
			300 &--- & $0.99126862663+0.00113322344i$ \\
			400 &--- & $0.99126859929+0.00113324918i$ \\
			500 &--- & $0.99126859618+0.00113325201i$ \\
			600 &--- & $0.99126859574+0.00113325236i$ \\
		\end{tabular}
		\caption{Convergence of the eigenvalue of the single unstable mode in two different formulations ($up$-$c$ and $\psi$-$g$) of the viscoelastic pipe flow at $Re=800$, $\alpha=1$, $Wi=65$ and $\beta=0.65$.}
		\label{Tab:convergence-Garg}
	\end{center}
\end{table}

In addition to the convergence check of the unstable mode in the linear eigenvalue problem, we have also examined the convergence of the group velocity $c_g$ in equation \eqref{eq:group_velocity} and the coefficients $a_1$, $a_2$, $a_3$ in equation \eqref{eq:Landau-coefficients}. To calculate these coefficients, the critical parameter needs to be determined first. As will be presented in the next subsection, the neutral curve (see figure \ref{Fig:linear_Wi65beta65}$(a)$ for an example at $Wi=65,\ \beta=0.65$) is in the form of a closed loop and we only consider the critical condition at the left end where transition from laminar to turbulent happens upon increasing $Re$. At $Wi=65,\ \beta=0.65$, the left-end critical parameters are $Re_{c}=265.572335, \ \alpha_{c}=0.515525$. The convergence of the critical eigenvalue $\mu_{c}$, the group velocity $c_{g}$ and the three coefficients $a_{1}$, $a_{2}$ and $a_{3}$ in the GLE are examined using the $\psi$-$g$ formulation as shown in table \ref{Tab:convergence-pipe-psig}. We can see that the resolution $N=400$ is high enough to get well-converged results while $N=200$ also gives results with little loss of accuracy. In addition, the results obtained from the $up$-$c$ formulation are shown in table \ref{Tab:convergence-pipe-upc}. It is clear that $c_{g}$, $a_{1}$, $a_{2}$ and $a_{3}$ agree well with those obtained from the $\psi$-$g$ formulation where the first three to six non-zero digits are identical. In this case, the real part of the Landau coefficient $a_3$ is negative, indicating a supercritical bifurcation at these parameters. There is a residual on the order of $10^{-6} \sim 10^{-7}$ in the imaginary part of the group velocity $c_g$ which is supposed to be a real number, and the reason may be attributed to numerical discretization errors. Such kind of residual has also been reported in weakly nonlinear analyses of other fluid systems (for example, the residual $O(10^{-7})$ in \cite{Kolyshkin2003Stability} and the residual $O(10^{-5})$ in \cite{Gao2013Transition} and \cite{Zhang2016Weakly}). By comparing the inner product $\langle \boldsymbol{\gamma}_{1},\boldsymbol{M}_{0} \boldsymbol{\gamma}_{1} \rangle$ in the $up$-$c$ (equation \ref{upc_inner}) and $\psi$-$g$ (equations \ref{kineticenergy},\ref{elastic_inner}) formulations, one can see that the $\psi$-$g$ formulation has clear physical meanings whereas the $up$-$c$ formulation does not bear any physical significance. Nevertheless, the results of the Landau coefficient are consistent. We have additionally formulated the problem (details not shown in this paper) with variables $u_{r}$, $u_{z}$, $p$, $g_{rr}$, $g_{rz}$, $g_{\theta\theta}$ and $g_{zz}$. All these three formulations give the same results, indicating that the specific definition of the inner product does not affect the values of Landau coefficients in the present weakly nonlinear stability analysis. In the following result section, the $up$-$c$ formulation will be used to generate most of the results.

\begin{table}
	\begin{center}
		\begin{tabular}{c c c c}
			& N=200 & N=400 & N=600 \\
			$\mu_r$ & $-1.352\times{10}^{-7}$ & $-2.176\times{10}^{-10}$ & $8.269\times{10}^{-11}$ \\
			$\mu_i$ & $-0.5114351880$ & $-0.5114350995$ & $-0.5114350986$ \\
			$c_{g}$ & $1.012782769+0.000000240i$ & $1.012781496+0.000000928i$ & $1.012781495+0.000000931i$ \\
			$a_{1}^*$ & $2.1696801+1.7383761i$ & $2.1697084+1.7383569i$ & $2.1697086+1.7383573i$ \\
			$a_{2}$ & $0.033028746-0.015449716i$ & $0.033028180-0.015449435i$ & $0.033028188-0.015449439i$ \\
			$a_{3}$ & $-154.33462+970.50769i$ & $-154.24267+970.55392i$ & $-154.24266+970.55740i$ \\
		\end{tabular}
	\caption{Convergence of the Landau coefficients of the viscoelastic pipe flow of Oldroyd-B fluids at $Wi=65$, $\beta=0.65$, ${Re}_{c}=265.572335$ and $\alpha_{c}=0.515525$ using the $\psi$-$g$ formulation. Note that $a_{1}^* = a_{1} \times {10}^{5}$.}
	\label{Tab:convergence-pipe-psig}
	\end{center}
\end{table}

\begin{table}
	\begin{center}
		\begin{tabular}{c c c c}
			& N=100 & N=150 & N=200 \\
			$\mu_r$ & $-2.829\times{10}^{-12}$ & $-5.236\times{10}^{-12}$ & $-5.234\times{10}^{-12}$ \\
			$\mu_i$ & $-0.511435094989528$ & $-0.511435094988327$ & $-0.511435094988327$ \\
			$c_{g}$ & $1.0128031-0.0000155i$ & $1.0127911-0.0000064i$ & $1.0127869-0.0000032i$ \\
			$a_{1}^*$ & $2.169009+1.738026i$ & $2.169397+1.738209i$ & $2.169533+1.738274i$ \\
			$a_{2}$ & $0.032936-0.015260i$ & $0.033035-0.015490i$ & $0.033033-0.015465i$ \\
			$a_{3}$ & $-154.283+1031.755i$ & $-154.299+1031.817i$ & $-154.302+1031.835i$ \\
		\end{tabular}
		\caption{Convergence of the Landau coefficients of the viscoelastic pipe flow of Oldroyd-B fluids at $Wi=65$, $\beta=0.65$, ${Re}_{c}=265.572335$ and $\alpha_{c}=0.515525$ using the $up$-$c$ formulation. Note that $a_{1}^* = a_{1} \times {10}^{5}$.}
		\label{Tab:convergence-pipe-upc}
	\end{center}
\end{table}

\begin{figure}
	\centering
	\includegraphics[width=0.47\textwidth,trim= 0 0 0 0,clip]{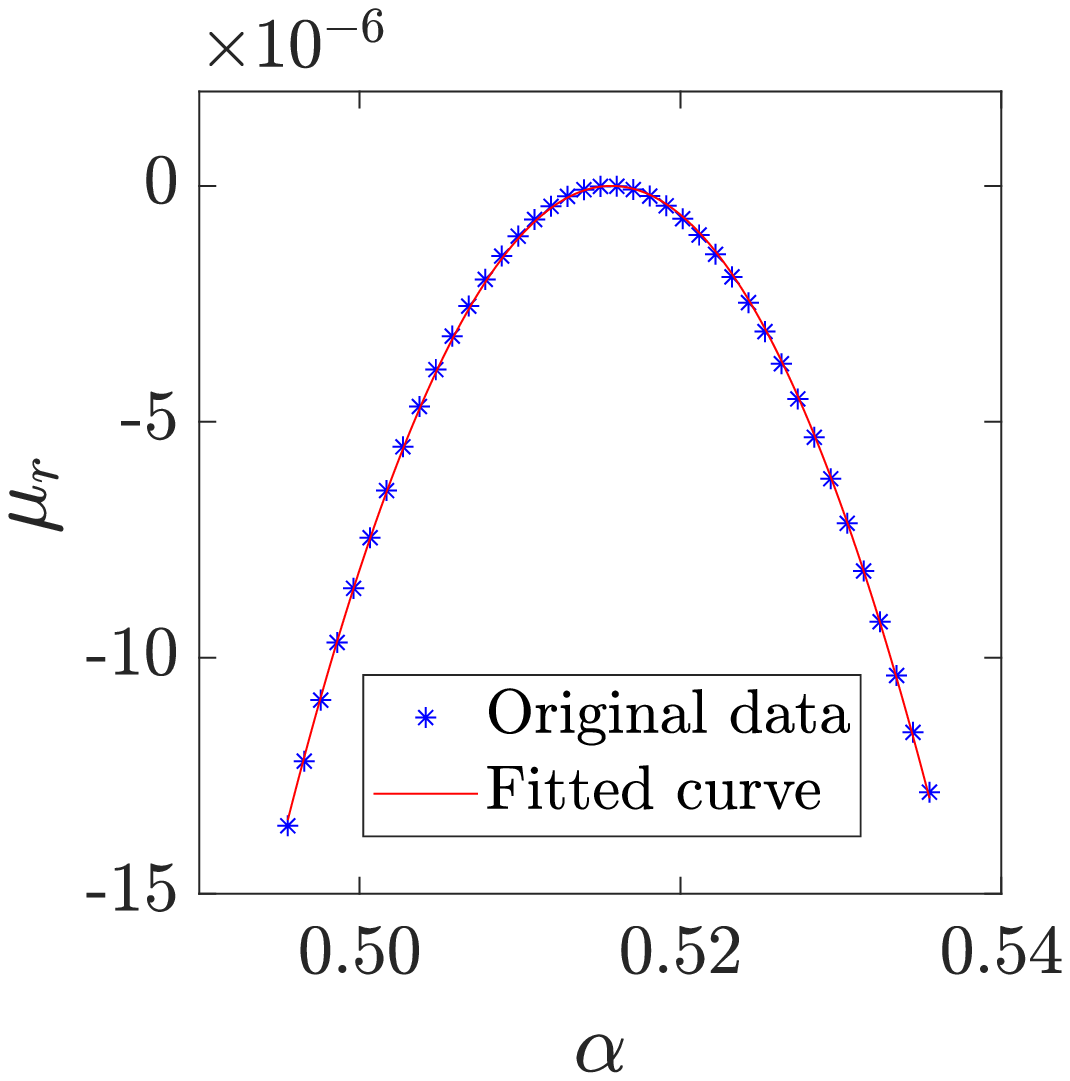}
	\put(-163,120){$(a)$}
	\includegraphics[width=0.47\textwidth,trim= 0 0 0 0,clip]{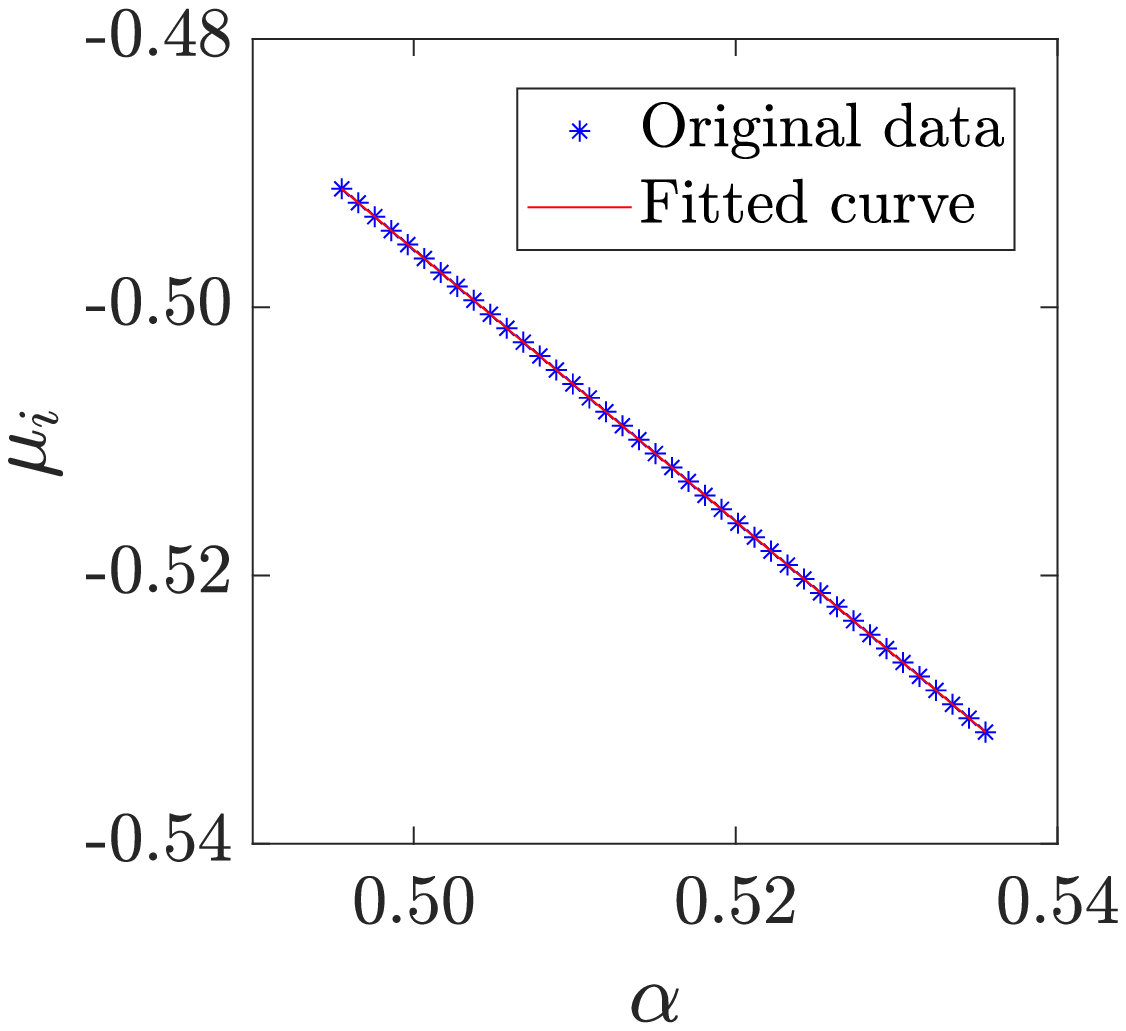}
	\put(-172,120){$(b)$}
	\caption{Dispersion relation at $Re_{c}=265.572335$, $\alpha_{c}=0.515525$, $Wi=65$ and $\beta=0.65$: $(a)$ $\mu_r$ and $(b)$ $\mu_i$ as functions of $\alpha$.}
	\label{Fig:dispersion}
\end{figure}

Next, we validate the group velocity $c_{g}$ and the coefficient $a_{2}$ (listed in tables \ref{Tab:convergence-pipe-psig} and \ref{Tab:convergence-pipe-upc}) in another way by examining the dispersion relation between the eigenvalue $\mu = \mu_{r} + i \mu_{i}$ and the wavenumber $\alpha$ near the critical wavenumber $\alpha_{c}$. Firstly, the growth rate $\mu_{r}(\alpha)$ can be approximated by a parabola function of $\alpha$ around the critical parameter $(Re_{c},\alpha_{c})$ \citep{Huerre1998Hydrodynamic}. In the present case, figure \ref{Fig:dispersion}$(a)$ indeed presents such a parabola and the fitted function is shown in equation \eqref{eq:dispersion-mu_r} below. Secondly, the relationship $c_{g}=-\partial \mu_{i}/ \partial \alpha$ holds at $\alpha_{c}$. To verify this, $\mu_{i}$ as a function of $\alpha$ is plotted in figure \ref{Fig:dispersion}$(b)$ and a second-order polynomial is again used to fit the data (even though the curve looks like a straightline); evaluating $-\partial \mu_{i}/ \partial \alpha$ at $\alpha_{c}$ from function \eqref{eq:dispersion-mu_i} gives $1.012771$ which is quite close to the result $c_{g} = 1.012781 - 0.000001i$ in table \ref{Tab:convergence-pipe-psig} and $c_{g} = 1.012787 - 0.000003i$ in table \ref{Tab:convergence-pipe-upc}.
\begin{subeqnarray}\label{eq:dispersion}
\mu_{r,fit} &=& -0.033015 \alpha^{2} + 0.034052 \alpha - 0.008780, \label{eq:dispersion-mu_r} \\
\mu_{i,fit} &=& 0.015435 \alpha^{2} - 1.028685 \alpha - 0.014776. \label{eq:dispersion-mu_i}
\end{subeqnarray}
Thirdly, following \cite{Stewartson1971Non-linear}, the coefficient $a_{2}$ is related to the eigenvalue $\mu$ in the form of
\begin{equation}\label{eq:a2_mu}
a_{2} = - \frac{1}{2} \frac{\partial^2 \mu_{r}}{\partial \alpha^{2}} - \frac{1}{2} \frac{\partial^2 \mu_{i}}{\partial \alpha^{2}}i.
\end{equation}
Substitution of the fitted functions \eqref{eq:dispersion} into equation \eqref{eq:a2_mu} gives $a_{2,fit}=0.033015-0.015435i$, in good agreement with $a_{2}=0.033028-0.015449i$ listed in table \ref{Tab:convergence-pipe-psig} and $a_{2}=0.033033-0.015465i$ in table \ref{Tab:convergence-pipe-upc}. In Appendix \ref{Appendix_DNS_validation}, we additionally use DNS to qualitatively verify our calculations (in terms of the linear coefficient $a_1$ and Landau coefficient $a_3$)  to confirm that the bifurcation type is correct.

\subsection{Neutral stability curves and linear instability of UCM pipe flows}

\begin{figure}
	\centering
	\includegraphics[width=0.47\textwidth,trim= 20 0 30 0,clip]{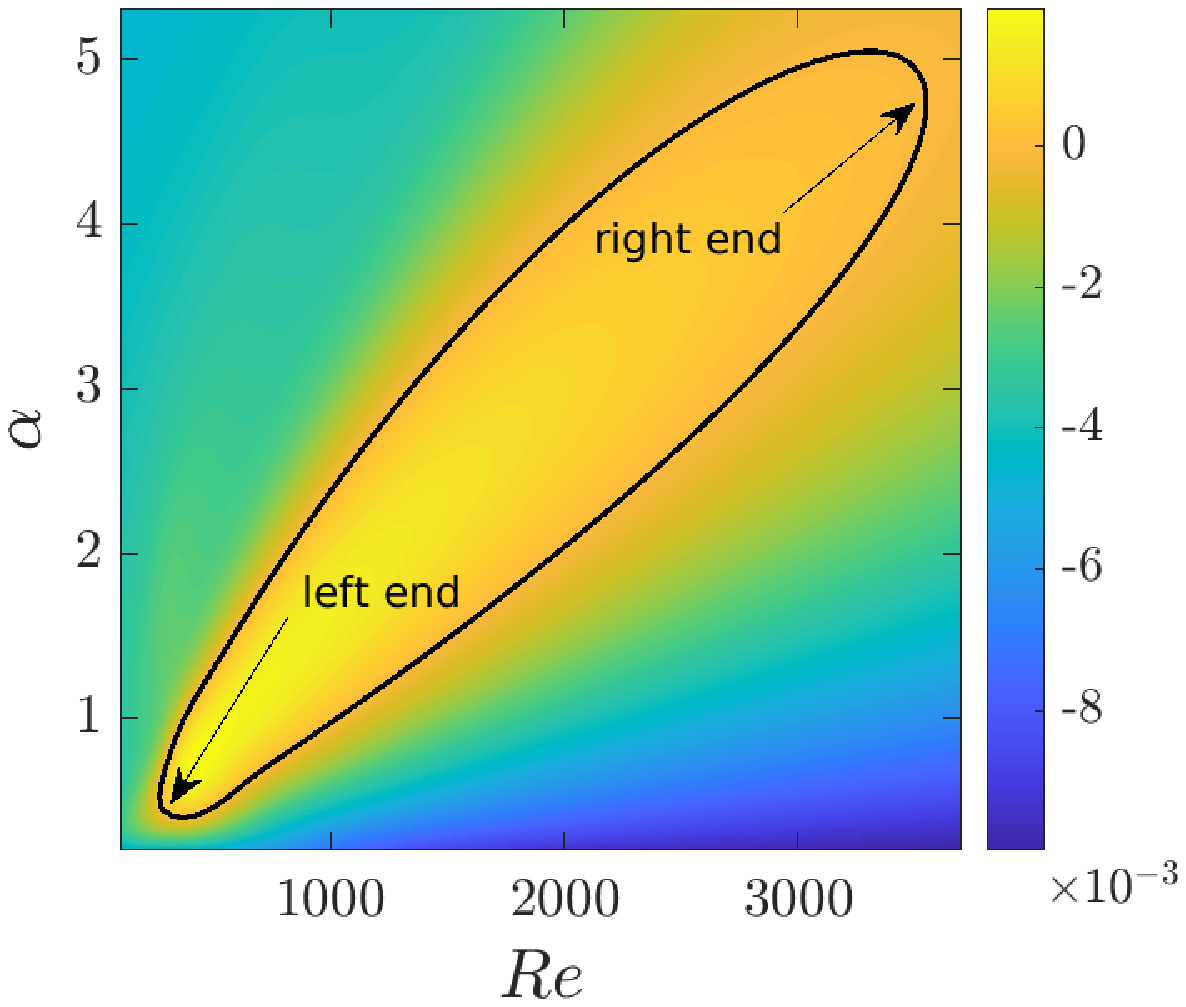}
	\put(-180,140){$(a)$}
	\includegraphics[width=0.47\textwidth,trim= 30 0 20 0,clip]{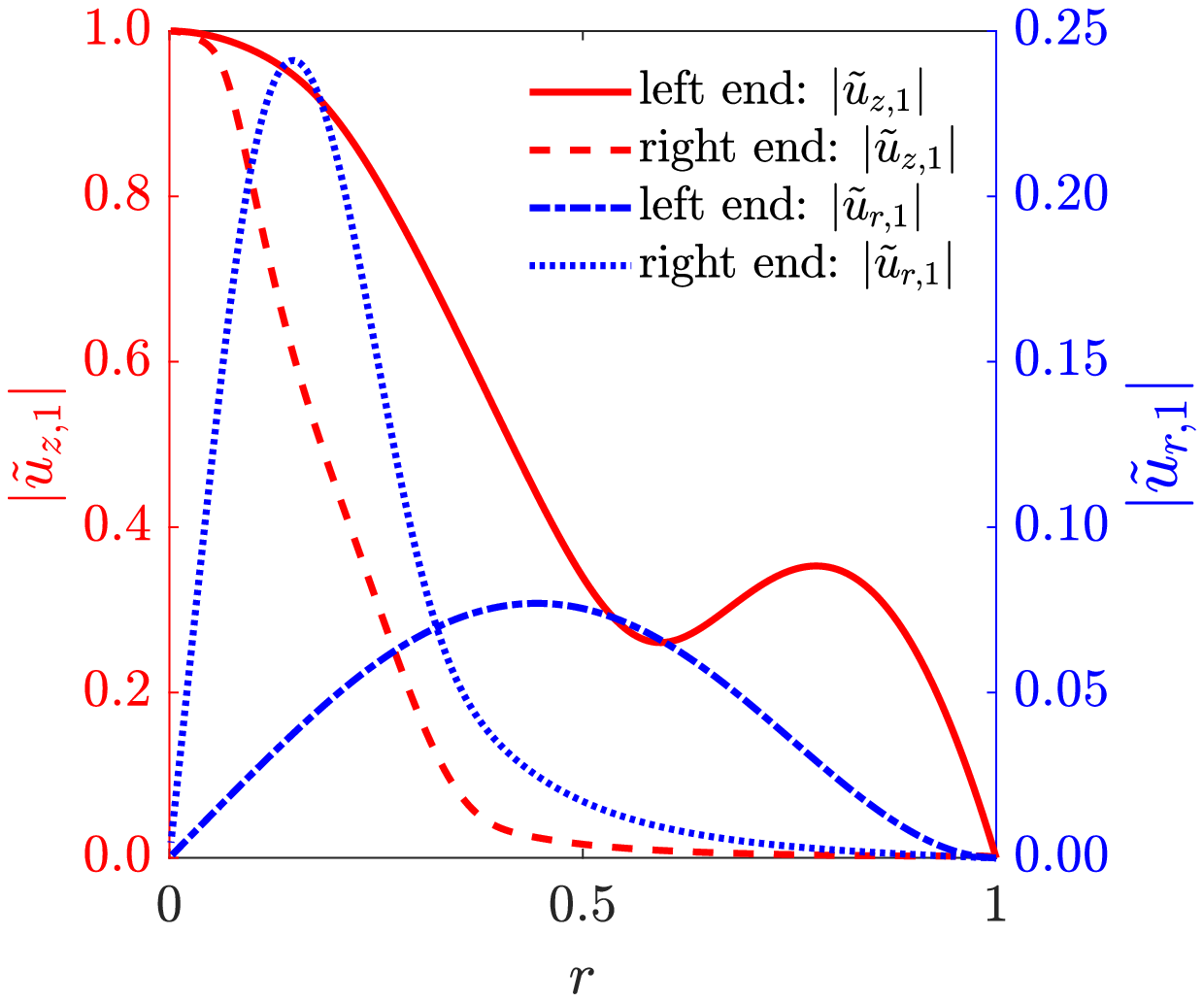}
	\put(-180,140){$(b)$}\\
	\includegraphics[width=0.48\textwidth,trim= 0 0 0 0,clip]{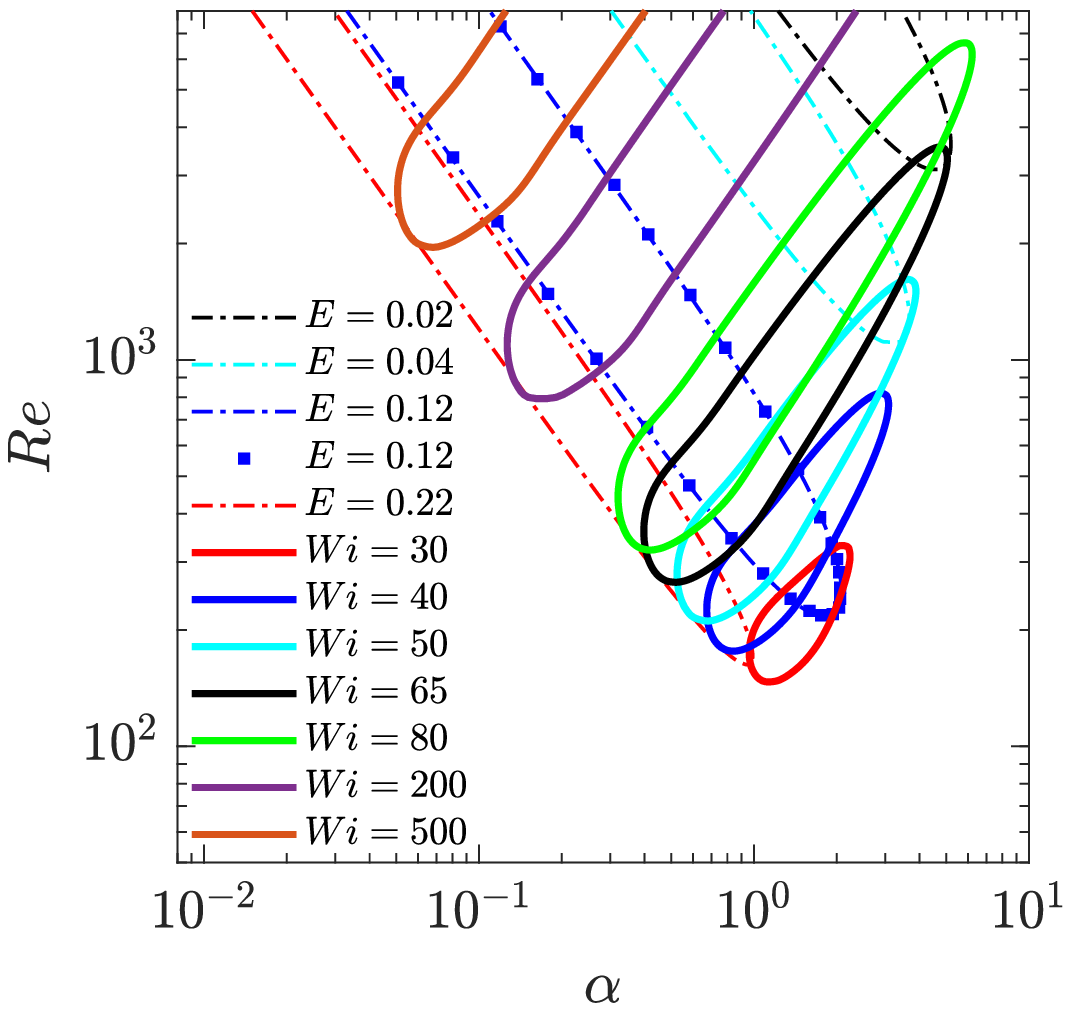}
	\put(-180,160){$(c)$}
	\includegraphics[width=0.48\textwidth,trim= 0 0 0 0,clip]{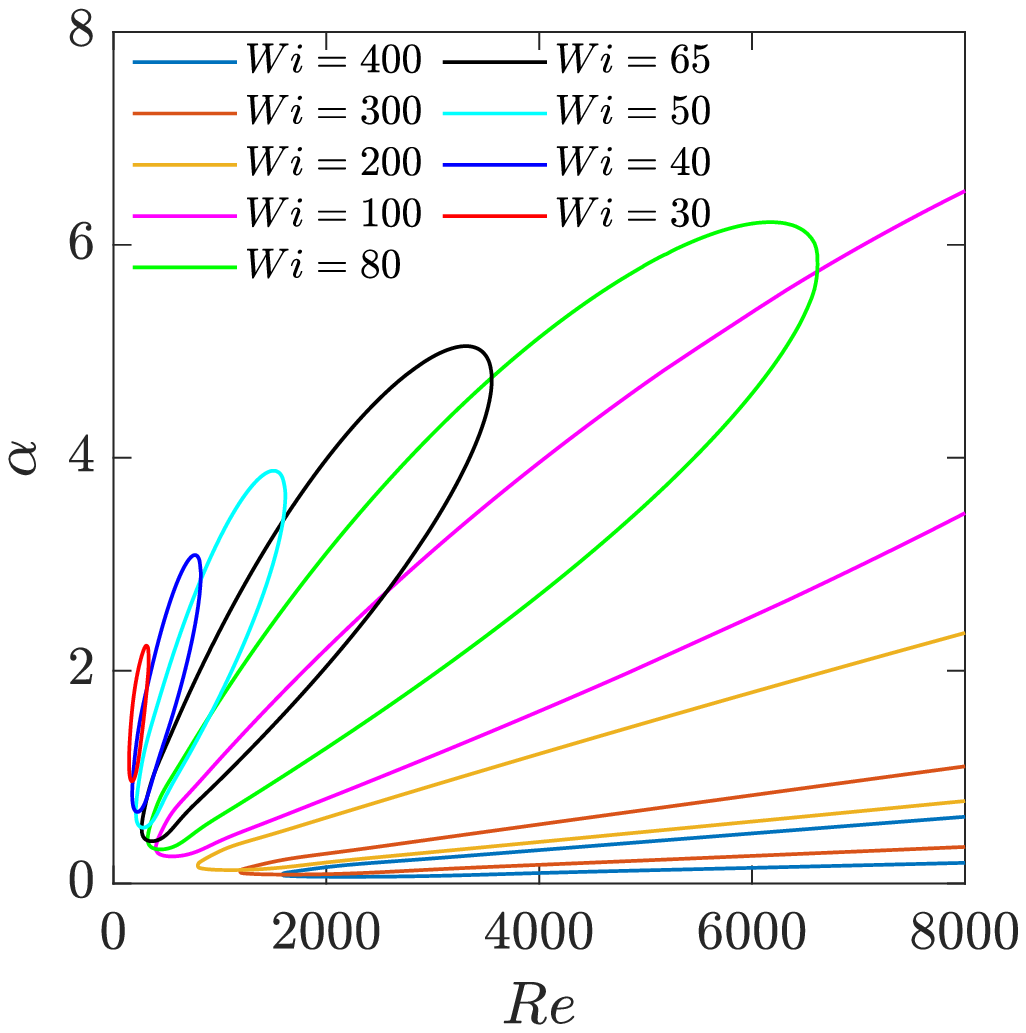}
	\put(-180,160){$(d)$}	
	\caption{$(a)$ Neutral curve in a $(\alpha,Re)$ plane with the colorbar for levels of growth rate and the area inside the loop is linear unstable. $(b)$ Amplitudes of velocity eigenfunctions, where solid and dash-dotted curves are for the left-end critical point $Re_{c}=265.572335$, $\alpha_{c}=0.515525$ and dashed and dotted curves for the right-end critical point $Re_{c}=3553.575$, $\alpha_{c}=4.7337$. $(c)$ Neutral curves in a log-log plot at a fixed viscosity ratio $\beta=0.65$ for various elasticity numbers $E$ and Weissenberg numbers $Wi$, where the discrete data points labelled by blue squares are manually digitised from figure 18$(a)$ in \cite{Chaudhary2021Linear}. $(d)$ Neutral curves in a linear-linear plot at various Weissenberg numbers $Wi$ for $\beta=0.65$.}
	\label{Fig:linear_Wi65beta65}
\end{figure}

Since the weakly nonlinear stability analysis is performed near the critical condition, it is necessary to calculate first the critical conditions and be aware of where the critical conditions are in the parameter space. Thus in this section, we present the neutral curves. In figure \ref{Fig:linear_Wi65beta65}$(a)$, a typical neutral curve at $\beta=0.65$ and $Wi=65$ is plotted, which appears in the form of a closed loop. In the study of viscoelastic channel flows by \cite{Chaudhary2019Elasto-inertial}, the neutral curves also appear in the form of loops with the interior area being linearly unstable. We look at the critical conditions at the two ends in terms of $Re$. The critical Reynolds number and wavenumber are $Re_{c}=265.572335$, $\alpha_{c}=0.515525$ at the left end, and $Re_{c}=3553.575$, $\alpha_{c}=4.7337$ at the right end. Although they are both centre modes, the corresponding eigenfunctions are different, as plotted in figure \ref{Fig:linear_Wi65beta65}$(b)$. It appears that the velocity eigenfunctions corresponding to the right critical point (see the dashed and dotted curves) are more localized near the pipe axis than those of the left one. In the remaining part of the paper, we will restrict our discussion to the left-end critical point, as it is related to the transition from a laminar state to turbulence when we increase $Re$. Figure \ref{Fig:linear_Wi65beta65}$(c)$ shows neutral curves in the $(Re,\alpha)$ plane for varying elasticity numbers $E$ at a fixed viscosity ratio $\beta=0.65$ in a log-log plot, following the same format in \cite{Chaudhary2021Linear}. We present four dot-dashed curves corresponding to $E=0.02,0.04,0.12,0.22$ and superpose the data points for $E=0.12$ (blue square) extracted from \cite{Chaudhary2021Linear} (see their figure 18$(a)$). Our results are in good agreement with theirs. Other cases of $E$ also show good agreement, but are not presented in order to not overload the figure. Similar to figure \ref{Fig:linear_Wi65beta65}$(a)$, the neutral curves can also be plotted at a constant $Wi$, which again exhibit in the form of a closed loop. In figure \ref{Fig:linear_Wi65beta65}$(d)$, the same neutral curves are displayed in a linear-linear plot. The purpose of showing these plots is to figure out where the critical conditions are, so that we can perform weakly nonlinear stability analysis around them.

\begin{figure}
	\centering
	\includegraphics[width=0.33\textwidth,trim=40 5 15 0,clip]{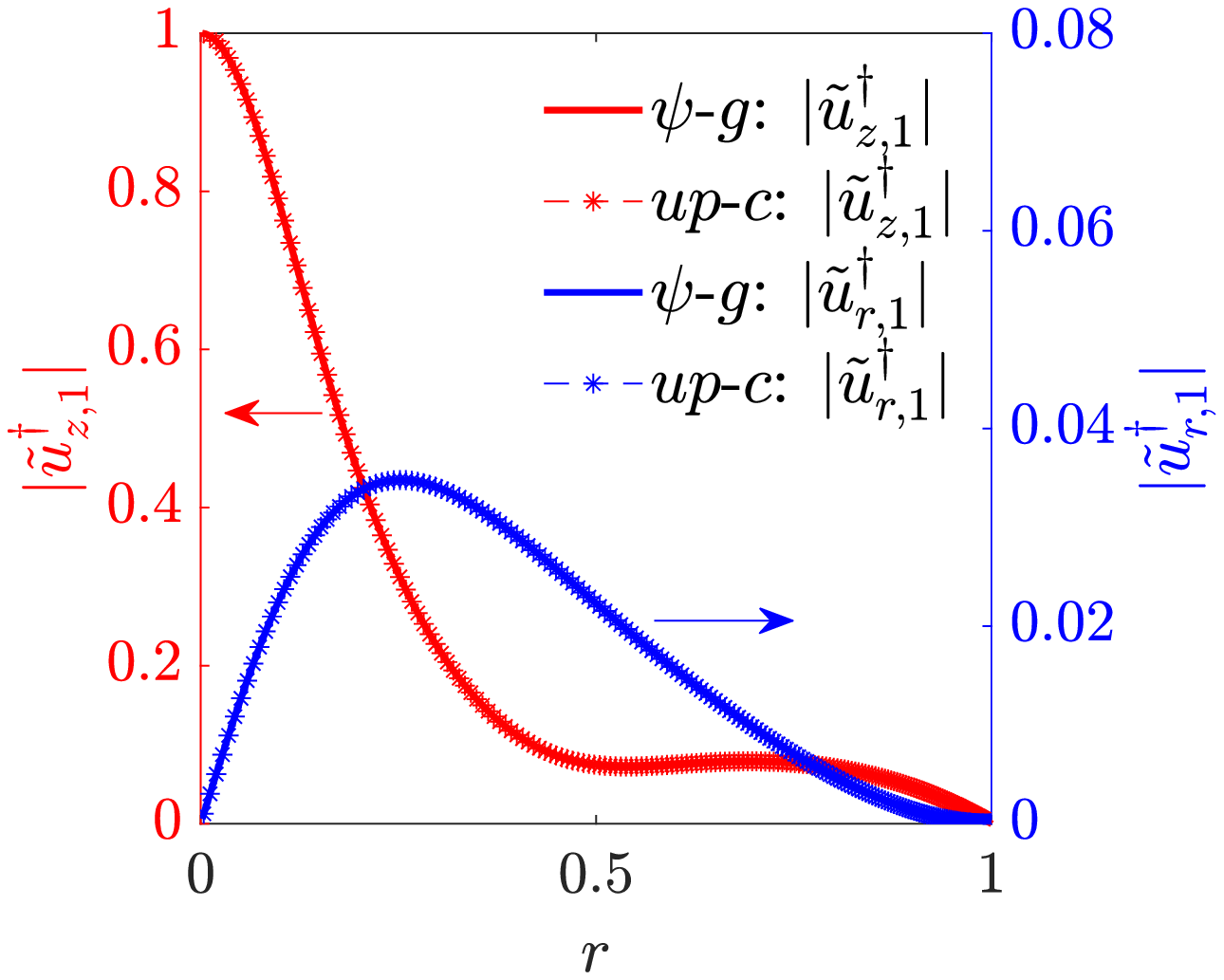}\put(-128,93){$(a)$}
	\includegraphics[width=0.33\textwidth,trim=30 5 30 0,clip]{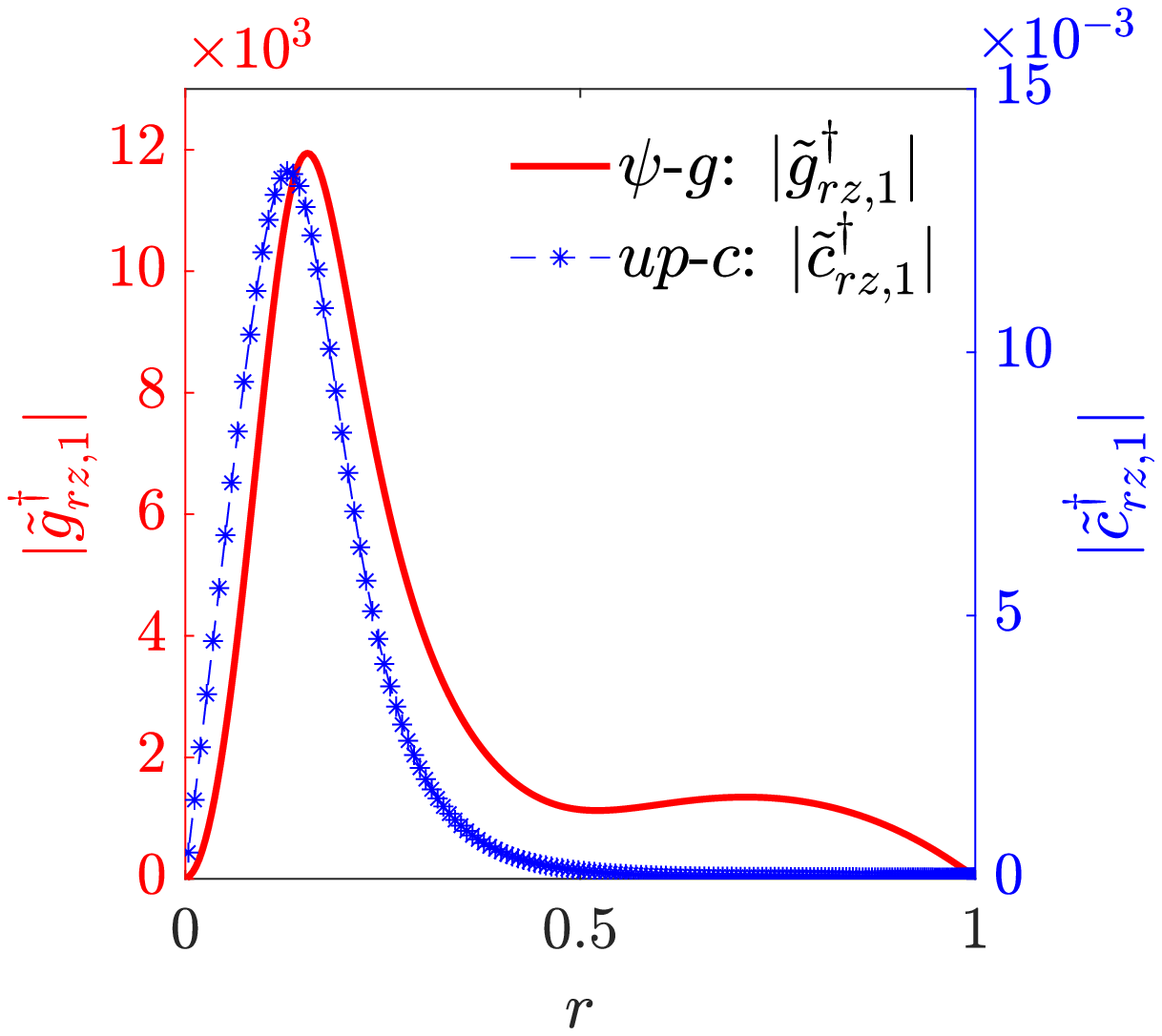}\put(-122,93){$(b)$}
	\includegraphics[width=0.33\textwidth,trim=30 5 30 0,clip]{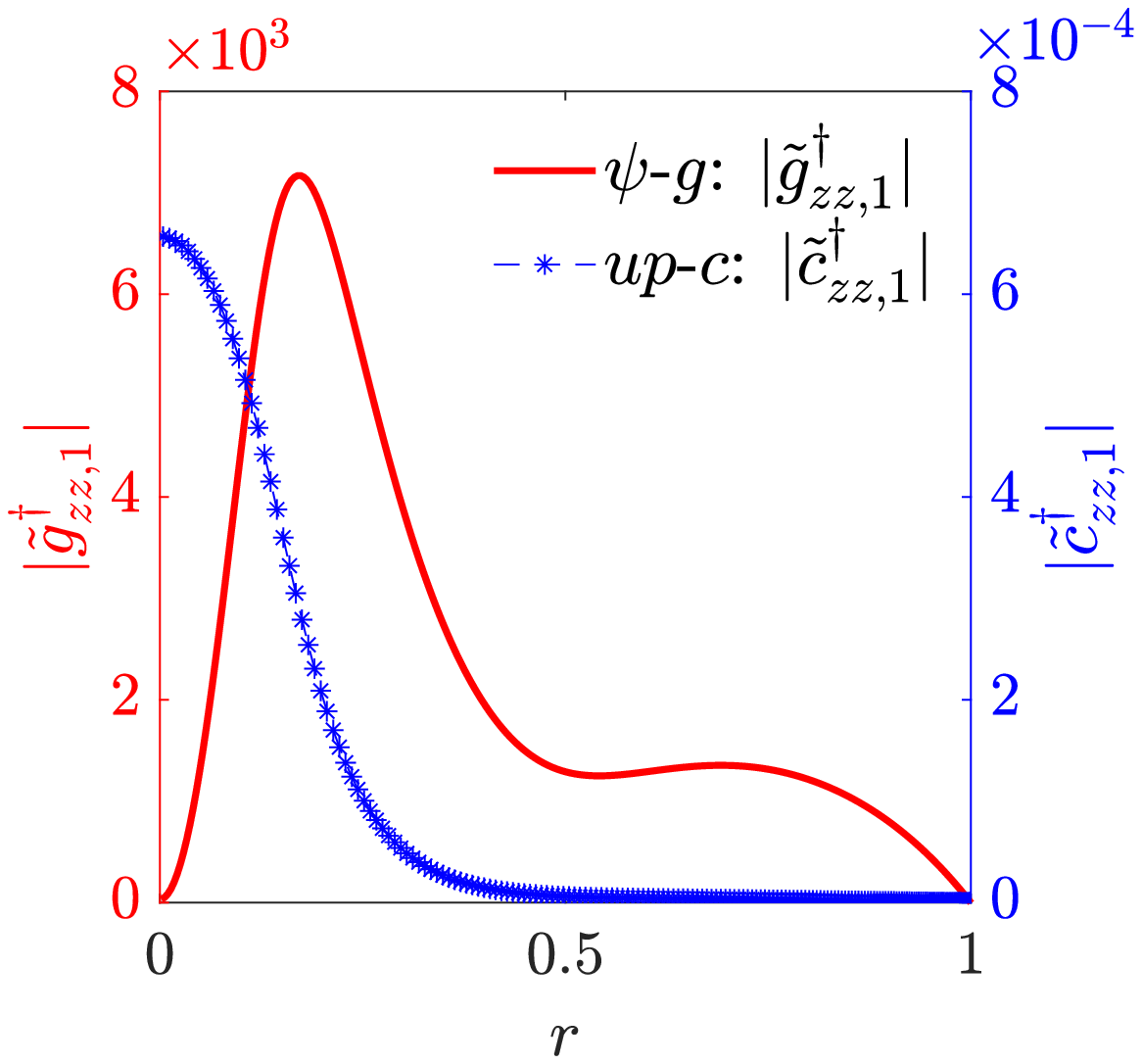}\put(-122,93){$(c)$} \\
	\includegraphics[width=0.33\textwidth,trim=40 5 15 0,clip]{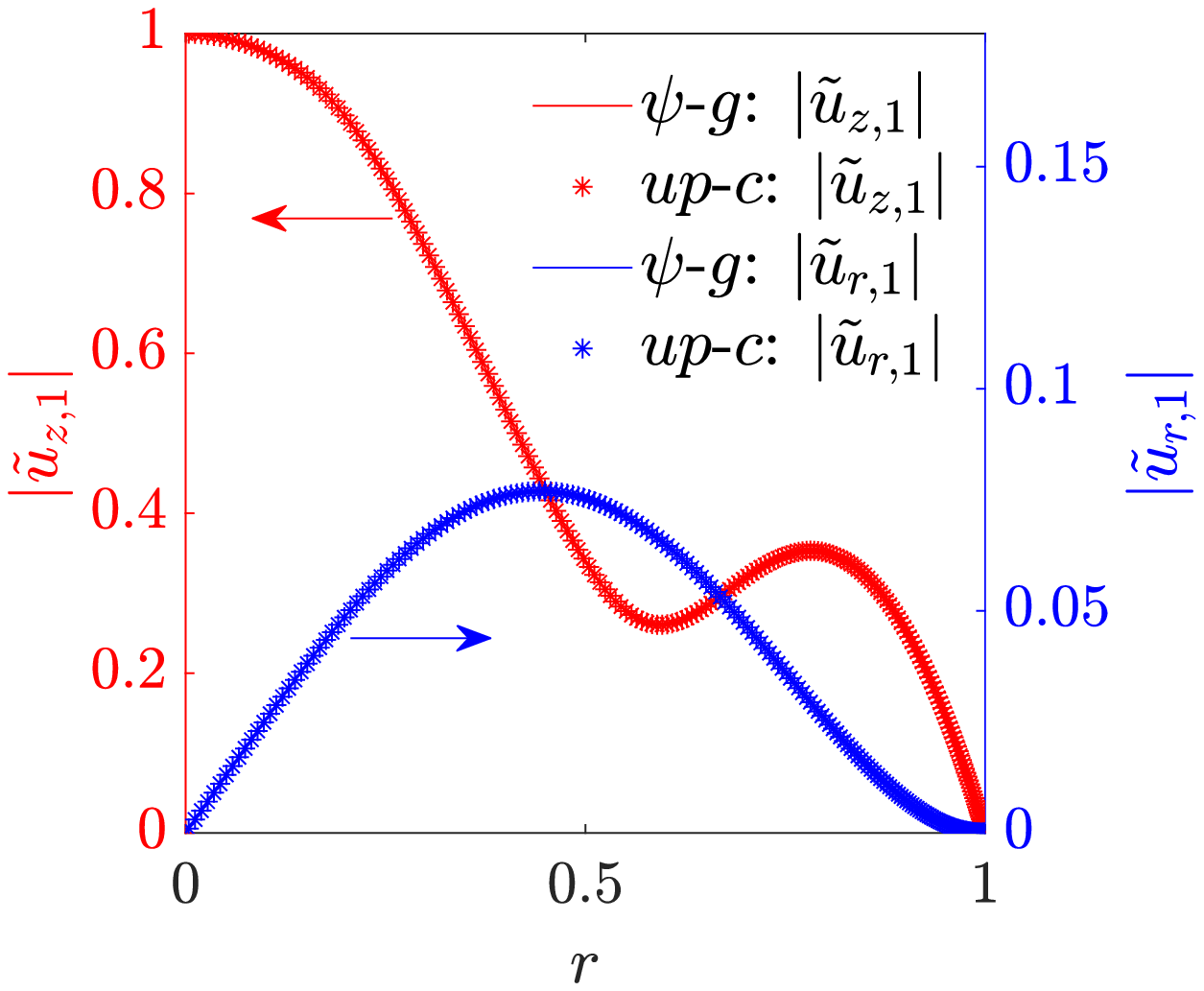}\put(-128,93){$(d)$}
	\includegraphics[width=0.33\textwidth,trim=30 5 30 0,clip]{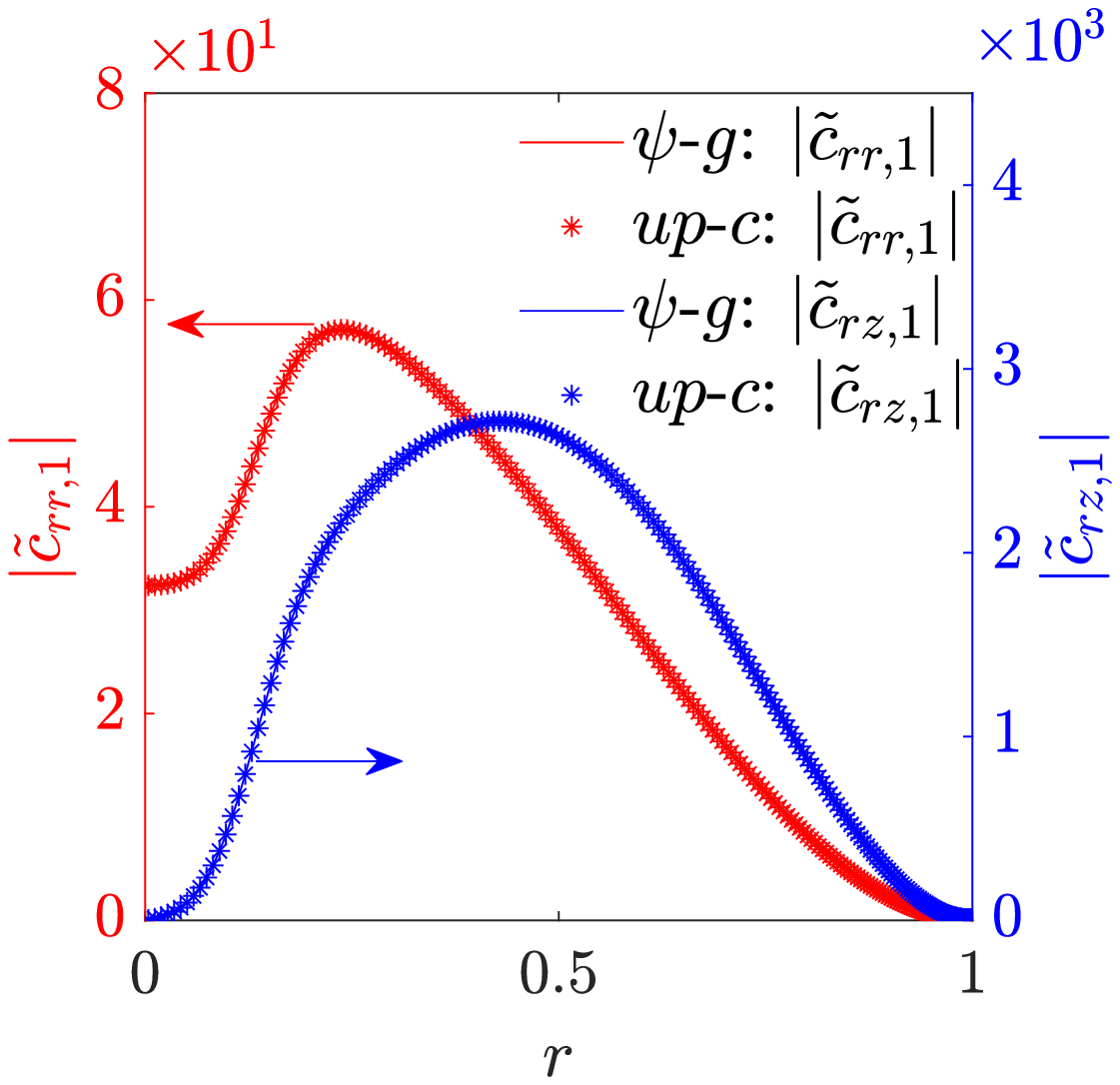}\put(-122,93){$(e)$}
	\includegraphics[width=0.33\textwidth,trim=30 5 30 0,clip]{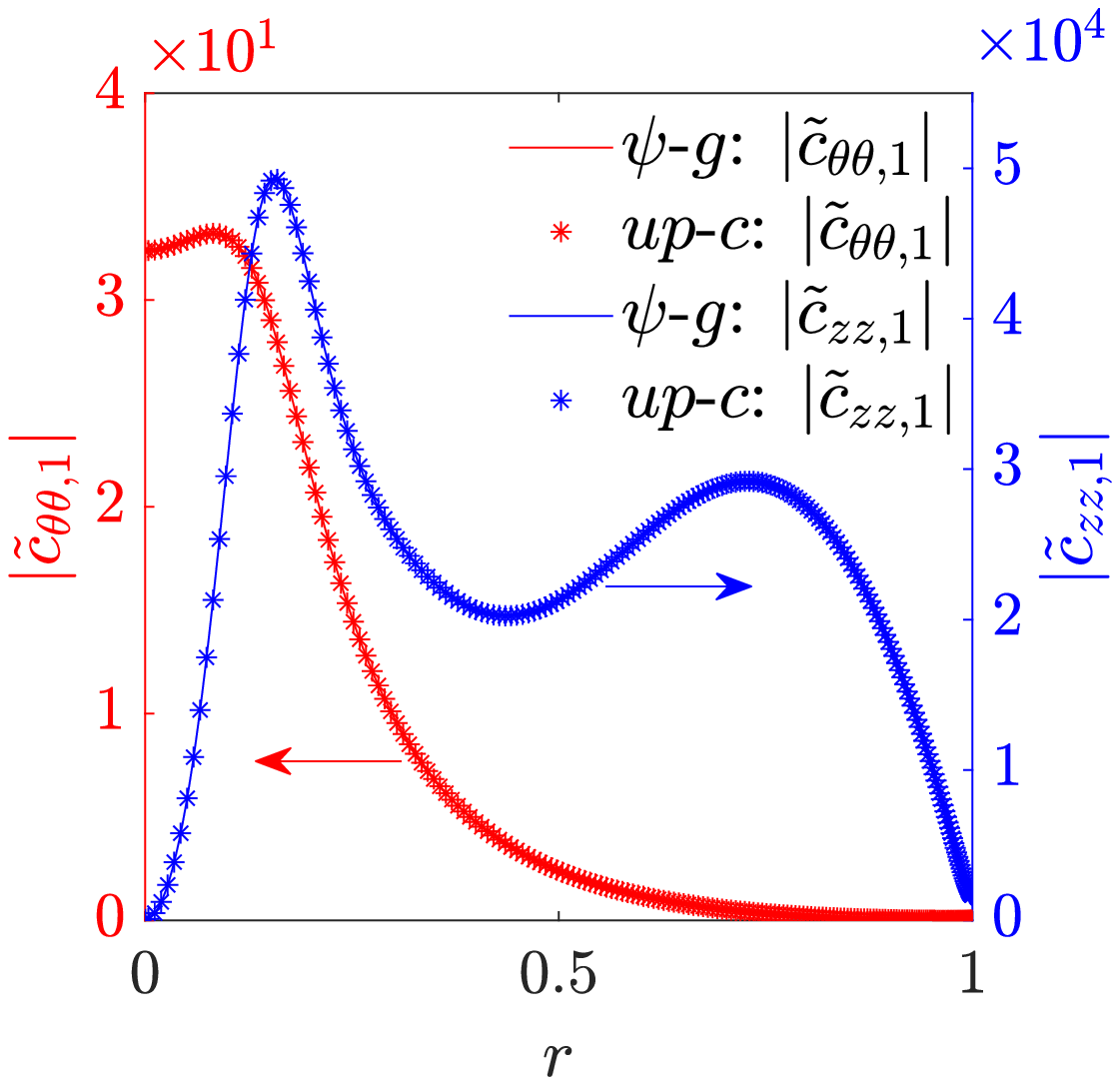}\put(-122,93){$(f)$}
	\caption{Adjoint and direct eigenfunctions at $Re_{c}=265.572335$, $\alpha_{c}=0.515525$, $Wi=65$ and $\beta=0.65$, corresponding to the left-end critical point in figure \ref{Fig:linear_Wi65beta65}$(a)$: $(a)$ adjoint velocity components $\tilde{u}_{z,1}^{\dagger}$ and $\tilde{u}_{r,1}^{\dagger}$; $(b)$ adjoint conformation tensor components $\tilde{g}_{rz,1}^{\dagger}$ and $\tilde{c}_{rz,1}^{\dagger}$; $(c)$ adjoint conformation tensor components $\tilde{g}_{zz,1}^{\dagger}$ and $\tilde{c}_{zz,1}^{\dagger}$; $(d)$ velocity components $\tilde{u}_{z,1}$ and $\tilde{u}_{r,1}$; $(e)$ conformation tensor components $\tilde{c}_{rr,1}$ and $\tilde{c}_{rz,1}$; $(f)$ conformation tensor components $\tilde{c}_{\theta\theta,1}$ and $\tilde{c}_{zz,1}$. The conformation tensor components $\tilde{c}_{rr,1}$, $\tilde{c}_{rz,1}$, $\tilde{c}_{\theta\theta,1}$, $\tilde{c}_{zz,1}$ in the $\psi$-$g$ formulation are transformed from $\tilde{g}_{rr,1}$, $\tilde{g}_{rz,1}$, $\tilde{g}_{\theta\theta,1}$, $\tilde{g}_{zz,1}$ according to equation \eqref{eq:cg_transformation}.}
	\label{Fig:adjoint_eigen}
\end{figure}

For the left-end critical point, the corresponding adjoint eigenfunctions of the linear problem are shown in figures \ref{Fig:adjoint_eigen}$(a)-(c)$. As the adjoint problem is linear and can be arbitrarily scaled, we normalize the adjoint eigenfunction so that the maximum amplitude of the adjoint axial velocity $max(|u_{z,1}^{\dagger}|)=1$. With this normalisation, the adjoint eigenfunction can be uniquely determined and it is noted that this normalization process does not affect the value of the Landau coefficient $a_3$ (see equation \ref{eq:Landau-coefficients}c). In figure \ref{Fig:adjoint_eigen}, we show the adjoint modes in the first row and the corresponding direct modes in the second row for a comparison. As shown in panel $(a)$, the adjoint velocity components $|u_{r,1}^{\dagger}|, |u_{z,1}^{\dagger}|$ obtained from the $up$-$c$ and $\psi$-$g$ formulations agree well with each other. On the other hand, figures \ref{Fig:adjoint_eigen}$(b)$ and $(c)$ plot respectively the adjoint conformation tensor components $g_{rz,1}^{\dagger}$ versus $c_{rz,1}^{\dagger}$ and $g_{zz,1}^{\dagger}$ versus $c_{zz,1}^{\dagger}$ in the two formulations. We find that even though for the linear variables, we have $\bc=\boldsymbol{Q}\boldsymbol{g}$ (as introduced in equation \eqref{eq:cg_transformation} and plotted in figures \ref{Fig:adjoint_eigen}$(d)-(f)$), there is no simple relation between $\bc^{\dagger}$ and $\boldsymbol{g}^{\dagger}$, probably due to the non-normality of $\boldsymbol{Q}$ and the linear operators. Still, the Landau coefficients obtained from these two different formulations are in good agreement as discussed in the validation subsection, suggesting that the solvability conditions described in equation \eqref{eq:solvability} have been correctly applied with the aid of the adjoint variables.

\begin{figure}
	\centering
	\includegraphics[width=0.48\textwidth,trim= 38 0 62 0,clip]{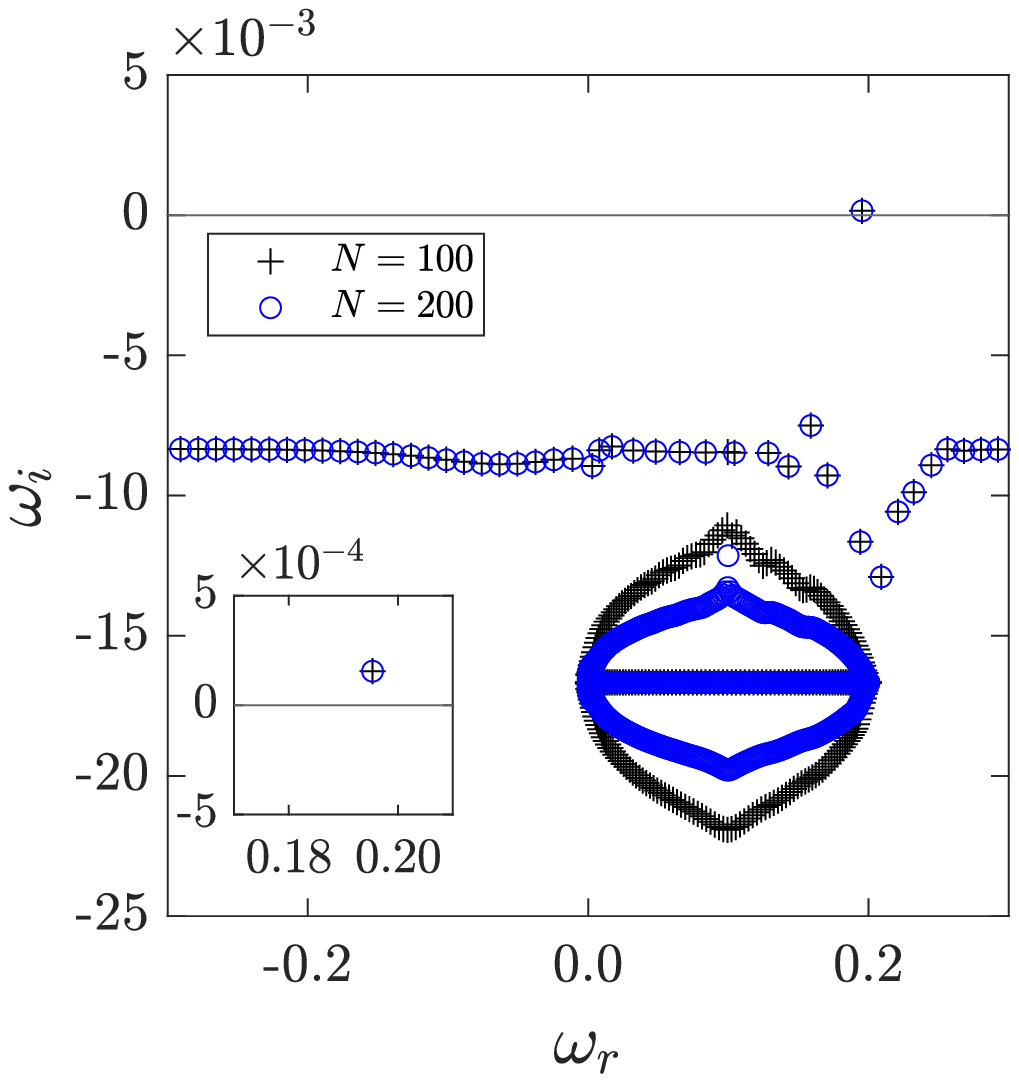}
	\put(-180,170){$(a)$}
	\includegraphics[width=0.48\textwidth,trim= 38 0 62 0,clip]{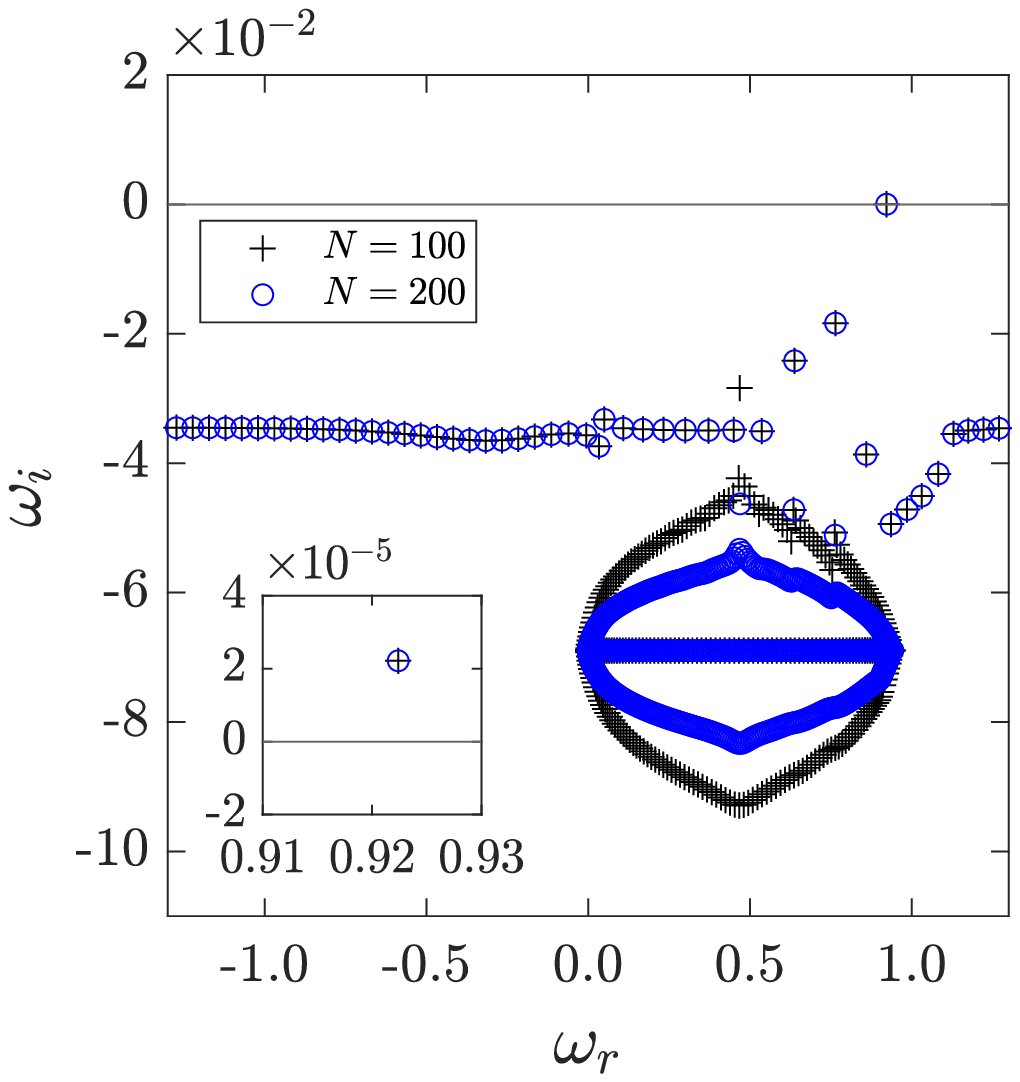}
	\put(-180,170){$(b)$}
	\caption{Eigenspectra obtained from a linear stability analysis of pipe Poiseuille flows of UCM fluids: $(a)$ at $Re=1000$, $\alpha=0.2$ and $Wi=60$; $(b)$ at $Re=260$, $\alpha=0.94$ and $Wi=14.5$.}
	\label{Fig:UCM_eigens}
\end{figure}

Before ending this section, we mention that pipe flow of UCM fluids can be linearly unstable. The UCM fluids can be obtained from the Oldroyd-B fluids with $\beta=0$ (zero solvent viscosity) and thus the flow is dominated by inertia, polymer viscosity and elasticity. To the best of our knowledge, no previous works have reported linear instability in pipe flows of UCM fluids. After exploring a large parameter space of axisymmetric UCM flows covering $0.5<\alpha<3$, $100<Re<10000$ and $0<E<1$, \cite{Chaudhary2021Linear} found stable modes in this flow. They concluded that solvent viscosity is important in constituting the linear instability in viscoelastic pipe flows along with the elasticity and flow inertia. However, we find that the viscoelastic pipe flow of UCM fluids can be linearly unstable. Figure \ref{Fig:UCM_eigens} shows the eigenspectra at two sets of  parameters: one is at $Re=1000$, $\alpha=0.2$, $Wi=60$ and the other is at $Re=260$, $\alpha=0.94$, $Wi=14.5$. For both of them, there is a single unstable mode and the convergence has been examined by increasing the total number of nodes $N$ (we have additionally used the \Dongdong{Petrov-Galerkin } method \citep{Meseguer2003Linearized} to double-check the existence and convergence of this mode using the code in \cite{Zhang2021}; the details are not shown). Because this finding is only a byproduct of our numerical investigation of weakly nonlinear stability of viscoelastic pipe flows, we place the discussions of the linear instability of UCM fluids in Appendix \ref{app_UCM}.

\subsection{Bifurcation studies in a weakly nonlinear framework}

The present weakly nonlinear stability analysis is performed around the linear critical condition. For all the results presented here, the neutral mode with almost zero growth rate is well separated from the continuous spectrum (by using a sufficiently large $N$) and thus its eigenfunction can be safely used to construct the higher-order solutions for the evaluation of the Landau coefficient. In what follows, we will successively discuss supercritical/subcritical bifurcations and a scaling law for the Landau coefficient. We find that the main parameter that differentiates the bifurcation types appears to be $\beta$. Thus we will first consider different ranges of $\beta$ in these two scenarios and then delimit the boundary between them.

\subsubsection{Supercritical bifurcations at small $\beta$}

\begin{figure}
	\centering
	\includegraphics[width=0.45\textwidth,trim=10 22 50 0,clip]{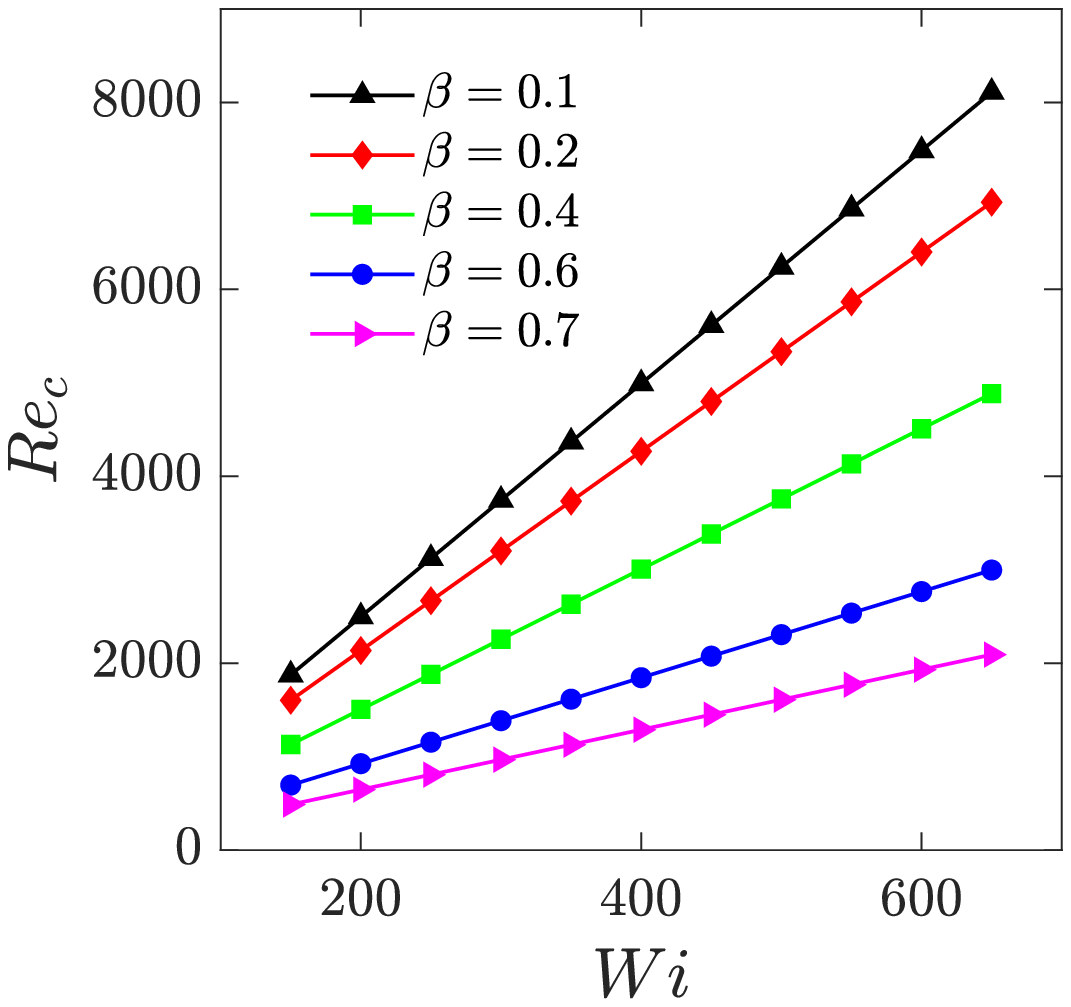}\put(-163,123){$(a)$}
	\includegraphics[width=0.45\textwidth,trim=10 22 50 0,clip]{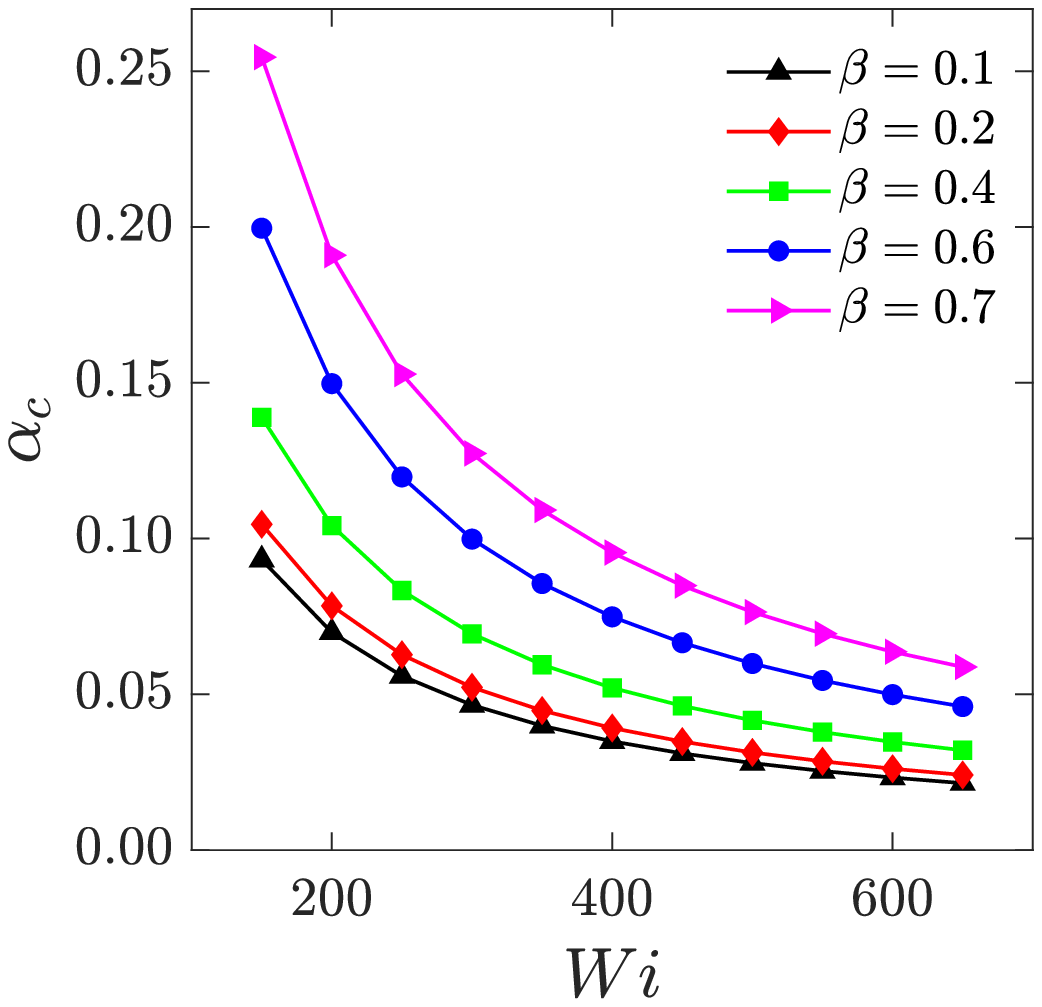}\put(-163,123){$(b)$}\\
	\includegraphics[width=0.45\textwidth,trim=10 0 50 0,clip]{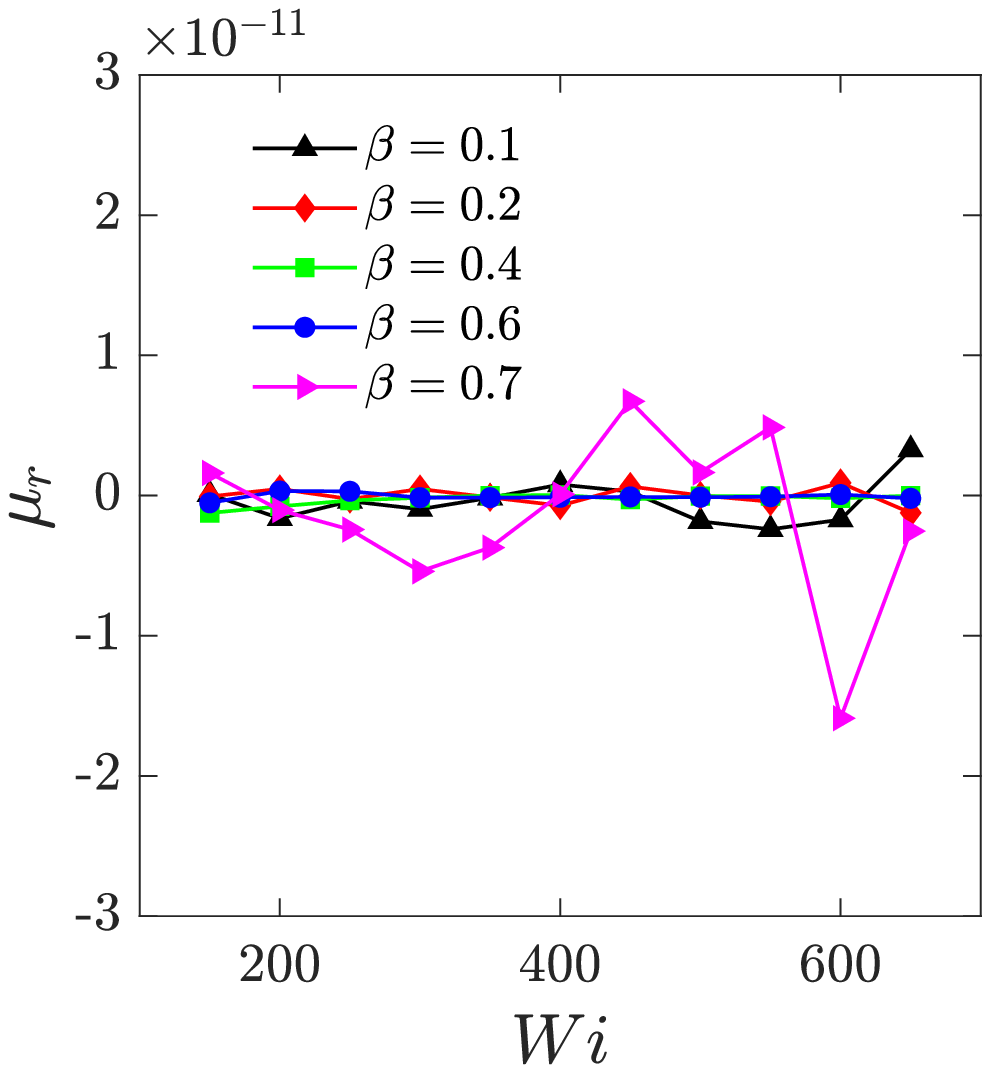}\put(-163,133){$(c)$}
	\includegraphics[width=0.45\textwidth,trim=10 0 50 0,clip]{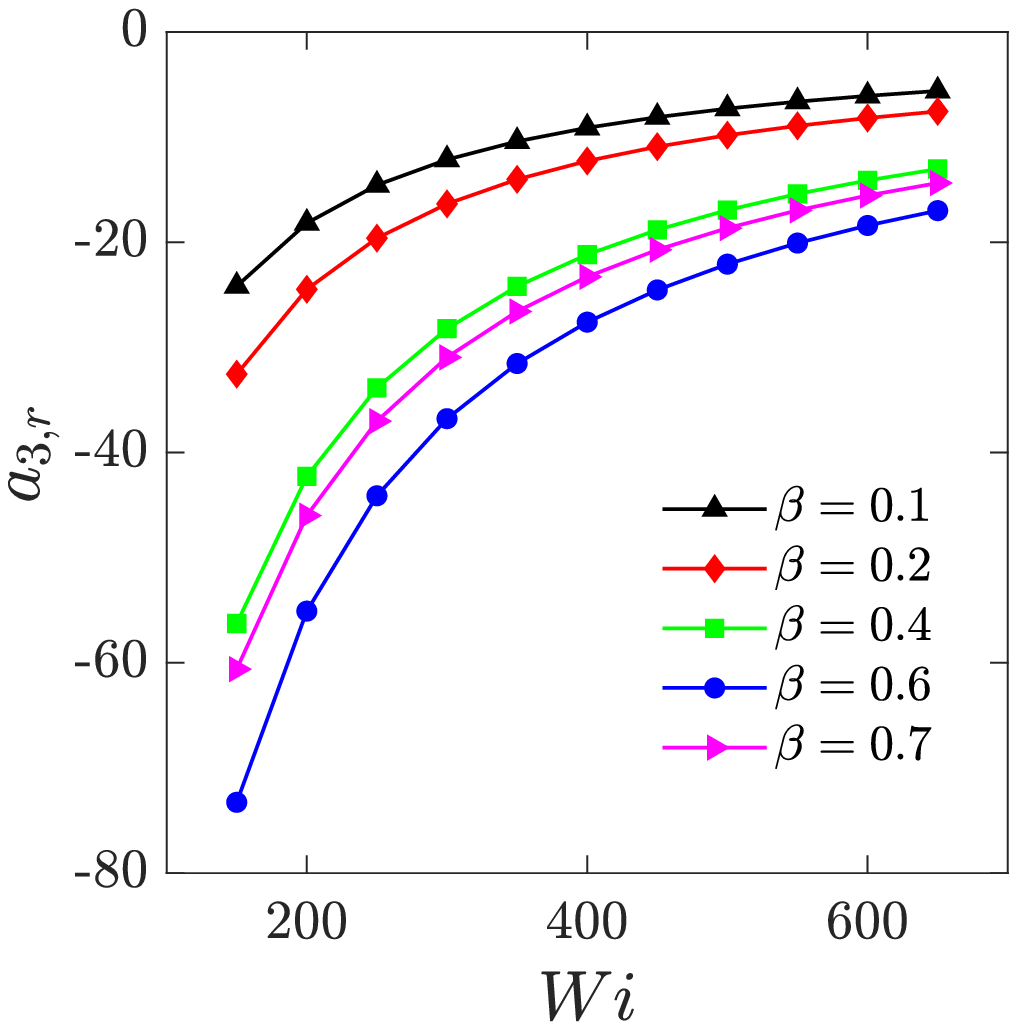}\put(-163,133){$(d)$}
	\caption{Critical conditions and the Landau coefficient for pipe flow of Oldroyd-B fluids at various $Wi$ and $\beta$ (small and moderate values): $(a)$ critical Reynolds number $Re_{c}$; $(b)$ critical wavenumber $\alpha_{c}$; $(c)$ linear growth rate $\mu_{r}$; $(d)$ Landau coefficient $a_{3,r}$. All the cases here are supercritical as $a_{3,r}$ is negative.}
	\label{Fig:super_betas}
\end{figure}

As mentioned in the previous sections, the real part of the Landau coefficient $a_3$ (denoted as $a_{3,r}$) determines the bifurcation type: a negative (positive) value indicates a supercritical (subcritical) bifurcation, see equation \eqref{eq:GL-t2 scale}. In this subsection, we consider relatively small values of $\beta$. Firstly, the critical Reynolds number $Re_c$ and wavenumber $\alpha_c$ at which the Landau coefficient $a_3$ is evaluated are shown in figures \ref{Fig:super_betas}$(a)$ and $(b)$. On the one hand, at a fixed $\beta$, increasing the Weissenberg number $Wi$ leads to high $Re_c$ and low $\alpha_c$, implying a strong stabilizing effect of polymer elasticity on the laminar base flow. On the other hand, at a fixed $Wi$, increasing polymer concentrations (decreasing $\beta$ from approximately 0.7) also results in an increase of $Re_c$ and a decrease of $\alpha_c$. Secondly, figure \ref{Fig:super_betas}$(c)$ shows the corresponding real parts (growth rate) of the eigenvalue $\mu_r$ of the almost neutral modes. For all the five values of $\beta$, the growth rates are all indeed very small on the order of $O(10^{-11})$, around the linear instability onset. Thirdly, figure \ref{Fig:super_betas}$(d)$ illustrates the corresponding values of $a_{3,r}$ for different values of $Wi$ and $\beta$. Clearly, all the curves fall in the negative range of $a_{3,r}$, suggesting that supercritical bifurcations exist in the parameter space explored here. Thus the lowest-order nonlinear term in the governing equation stabilizes the disturbances and drives them to saturate. Besides, increasing $Wi$ or decreasing $\beta$ generally corresponds to a smaller amplitude of $a_{3,r}$, indicating that the degree of supercriticality is reduced. Thus, the stabilising effect of nonlinearity becomes weaker when increasing $Wi$ or decreasing $\beta$ in their respective ranges, as shown in figure \ref{Fig:super_betas}$(d)$. From the slopes of these curves, one may infer that the degree of supercriticality probably approaches an asymptote at sufficiently high $Wi$ for each $\beta$. 

\begin{figure}
	\centering
	\includegraphics[width=0.45\textwidth,trim=30 23 60 10,clip]{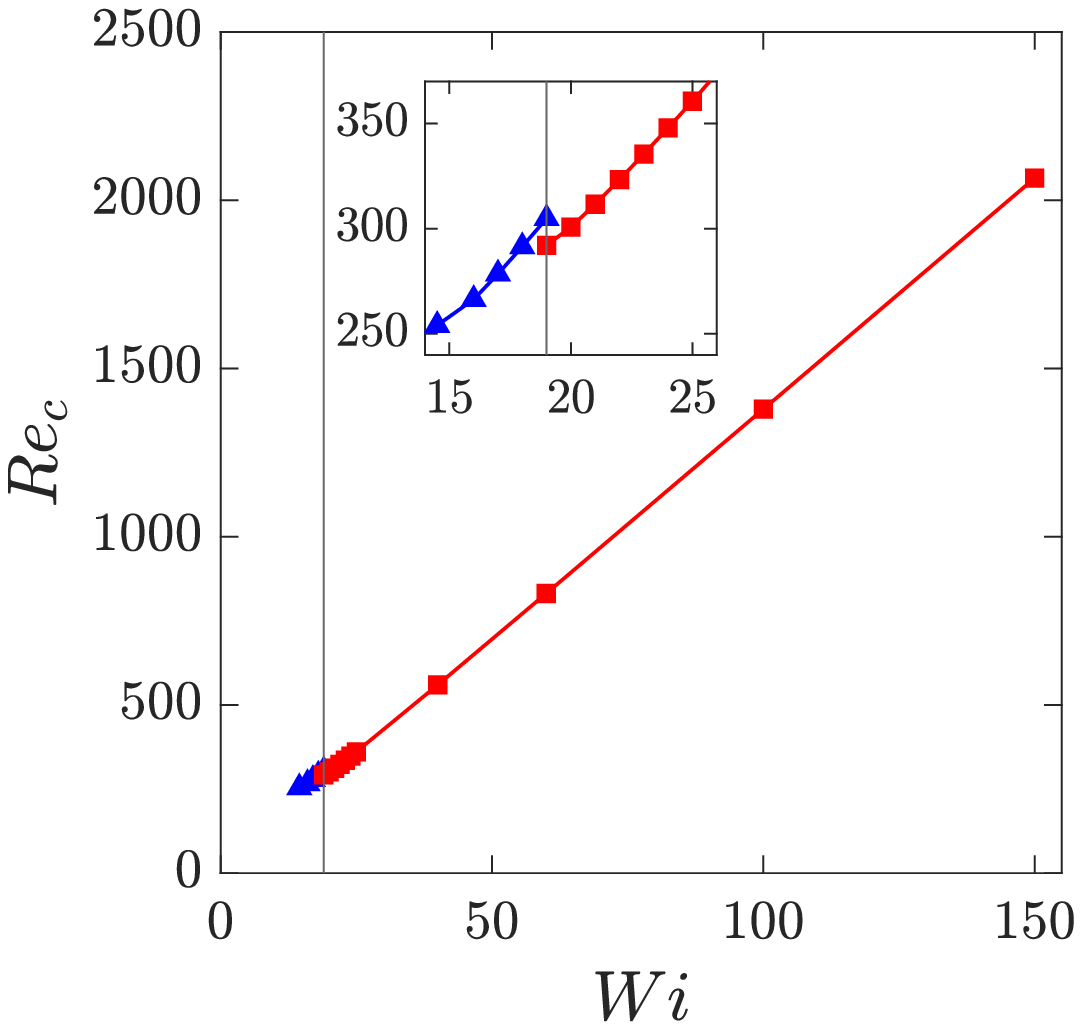}\put(-173,133){$(a)$}
	\includegraphics[width=0.45\textwidth,trim=30 23 60 10,clip]{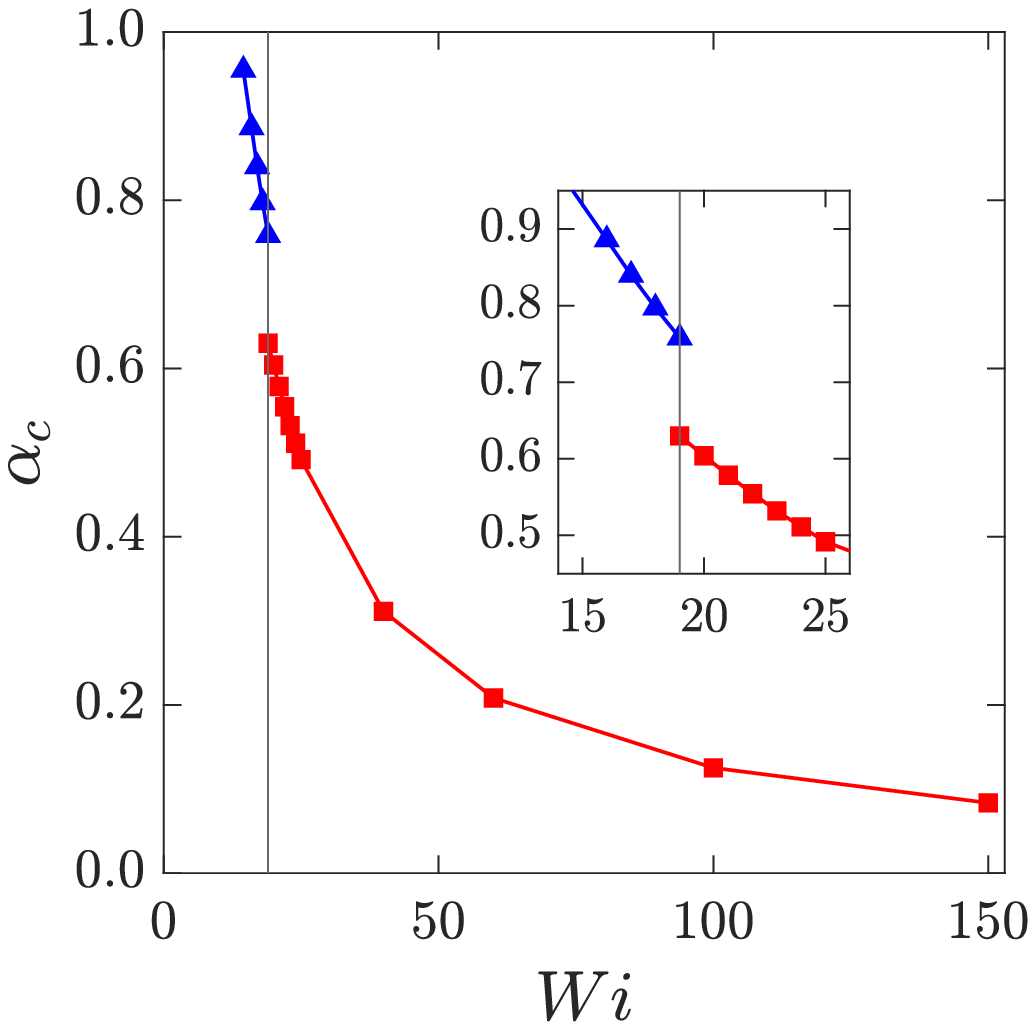}\put(-170,133){$(b)$}\\
	\includegraphics[width=0.45\textwidth,trim=30 5 60 10,clip]{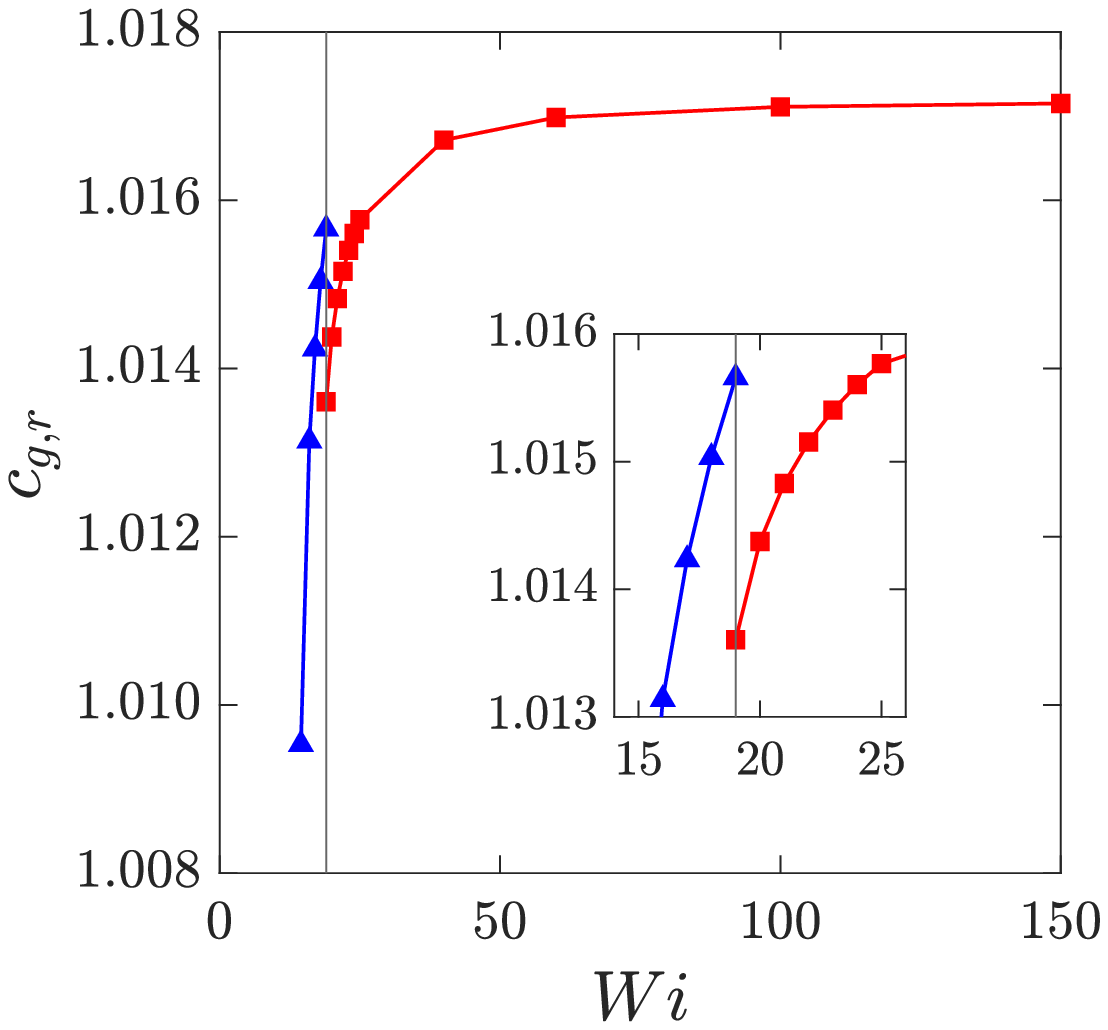}\put(-173,143){$(c)$}
	\includegraphics[width=0.45\textwidth,trim=30 5 60 10,clip]{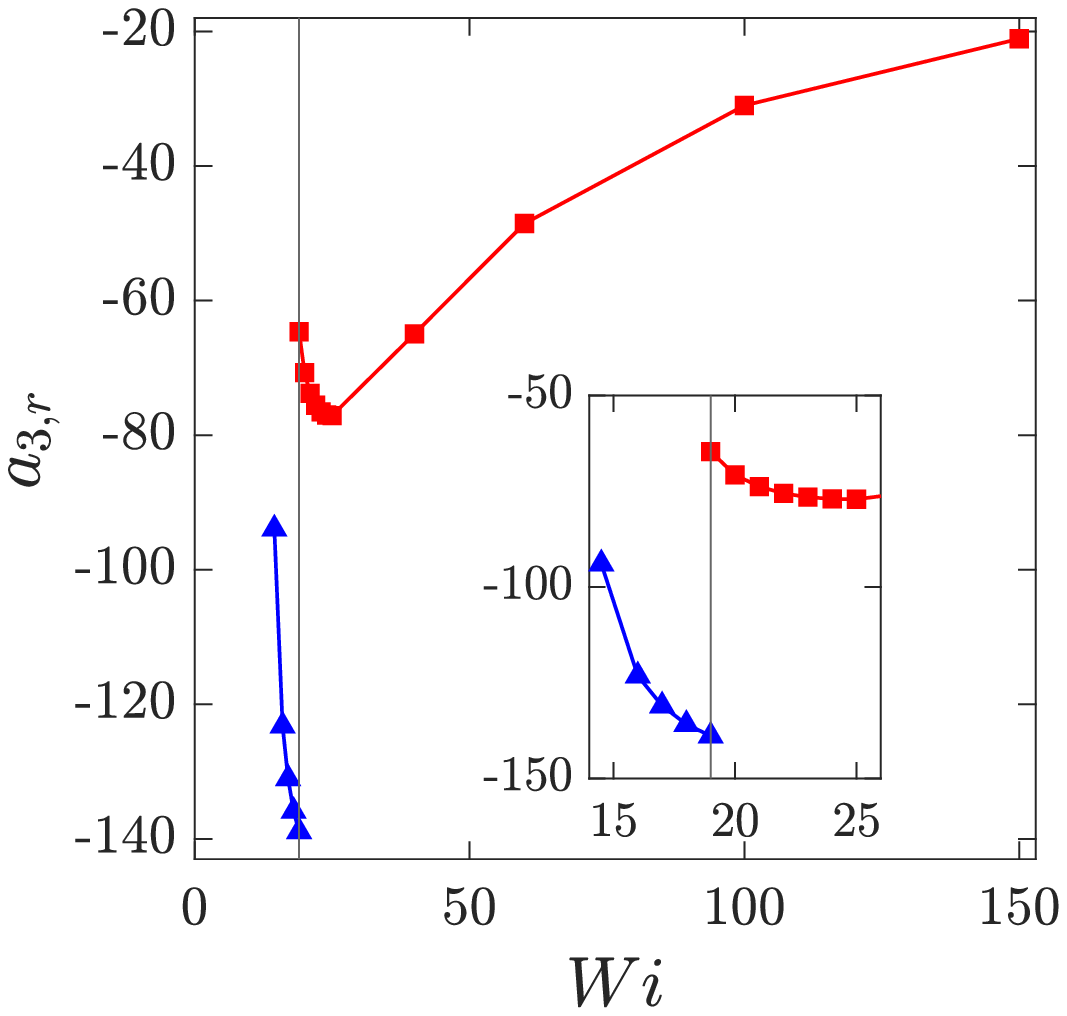}\put(-170,143){$(d)$}
	\caption{Critical conditions and the Landau coefficient at various $Wi$ for pipe flow of UCM fluids ($\beta=0$): $(a)$ critical Reynolds number $Re_{c}$; $(b)$ critical wavenumber $\alpha_{c}$; $(c)$ group velocity $c_{g,r}$; $(d)$ Landau coefficient $a_{3,r}$. All the cases here are supercritical as $a_{3,r}$ is negative.}
	\label{Fig:UCM_super}
\end{figure}

In the UCM limit $\beta=0$, we have reported the linear instability in the last subsection. From the neutral curves (see figure \ref{Fig:NC-UCM-Wi} in Appendix \ref{app_UCM}), the critical condition $Re_{c}$ and $\alpha_{c}$ at the leftmost end of the multiple loops can be determined, as shown in figures \ref{Fig:UCM_super}$(a)$ and $(b)$. Basically, decreasing $Wi$ leads to smaller $Re_c$ and larger $\alpha_c$. One can observe a discontinuity in the curve around $Wi=19$ whose origin results from the complexity of the neutral curves at low $Wi$, as we explain now. The neutral curve at $Wi=19$ (shown in figure \ref{Fig:NC-UCM-Wi}$(b)$) consists of two separate loops. The left tips of them are denoted as point A (for the smaller loop) and point B (for the larger loop), respectively. In figure \ref{Fig:UCM_super}, the corresponding $Re$ and $\alpha$ at point A and point B are represented by red filled squares and blue filled triangles at $Wi=19$ (see the vertical lines in panels $(a)$ and $(b)$ and their insets). When $Wi>19$, we use the critical point A in the weakly nonlinear analysis because it is the lowest $Re$ which is the critical $Re_c$ in a global sense. With $Wi$ reducing to $18$ and smaller, the smaller loop disappears and we thus use the critical point B in the weakly nonlinear stability analysis, resulting in the apparent jump of $\alpha_{c}$ by about 0.1. Such discontinuity can also be seen from the group velocity $c_{g,r}$ in figure \ref{Fig:UCM_super}$(c)$, where with decreasing $Wi$ from 150 the group velocity $c_{g,r}$ firstly \Dongdong{goes down and then suddenly increases at $Wi=19$ and decreases again}. As for the Landau coefficient, $a_{3,r}$ undergoes a minimum at about $Wi=25$ and also there is a clear discontinuity at $Wi=19$. In spite of such complex behaviour, all the values of $a_{3,r}$ for $Wi$ ranging from 14.5 to 150 are negative, again suggesting supercritical bifurcations for UCM pipe Poiseuille flow at the inertial regime. In the literature, \cite{Meulenbroek2003Intrinsic} reported subcritical bifurcations in the inertialess regime in their weakly nonlinear analysis of the viscoelastic pipe Poiseuille flow of UCM fluids. The weakly nonlinear analysis was also extended to plane Poiseuille flow \citep{Meulenbroek2004Weakly} and plane Couette flow \citep{Morozov2005} of UCM fluids, both resulting subcritical bifurcations in the inertialess regime. It is thus interesting as a future work to conduct weakly nonlinear stability analysis of these flows with relatively strong inertia to assess its effect.

Next, we discuss the possible link between our theoretical prediction of flow bifurcations and the flow transitions observed in experiments. As briefly reviewed in the introduction, \cite{Samanta2013Elasto-inertial} observed no hysteresis of pressure fluctuations in pipe Poiseuille flow at high polymer concentrations of 500 ppm solution (approximately equivalent to $\beta=0.69$), suggesting the existence of supercritical bifurcations (see figure 2B in their paper). In this case, the transitional Reynolds number is $Re_t \approx 800$ at which our weakly nonlinear analysis also predicts a supercritical bifurcation but at a high Weissenberg number $Wi \approx 250$ (at which the linear growth rate is almost zero). Note that we confine ourselves to the linear condition in our calculation in order that weakly nonlinear results are guaranteed to converge. If we use $Wi=15$ (the maximum value of transition $Wi$ in \cite{Samanta2013Elasto-inertial}) in our weakly nonlinear analysis anyway, we can formally also get a negative $a_{3,r}$ (again the convergence of this result may be questionable). In any case, in experiments, infinitesimal perturbations are difficult to realise and realistic disturbances are always to some extent of finite amplitude, so it is difficult to differentiate the genuine supercriticality from \Dongdong{the subcriticality which is very sensitive to noise-level disturbances (see Appendix \ref{Appendix_subsuper} for more discussions)}. But the present theoretical prediction seems to lend some supports to a supercritical bifurcation to a certain degree.

\begin{figure}
	\centering
	\includegraphics[width=0.99\textwidth,trim=50 10 10 10,clip]{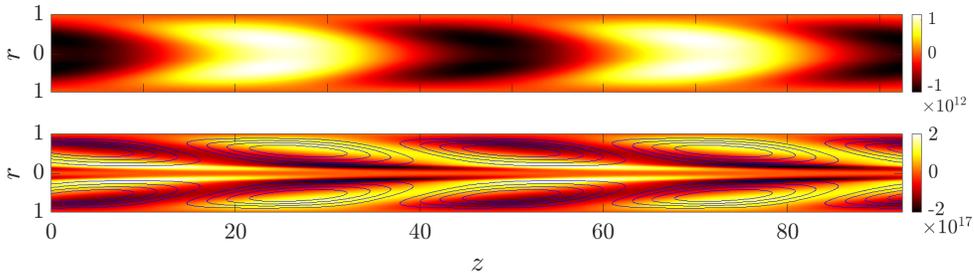}
	\caption{Flow pattern of the second order solution at $Re_c = 1054.35$, $\alpha_{c}=0.1359$, $Wi=203$ and $\beta=0.56$. Upper panel for $\tilde{u}_{z,21}/U_z$; lower panel for $tr(\tilde{\boldsymbol{c}}_{21})$ along with the blue curves for streamlines.}
	\label{Fig:flow_pattern}
\end{figure}

Another experimental study conducted very recently by \cite{Choueiri2021Experimental} showed that in the pipe flow of 600 ppm polymer solution, the measured pressure fluctuation amplitude $p_m$ grows continuously with increasing $Re$ near the instability onset $Re_t\approx 18$, following $p_m \propto \sqrt{Re-Re_t}$ which signifies a supercritical transition. In this experiment, the viscosity ratio $\beta$ is about 0.56, at which our weakly nonlinear analysis predicts supercritical bifurcations in a wide range of $Wi$ but at a higher critical Reynolds number $Re_c \sim 150$ than the instability threshold $Re_t=18$ observed in experiments. Concerning this difference, it should be noted that, in view of the experimental fact that the onset of structural turbulence is characterized by a critical shear rate instead of a critical Reynolds number \citep{Samanta2013Elasto-inertial}, the critical Reynolds number actually varies with pipe radius as $Re_c \sim R^2$ \citep{Ram1964Structural}. Therefore, the low Reynolds number $Re_t=18$ is expected to become higher (may exceed $Re \sim O(150)$ predicted by linear theory) should a pipe with larger radius is used. Despite the dependency of $Re_c$ on the flow geometry, our prediction of the intrinsic bifurcation type seems to be consistent with their observations. In fact, when \cite{Choueiri2021Experimental} investigated the flow pattern at various $Re$ (at a viscosity ratio of about $\beta=0.18$ at which our calculation also predicts supercritical bifurcations), they observed the scaling $p_m \propto \sqrt{Re-Re_t}$ near the onset, which seems to support a supercritical bifurcation at high polymer concentrations (even though the authors were cautious enough to not exclude possibilities of other scaling relations due to the \Dongdong{possibly sensitive nature of the viscoelastic pipe flow to the disturbances}. As for the flow pattern, they observed chevron shape streaks at low $Re$, which resembles the flow patterns predicted by linear stability theory. These flow structures correspond to our linear solution in equation \eqref{eq:gamma1} (and we have checked that they look similar). We now present the second-order solution in equation \eqref{eq:gamma2}. The axial velocity $\tilde{u}_{z,21}$ at the same parameter $Wi=203, \beta=0.56$ as in \cite{Choueiri2021Experimental} is shown in the upper panel of figure \ref{Fig:flow_pattern}. As one can see, they also exhibit such kind of chevron shape patterns. The same figure also shows the contours of the trace of the conformation tensor $tr(\tilde{\boldsymbol{c}}_{21})$ (lower panel). It is clear that these polymers are stretched along the streamwise direction.

\subsubsection{Subcritical bifurcations at large $\beta$}

In the last subsection, the viscosity ratio $\beta$ is restricted to small and moderate values and we found supercritical bifurcations in viscoelastic pipe flows in the weakly nonlinear stability analysis. In this subsection, our focus is on relatively larger values of $\beta$, corresponding to low polymer concentrations.

\begin{figure}
	\centering
	\includegraphics[width=0.32\textwidth,trim=0 5 0 10,clip]{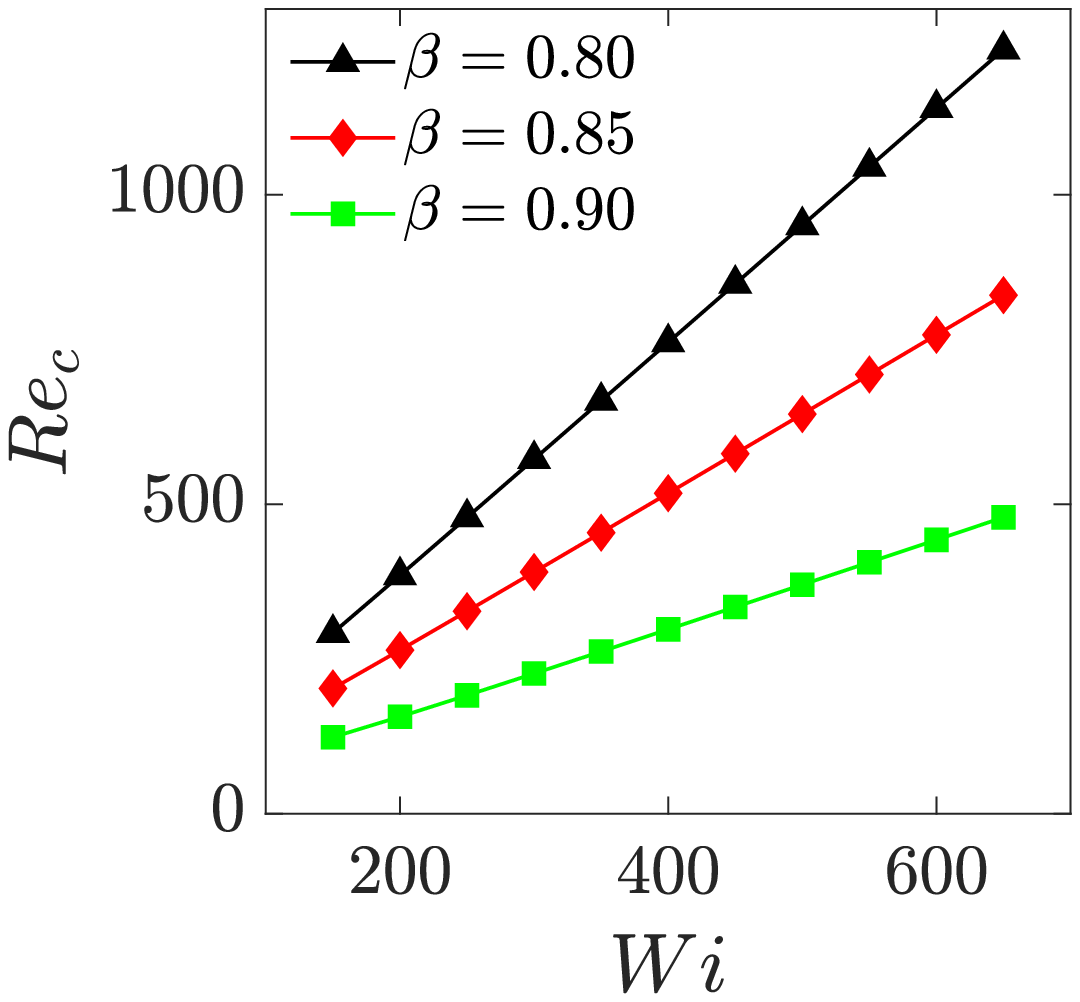}\put(-123,103){$(a)$}
	\includegraphics[width=0.32\textwidth,trim=-10 5 10 10,clip]{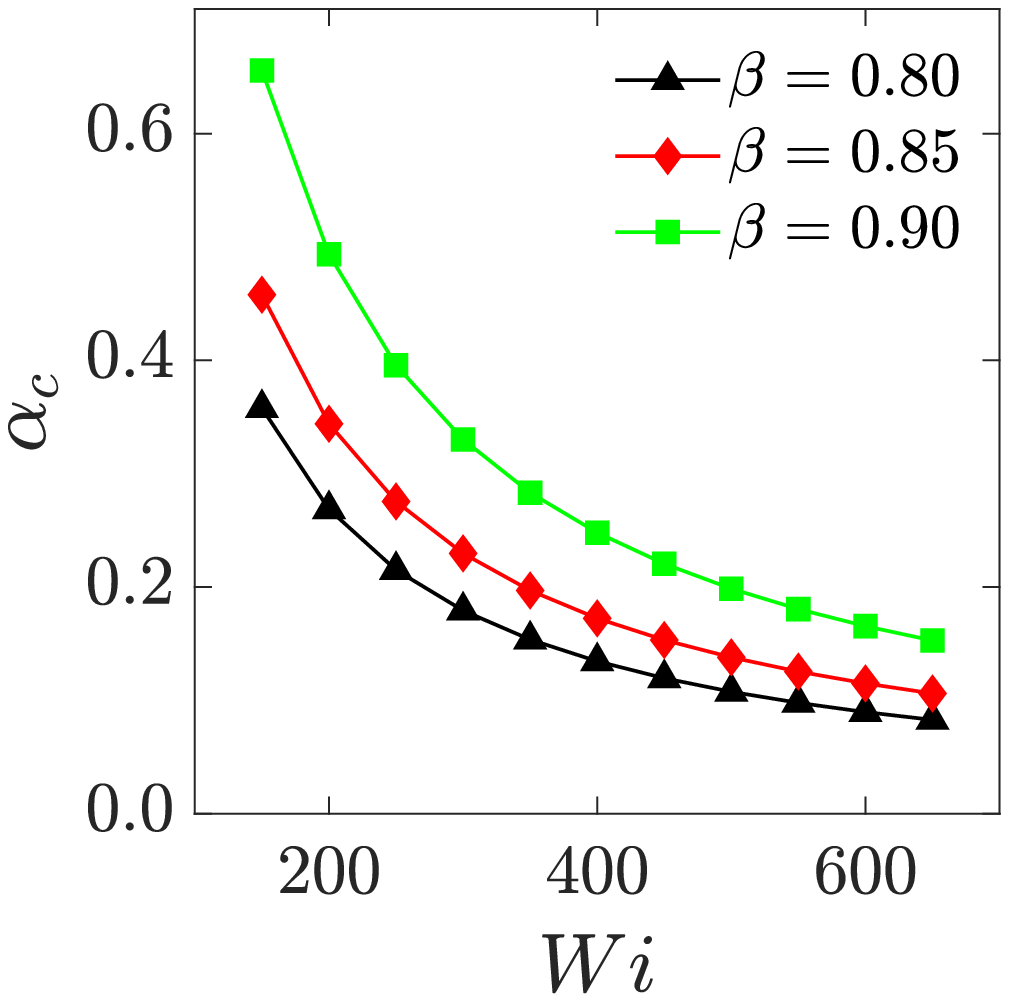}\put(-120,103){$(b)$}
	\includegraphics[width=0.32\textwidth,trim=0 5 0 10,clip]{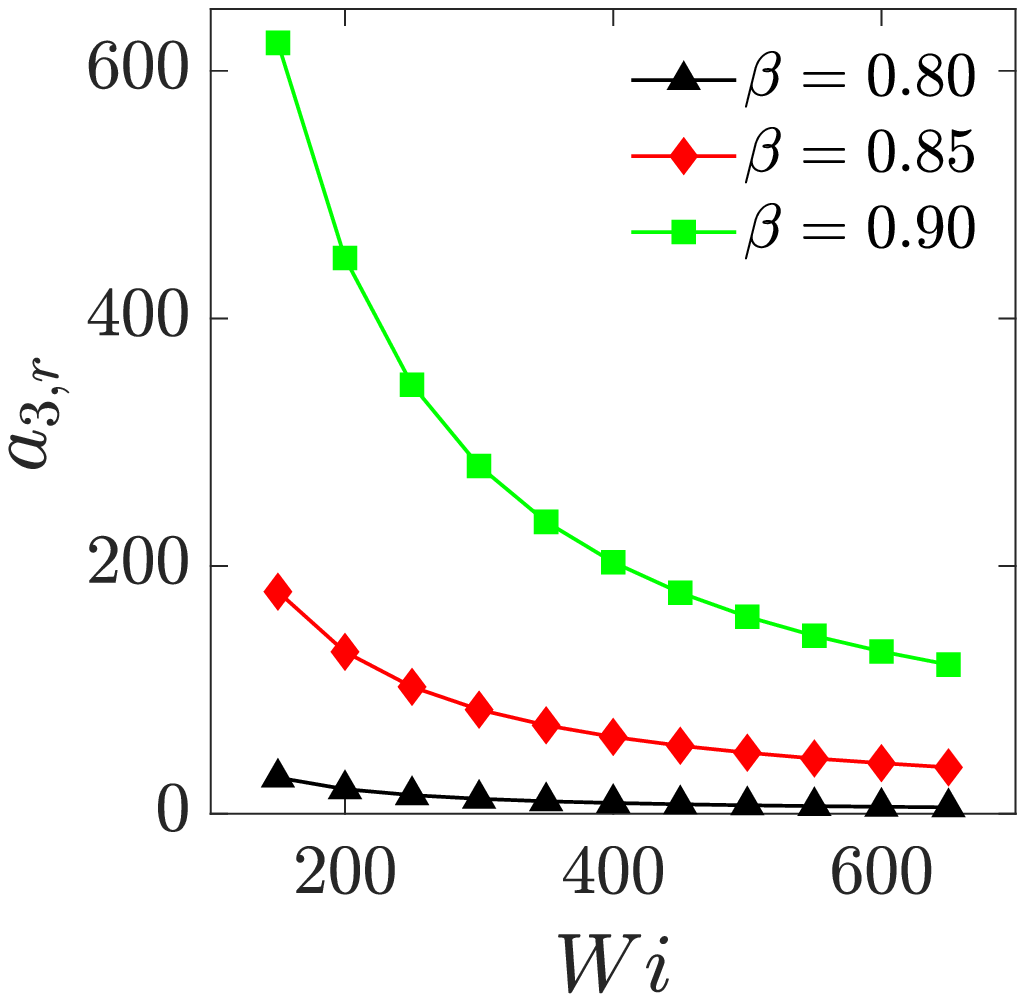}\put(-126,103){$(c)$}
	\caption{Critical conditions and the Landau coefficient for pipe flow of Oldroyd-B fluids at various $Wi$ and $\beta$ (large values): $(a)$ critical Reynolds number $Re_{c}$; $(b)$ critical wavenumber $\alpha_{c}$; $(c)$ Landau coefficient $a_{3,r}$. All the cases here are subcritical as $a_{3,r}$ is positive.}
	\label{Fig:critical_a3_sub}
\end{figure}

The effects of $\beta$ and $Wi$ on the critical Reynolds number and wavenumber, as shown in figures \ref{Fig:critical_a3_sub}$(a)$ and \ref{Fig:critical_a3_sub}$(b)$, are similar to those at low viscosity ratios: the flow is stabilized at higher values of $Wi$ and lower values of $\beta$ (both can be interpreted as strong viscoelastic effects due to polymers). The asymptotic trend shows that the critical $Re_c$ may be very large for sufficiently high $Wi$. As for the Landau coefficient $a_{3,r}$ shown in figure \ref{Fig:critical_a3_sub}$(c)$, it can be seen that its values are all positive, indicating subcritical bifurcations in the viscoelastic pipe flow with low polymer concentrations. The degree of subcriticality (characterized by the amplitude of $a_{3,r}$) becomes smaller for higher $Wi$ or lower $\beta$. At high $\beta$ (or low polymer concentrations), Newtonian-like regimes in the form of hysteresis and intermittency were observed in the experiments conducted by \cite{Samanta2013Elasto-inertial}. Hysteresis was observed in their experiment of viscoelastic pipe flows at a low polymer concentration of 100 ppm (equivalent to $\beta \approx 0.92$, see their figure 2A). Extrapolation of the Landau coefficient $a_{3,r}$ based on figure \ref{Fig:critical_a3_sub}$(c)$ suggests also subcritical bifurcations at $\beta \approx 0.92$ for a wide range of Weissenberg numbers, qualitatively consistent with the experimental observations. More specifically, at the two ends of $Wi$-range in figure \ref{Fig:critical_a3_sub}$(c)$, for $\beta=0.92$, the Landau coefficient at $Wi=600, Re_c=325.106379,\alpha_{c}=0.203808$ is $a_{3,r}=202$; the Landau coefficient at $Wi=70, Re_c=87.694576,\alpha_{c}=1.863525$ is $a_{3,r}=1655$, both showing subcritical bifurcations. Therefore, the current calculation supports the subcritical bifurcation in viscoelastic pipe flows at relatively high $\beta$.

Since both supercritical and subcritical bifurcations exist in the present viscoelastic pipe flows, there must be a boundary in the parameter space delimiting the bifurcation types, and this will be discussed in the following subsection.

\subsubsection{Bifurcation boundary delimited by $\beta$}

\begin{figure}
	\centering
	\includegraphics[width=0.45\textwidth,trim=0 5 0 10,clip]{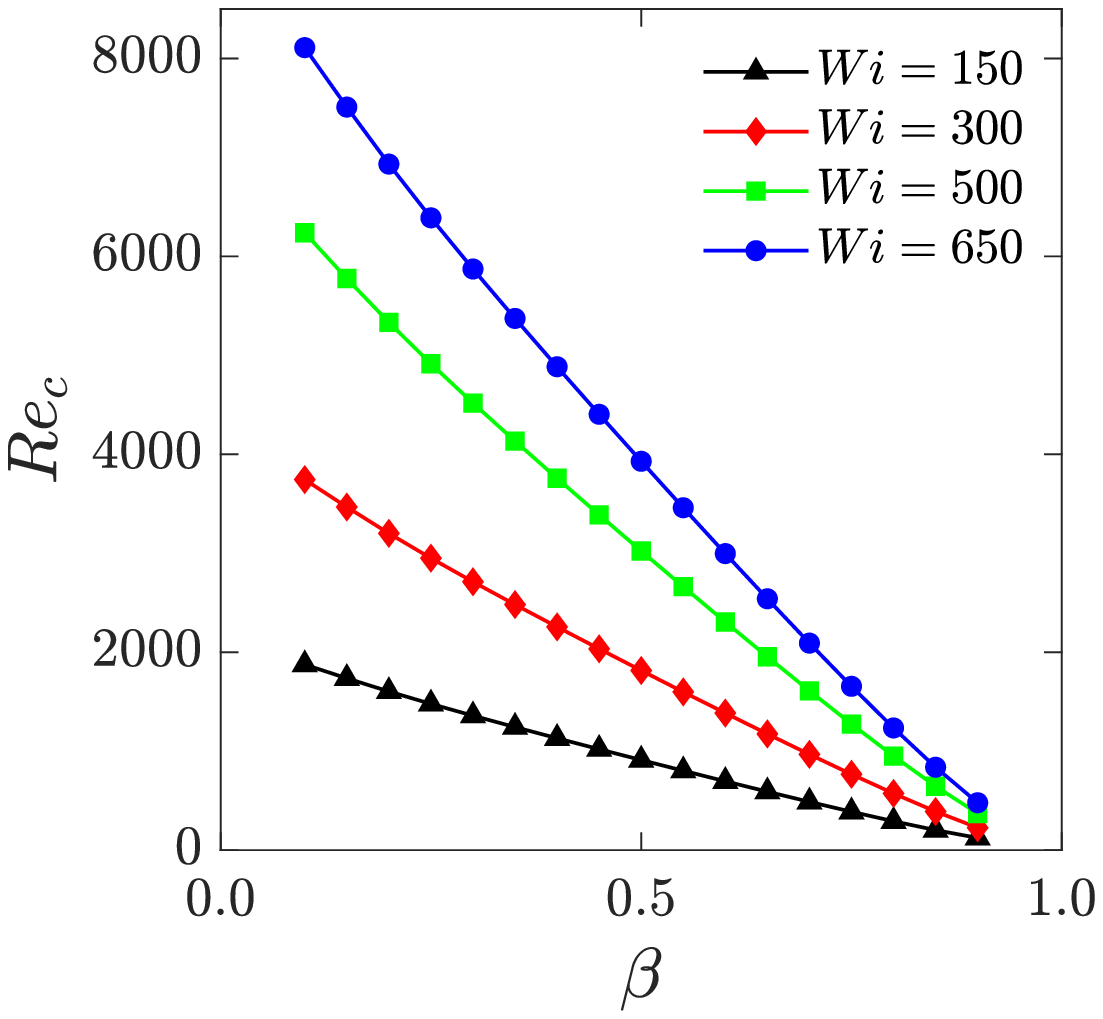}\put(-173,143){$(a)$}
	\includegraphics[width=0.45\textwidth,trim=0 5 0 10,clip]{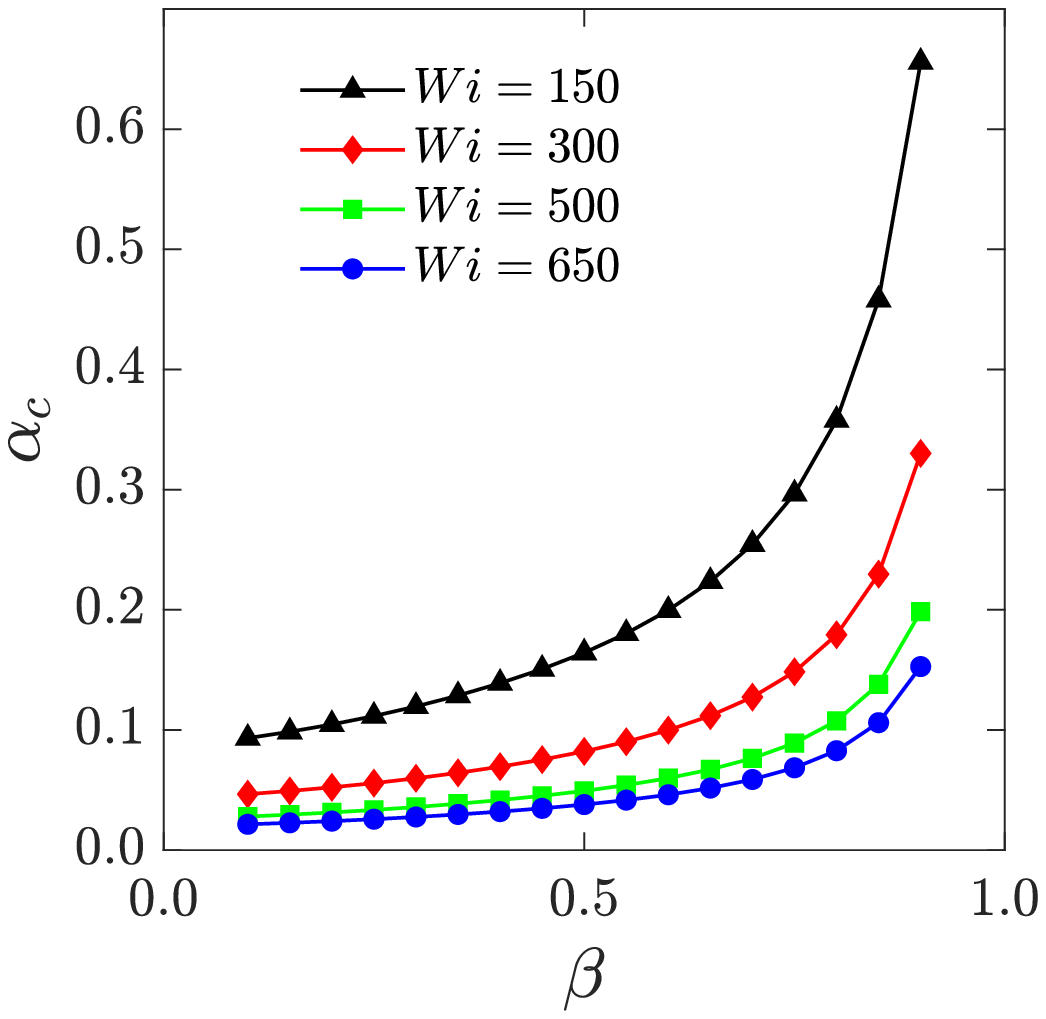}\put(-170,143){$(b)$}\\
	\includegraphics[width=0.45\textwidth,trim=30 5 65 10,clip]{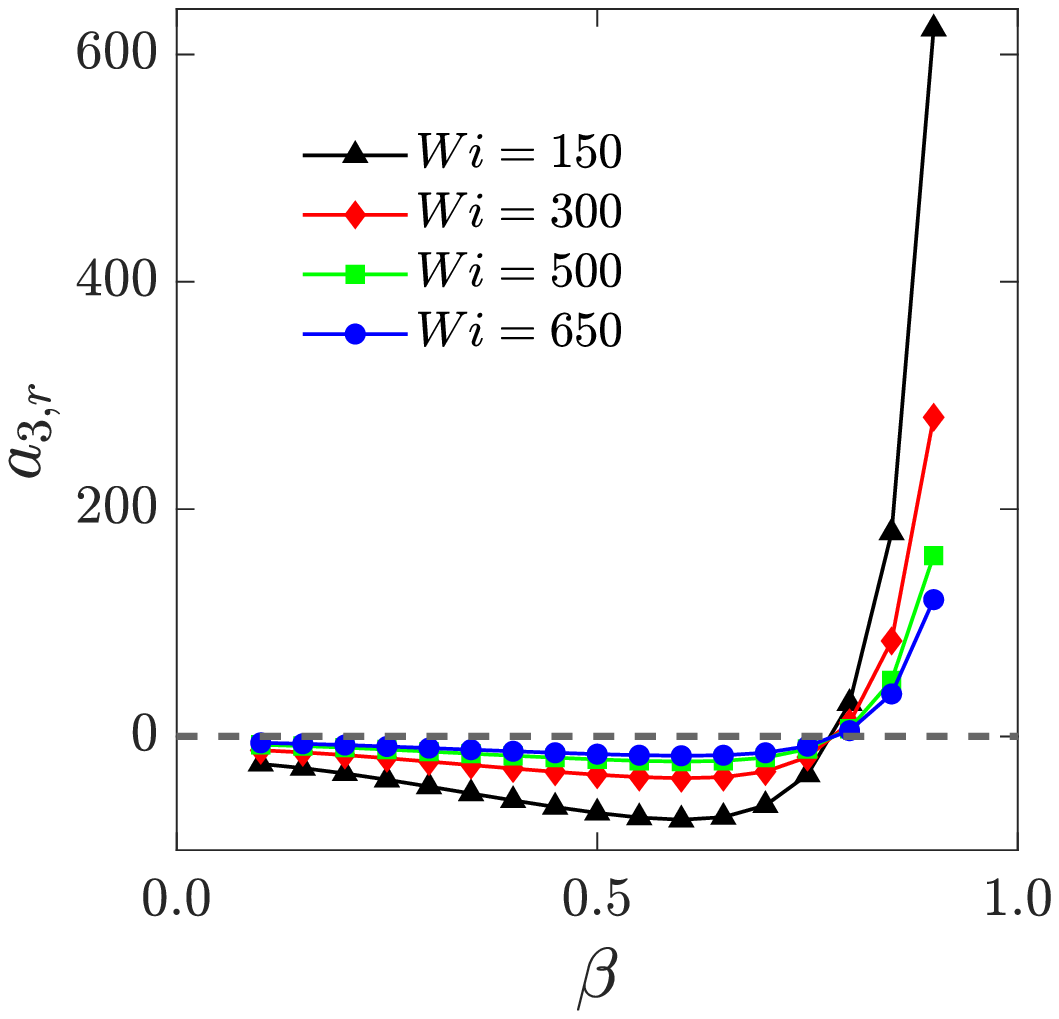}\put(-173,143){$(c)$}
	\includegraphics[width=0.45\textwidth,trim=35 5 60 10,clip]{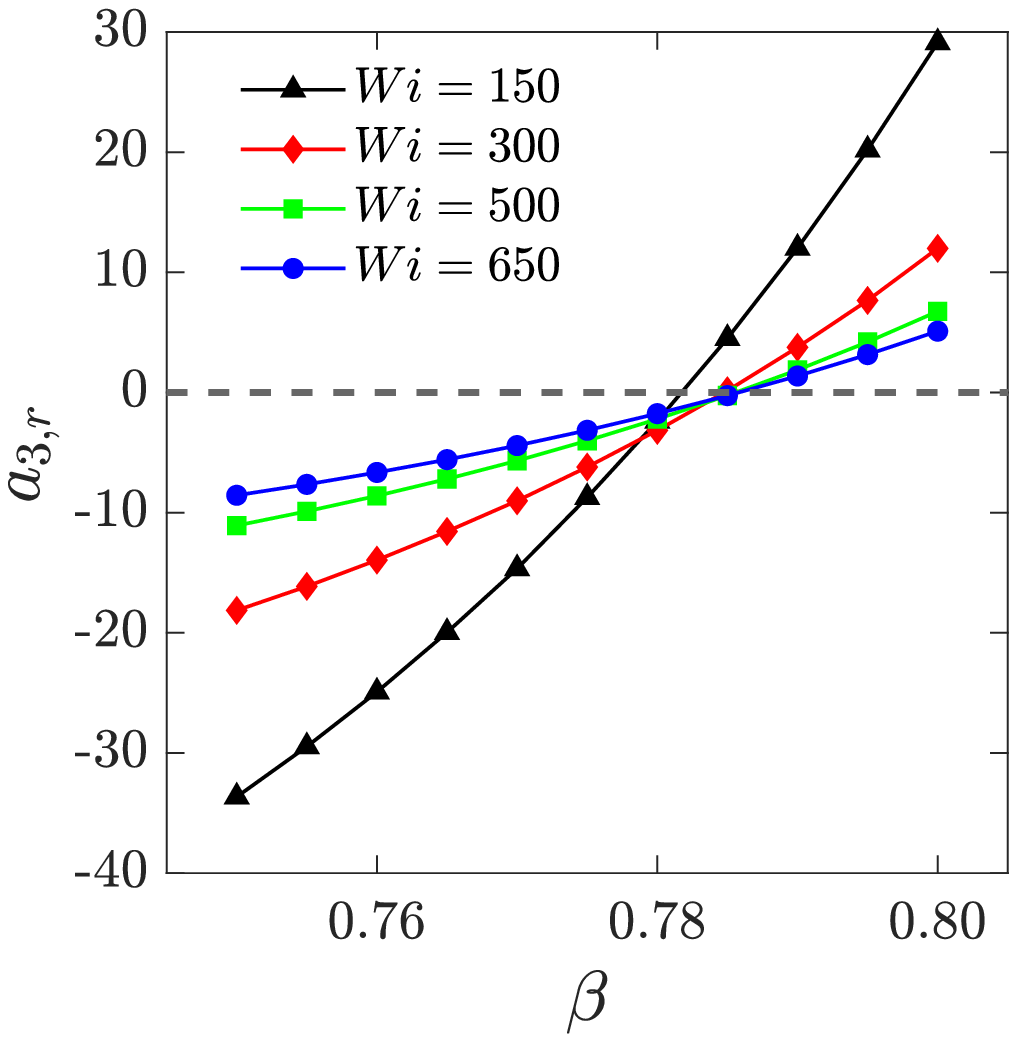}\put(-170,143){$(d)$}
	\caption{Continuous variation of the Landau coefficient with viscosity ratio $\beta$ at various Weissenberg numbers $Wi$: $(a)$ critical Reynolds number $Re_{c}$; $(b)$ critical wavenumber $\alpha_{c}$; $(c)$ Landau coefficient $a_{3,r}$; $(d)$ Landau coefficient $a_{3,r}$ near the bifurcation boundary. Data points below the horizontal dashed line in $(c)$ and $(d)$ denote supercritical bifurcations, while those above the line denote subcritical bifurcations.}
	\label{Fig:critical_a3_Wis}
\end{figure}

\begin{figure}
	\centering
	\includegraphics[width=0.48\textwidth,trim=40 30 40 10,clip]{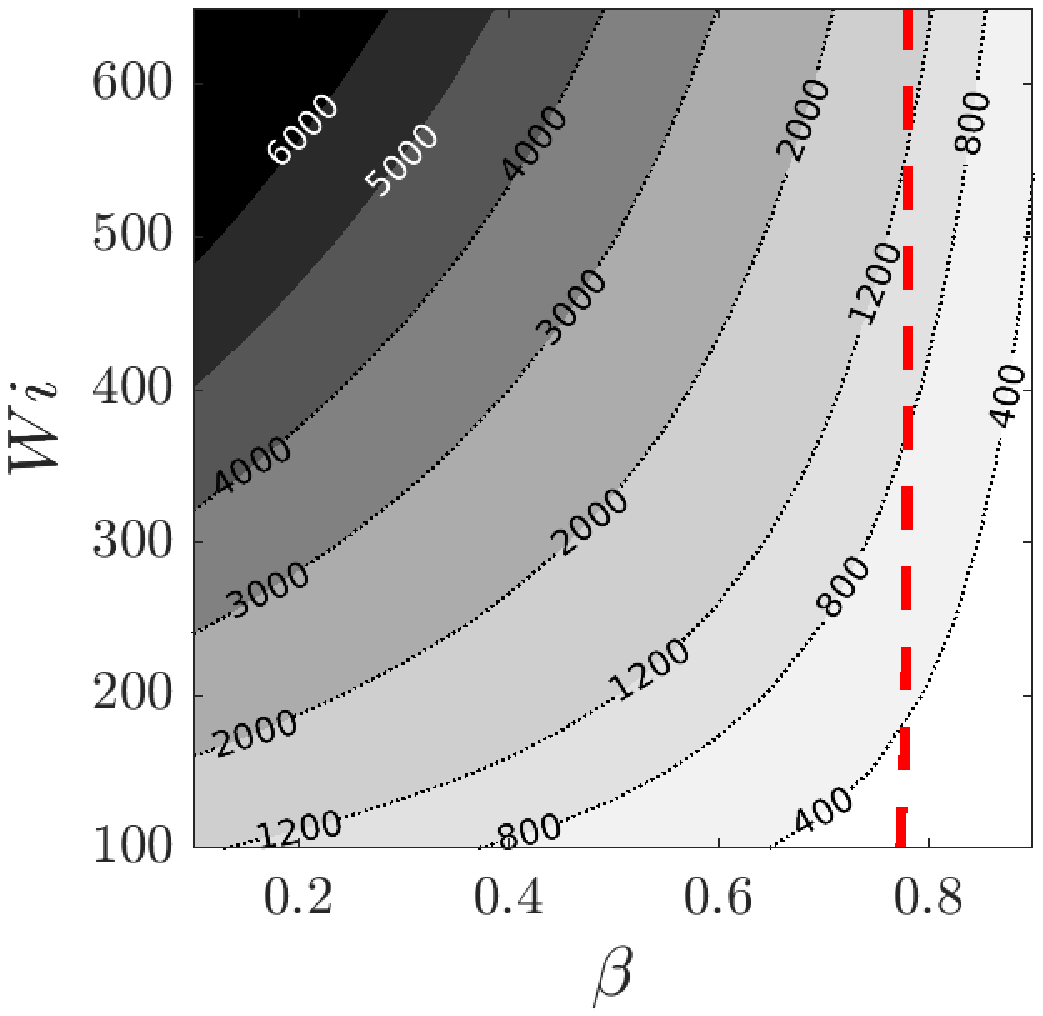}
	\put(-189,135){$(a1)$}
	\includegraphics[width=0.48\textwidth,trim=40 30 40 10,clip]{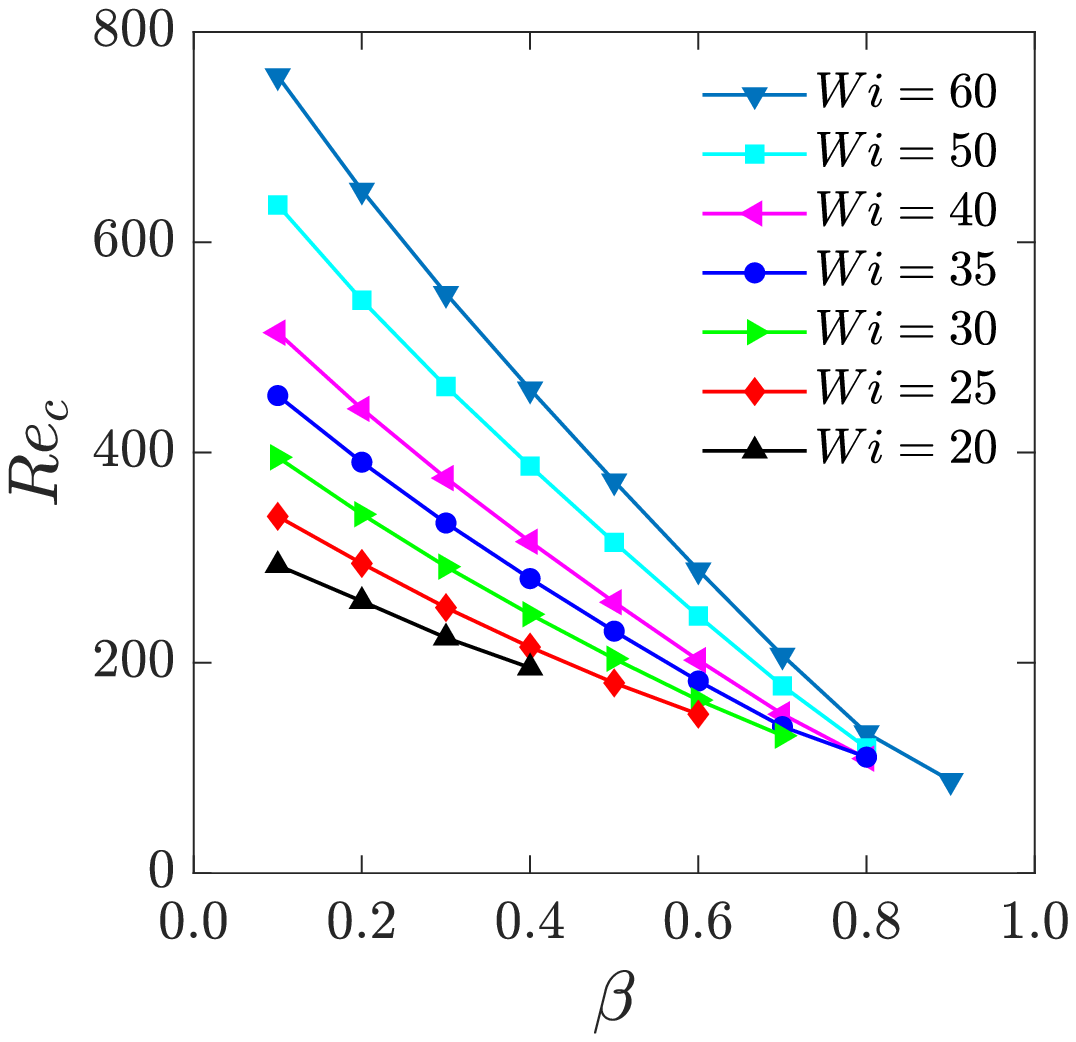}
	\put(-189,135){$(a2)$}\\
	\includegraphics[width=0.48\textwidth,trim=40 30 40 10,clip]{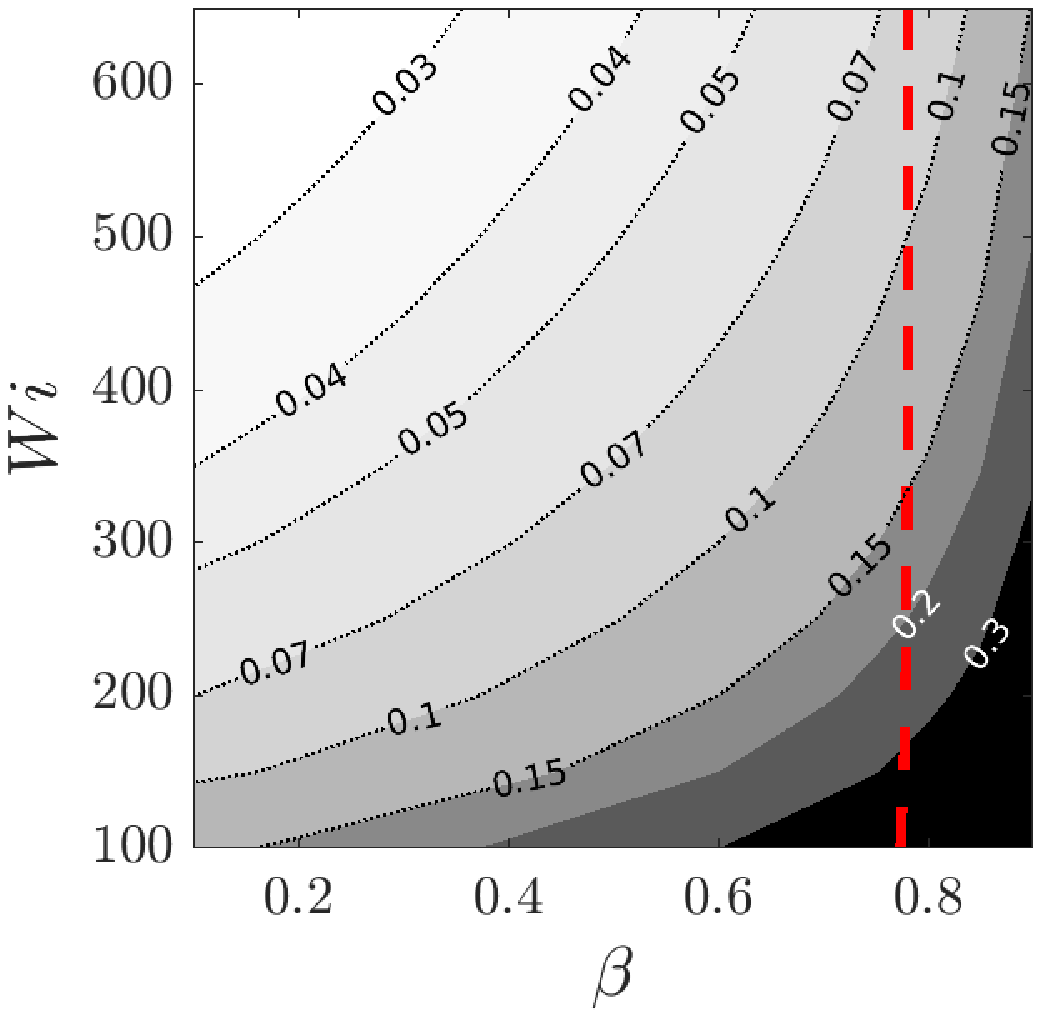}
	\put(-189,135){$(b1)$}
	\includegraphics[width=0.48\textwidth,trim=40 30 40 10,clip]{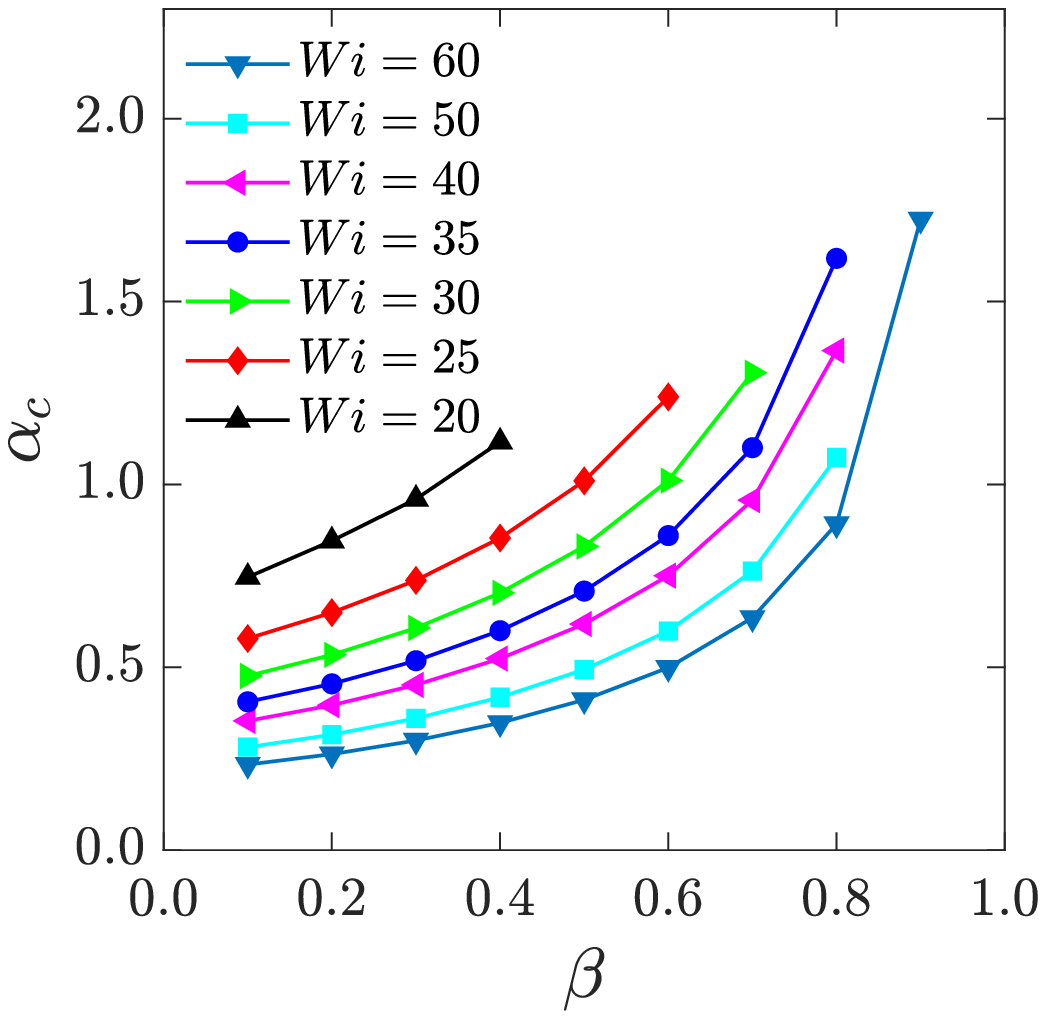}
	\put(-189,135){$(b2)$}\\
	\includegraphics[width=0.48\textwidth,trim=40 0 40 10,clip]{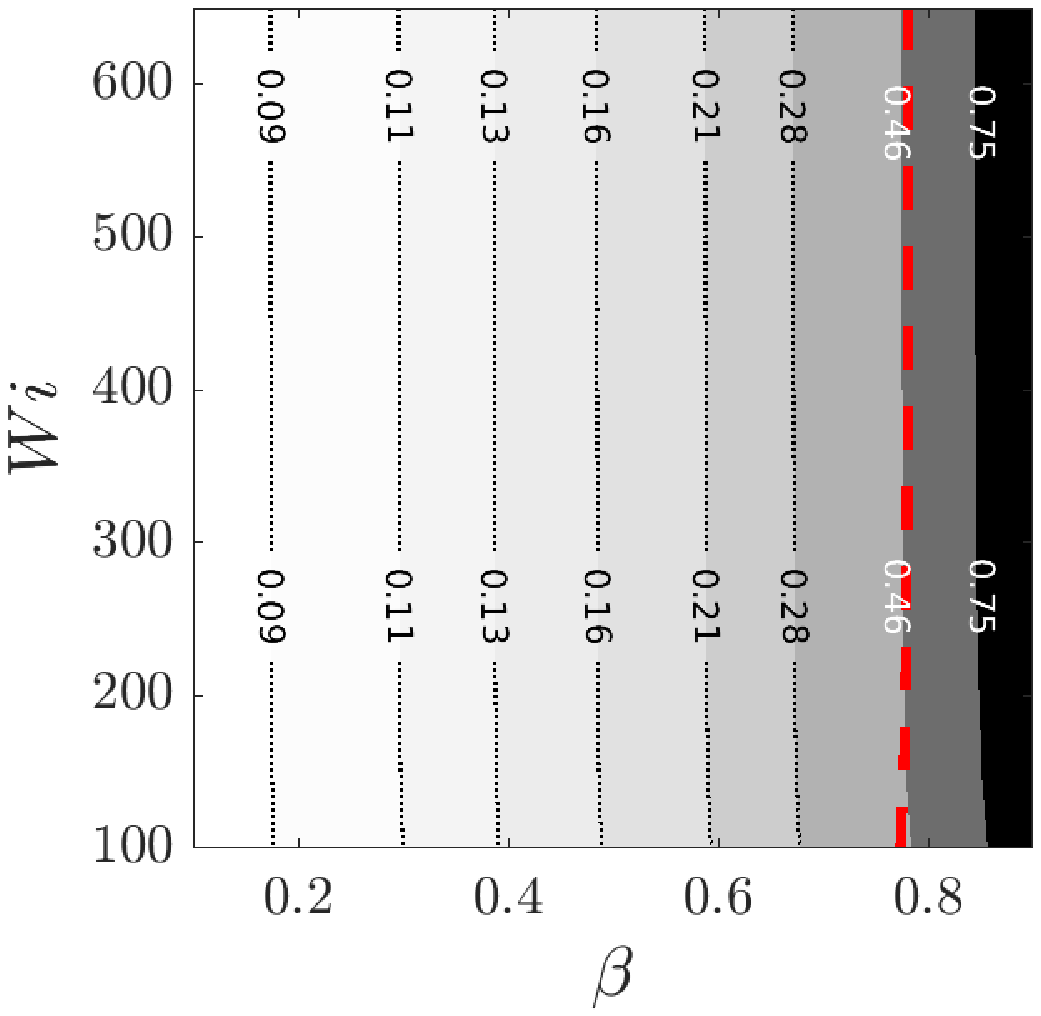}
	\put(-189,149){$(c1)$}
	\includegraphics[width=0.48\textwidth,trim=40 0 40 10,clip]{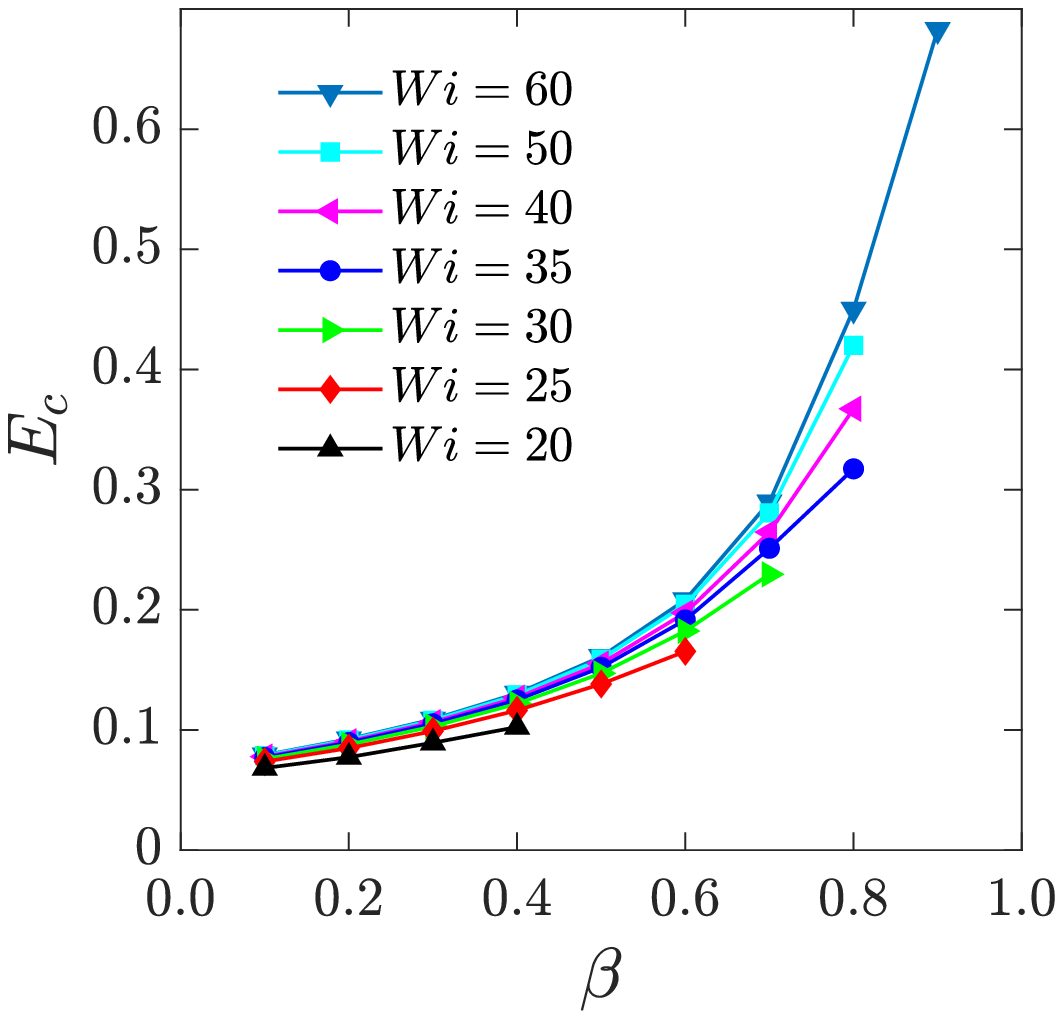}
	\put(-189,149){$(c2)$}\\		
	\caption{Critical conditions in a large $(Wi,\beta)$ plane. $(a1,a2)$ critical Reynolds number $Re_c$; $(b1,b2)$ critical wavenumber $\alpha_c$; $(c1,c2)$ Elasticity number at the critical condition $E_c=Wi/Re_c$. The right column is presented because for some values of $Wi,\beta$, there is no linear instability. The dashed red line labels the bifurcation boundary, see the discussion in figure \ref{Fig:El_2D_Wibeta}.}
	\label{Fig:Realp_2D_Wibeta}
\end{figure}

\begin{figure}
	\centering
	\includegraphics[width=0.48\textwidth,trim=40 0 40 10,clip]{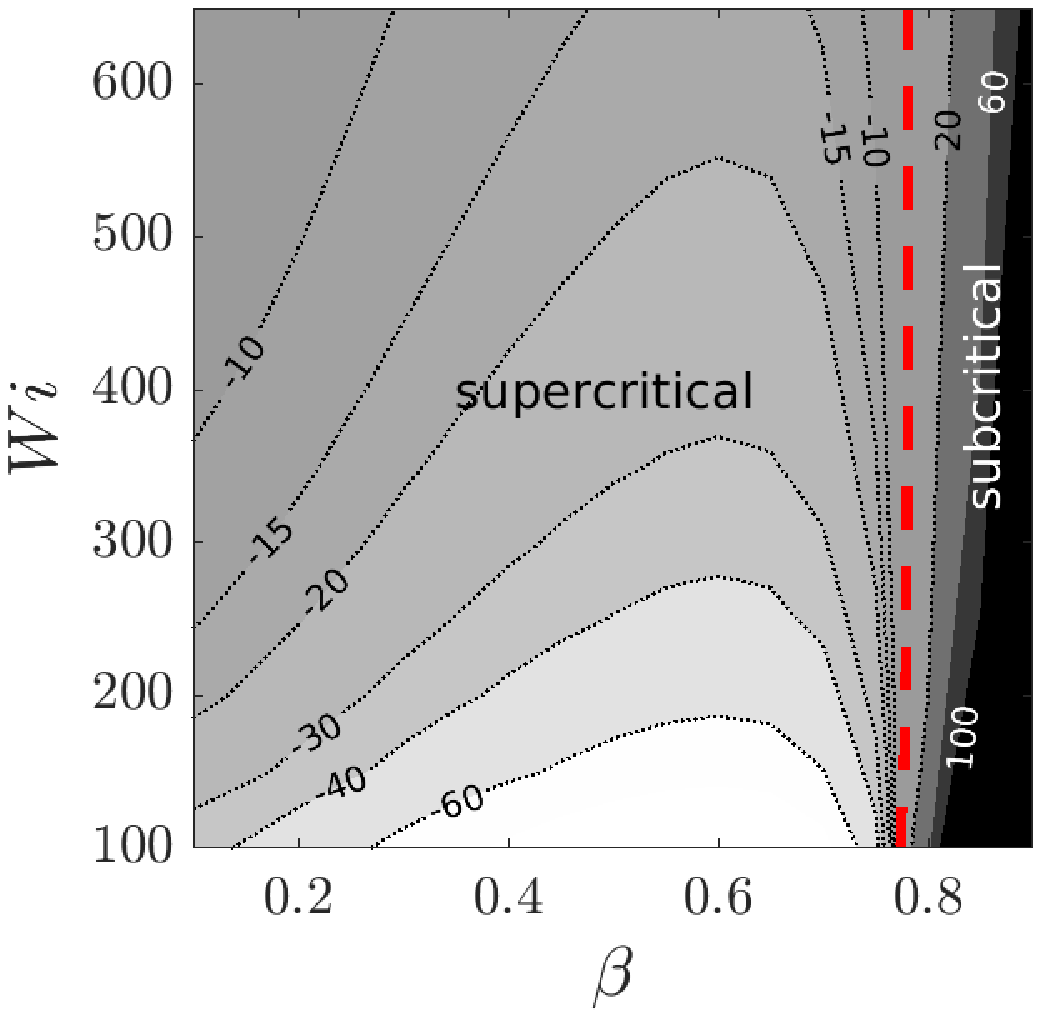}
	\put(-187,150){$(a)$}
	\includegraphics[width=0.48\textwidth,trim=40 0 40 10,clip]{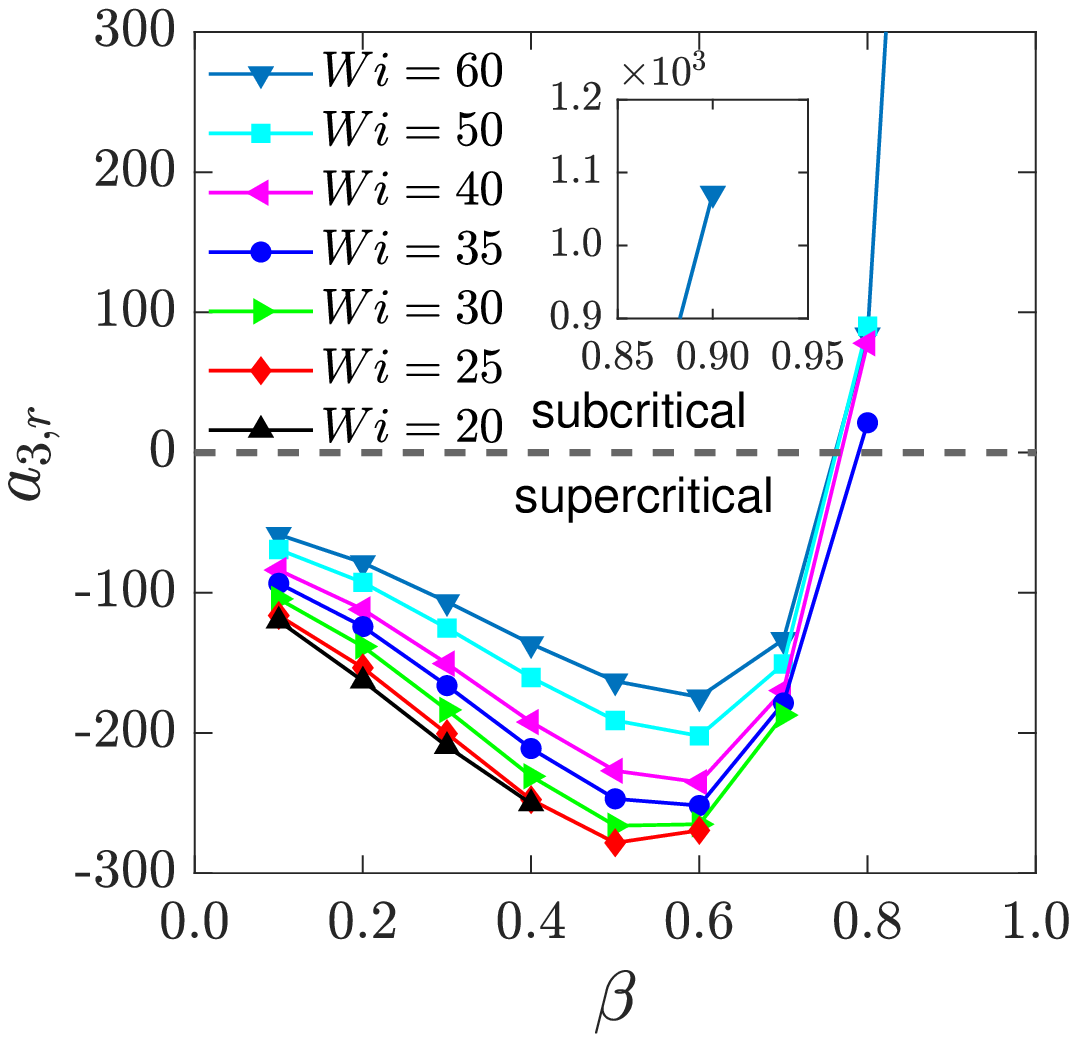}
	\put(-187,150){$(b)$}	
	\caption{Landau coefficient in a $(Wi,\beta)$ plane at linear critical conditions for high $Wi$ ($a$) and in a curve plot for low $Wi$ ($b$). The dashed red line representing $\beta\approx0.785$ in $(a)$ marks the bifurcation boundary, the same as those in figure \ref{Fig:Realp_2D_Wibeta}; this boundary almost coincides with the contour of $E_c \approx 0.46$ in figure \ref{Fig:Realp_2D_Wibeta}($c1$).}
	\label{Fig:El_2D_Wibeta}
\end{figure}

Figures \ref{Fig:critical_a3_Wis}$(a)$ and \ref{Fig:critical_a3_Wis}$(b)$ illustrate how the critical parameters $Re_c$ and $\alpha_c$ vary with the viscosity ratio $\beta$. This variation has been observed and discussed in the previous subsections and thus will not be repeated here. What is of significance here is the continuous but non-monotonic change of $a_{3,r}$ with the variation of $\beta$ from 0.1 to 0.9 as plotted in figure \ref{Fig:critical_a3_Wis}$(c)$. Two key observations can be made as follows. Firstly, at a fixed $Wi$, with $\beta$ increased, the Landau coefficient $a_{3,r}$ decreases first and reaches a minimum and then increases; at about $\beta\approx0.785$, the sign of $a_{3,r}$ changes from negative to positive, suggesting the change of bifurcation type from supercritical to subcritical. Figure \ref{Fig:critical_a3_Wis}$(d)$ shows a close-up of the critical $\beta$ at which the bifurcation type changes within \Dongdong{$0.75 \leqslant \beta \leqslant 0.80$}. It can be seen that for \Dongdong{$300\leqslant Wi\leqslant 650$ } the bifurcation boundary is almost fixed at $\beta\approx0.785$ while it slightly reduces for $Wi=150$. In general, the alteration of bifurcation type from supercritical to subcritical can be attributed to the interplay between the flow field and polymer dynamics (as $\beta$ is related to the ratio between solvent viscosity and polymer viscosity). At large $\beta$, viscous effects mainly result from the solvent and thus cause Newtonian-like subcriticality, while polymer dynamics becomes more significant at small $\beta$, resulting in the change to the supercritical bifurcation. Secondly, by comparing the curves at different $Wi$, one can find that the degrees of both supercriticality and subcriticality reduce at high Weissenberg numbers (e.g., the curve for $Wi=650$ is closer to the horizontal reference line $a_{3,r}=0$ than that for $Wi=150$). This indicates that the destabilising and stabilising of the nonlinear effects in viscoelastic pipe flow are more prone to the changes of polymer concentration ($1-\beta$) when the elastic effect is relatively small, whereas their degrees are less influenced by the polymer concentration when the elastic effect is strong. 
\Dongdong{The same data of $Re_c$ and $\alpha_c$ are further plotted in $Wi-\beta$ planes, as shown in figures \ref{Fig:Realp_2D_Wibeta}($a1,a2,b1,b2$). The panels ($a2,b2$) extend the range of $Wi$ to lower values where we cannot find linear instability for some $\beta\in[0.1,0.9]$. The meaning of the red dashed lines is the bifurcation boundary to be further discussed in figure \ref{Fig:El_2D_Wibeta} shortly. In panels ($c1,c2$), we present the data in terms of $E_c=Wi/Re_c$ (not to be confused with a general elastic number $E=Wi/Re$ and note that in this work we fix $Wi$ to perturb $Re$, instead of fixing $E$ to perturb $Re$). We can see that in the range of high $Wi$ (see figures \ref{Fig:Realp_2D_Wibeta}$(c1)$), these contours are almost vertical, showing that $E_c$ is basically independent of $Wi$ at the linear critical conditions. Moreover, the bifurcation boundary $\beta \approx 0.785$ almost coincides with the contour level of $E_c \approx 0.46$; high levels of $E_c$ are always related to high values of $\beta$.}

In figure \ref{Fig:El_2D_Wibeta}($a$), the values of $a_{3,r}$ are displayed and the bifurcation boundary within the parameter space of \Dongdong{$100\leqslant Wi\leqslant 650$ and $0.1 \leqslant \beta \leqslant 0.9$ } is illustrated in a $(Wi,\beta)$ plane. The dashed red line delimits the bifurcation types in this plane; it is almost a vertical line, with $\beta\approx0.785$. The highest degree of supercriticality (white area) is located at the smallest $Wi=100$ and moderate \Dongdong{$0.3 \lesssim \beta \lesssim 0.7$}, while the highest degree of subcriticality (black area) can be found at the largest $\beta$ and relatively lower $Wi$. When $Wi$ is less than about 60, linear instability ceases to exist for some values of $\beta$ and thus, we do not apply weakly nonlinear analysis to these cases. These results are presented in curves in panel ($b$). Within the parameter range of \Dongdong{$20\leqslant Wi\leqslant 60$ and $ 0.1\leqslant \beta \leqslant 0.9$}, data points not displayed correspond to the absence of linear instability. For example, figure \ref{Fig:Realp_2D_Wibeta}$(a2)$ shows that with the decreasing of $Wi$ down to 20, the linear instability only exists at low $\beta$, implying that polymer effects are essential to the linear instability. This scenario agrees with that in \cite{Chaudhary2021Linear} where a lower $\beta$ is needed to render the flow linearly unstable at smaller $Wi$ (see their figure 23$(b)$). As for the Landau coefficient $a_{3,r}$ shown in figure \ref{Fig:El_2D_Wibeta}($b$), its variation with $\beta$ at various $Wi$ is similar to that observed in figure \ref{Fig:critical_a3_Wis}$(c)$ with a minimum at approximately $\beta\in(0.5,0.6)$ and a change of bifurcation type at approximately $\beta\approx0.785$. 

Based on the above results, one can see that the key parameter governing the bifurcation type is the viscosity ratio $\beta$ (sweeping in $Wi$ does not change the bifurcation type in general). The change of the bifurcation type due to viscosity ratio $\beta$ seems to be supported by the experimental observations in \cite{Samanta2013Elasto-inertial}, where by increasing the polymer concentration (equivalent to decreasing $\beta$) from the Newtonian fluid, hysteresis behaviour can be firstly observed but then disappears when the polymer concentration is beyond 200 ppm ($\beta \approx 0.855$). Although their critical viscosity ratio $\beta \approx 0.855$ (for the existence of hysteresis or not) is slightly larger than our prediction $\beta \approx 0.785$, both their experiments and our predictions imply the existence of a bifurcation boundary defined by $\beta$. The reason for the discrepancy may be that they did not dedicate themselves in determining the critical value of $\beta$ for the bifurcation type and that our results are based on Oldroyd-B axisymmetric flows (other more realistic models such as FENE-P may improve the results). A similar scenario is observed in a more recent experiment by \cite{Choueiri2018Exceeding} where Newtonian-like transition (known to be subcritical) persists at low polymer concentrations but is replaced by a qualitatively different (likely to be supercritical) instability at high polymer concentrations. In addition, \cite{Chandra2020Early} has also suggested the possible crossover of transition type from subcritical to supercritical with the former at low polymer concentrations and the latter at high polymer concentrations in viscoelastic microtube flows. For viscoelastic channel flows, \cite{Graham2014Drag} depicted a phase diagram demonstrating that: subcritical bifurcations exist at small elasticity numbers $E$ where the turbulence is Newtonian-like; meanwhile, increasing $E$ to a sufficiently high value results in early turbulence. He speculated that there is a possible linear instability, which has been confirmed theoretically by \cite{Garg2018Viscoelastic} and \cite{Chaudhary2021Linear}. A similar diagram for viscoelastic pipe flow in \cite{Chaudhary2021Linear} showed that subcritical axisymmetric elasto-inertial structures may exist at low elasticity numbers $E$, while they \Dongdong{may change } to be supercritical at high $E$. In this diagram, the bifurcation boundary seems to be defined by $E$, instead of $\beta$. But it should be noted that their diagram is mainly for $\beta \sim 0.60$ and the situation is likely to be different for other values of $\beta$, according to our results here.

As elucidated above, a single parameter, viscosity ratio $\beta$ in the present case, can change the bifurcation type of viscoelastic pipe flows. Simple flows, such as Newtonian plane Poiseuille flow and pipe Poiseuille flow, are governed by a single control parameter $Re$ and there is only one type of bifurcation. As the flow system becomes complex, where, e.g., two and more competing mechanisms are present, the bifurcation type may be different in different parts of the parameter space. For example, \Dongdong{a weakly nonlinear analysis by \cite{Graham1998Effect} showed that, in circular Couette flows (with imposed axial Poiseuille flow) of UCM fluids, increasing the axial Weissenberg number can change the bifurcation type from subcritical to supercritical. Such change of bifurcation types also exists in Dean flows with an imposed axial Poiseuille flow of both UCM and Oldroyd-B fluids \citep{Ramanan1999Stability}}. \cite{Fujimura1997Degenerate} found that for stably stratified plane Poiseuille flows, the subcritical bifurcation exists in a wide range of Prandtl number while the bifurcation changes to be supercritical in a narrow range of $Pr<0.17$. In this case, the bifurcation boundary in the parameter space is defined by both the Prandtl number and the Richardson number (see their figure 5). For another example of a numerical study of electro-convection of viscoelastic fluids between two horizontal plates subjected to unipolar injection, \cite{Su2020Instability} found that, at $\beta=0.8$, the intrinsic subcritical bifurcation at $Wi=0.2$ changes to be supercritical when $Wi$ increases to 0.5 (see their figures 5 and 6). In this flow, it is $Wi$ that alters the bifurcation type, instead of $\beta$. In addition, it is worth mentioning that the supercritical bifurcations in \cite{Su2020Instability} are also caused by viscoelastic effects, consistent with our present finding about supercriticality at low and moderate $\beta$ where viscoelastic effects are strong. Another example is taken from the work by \cite{Wu2015Numerical}, where a multi-physics problem, i.e. the electro-thermo-convection between two infinite plane plate subjected to unipolar charge injection, was investigated. They found that the bifurcation type can be changed by varying the electric Rayleigh number or Prandtl number or ion mobility number (see the definitions of these dimensionless parameters in their paper). Such complex characteristics of the flow result from the superposition of two different systems: one is the Rayleigh-B\'{e}nard convection, well known to be supercritical; the other is the unipolar charge injection induced convection, which is subcritical \citep{Zhang2016Weakly}. From this point of view, our present results of the change of bifurcation type can also be interpreted as an intermediate state between the subcritical Newtonian pipe flow and the supercritical UCM pipe flow. Thus, at finite viscosity ratios, there exists a change of the bifurcation type.

\Dongdong{\subsubsection{A scaling law of the Landau coefficient $a_3$}}

\begin{figure}
	\centering
	\includegraphics[width=0.50\textwidth,trim=30 5 30 0,clip]{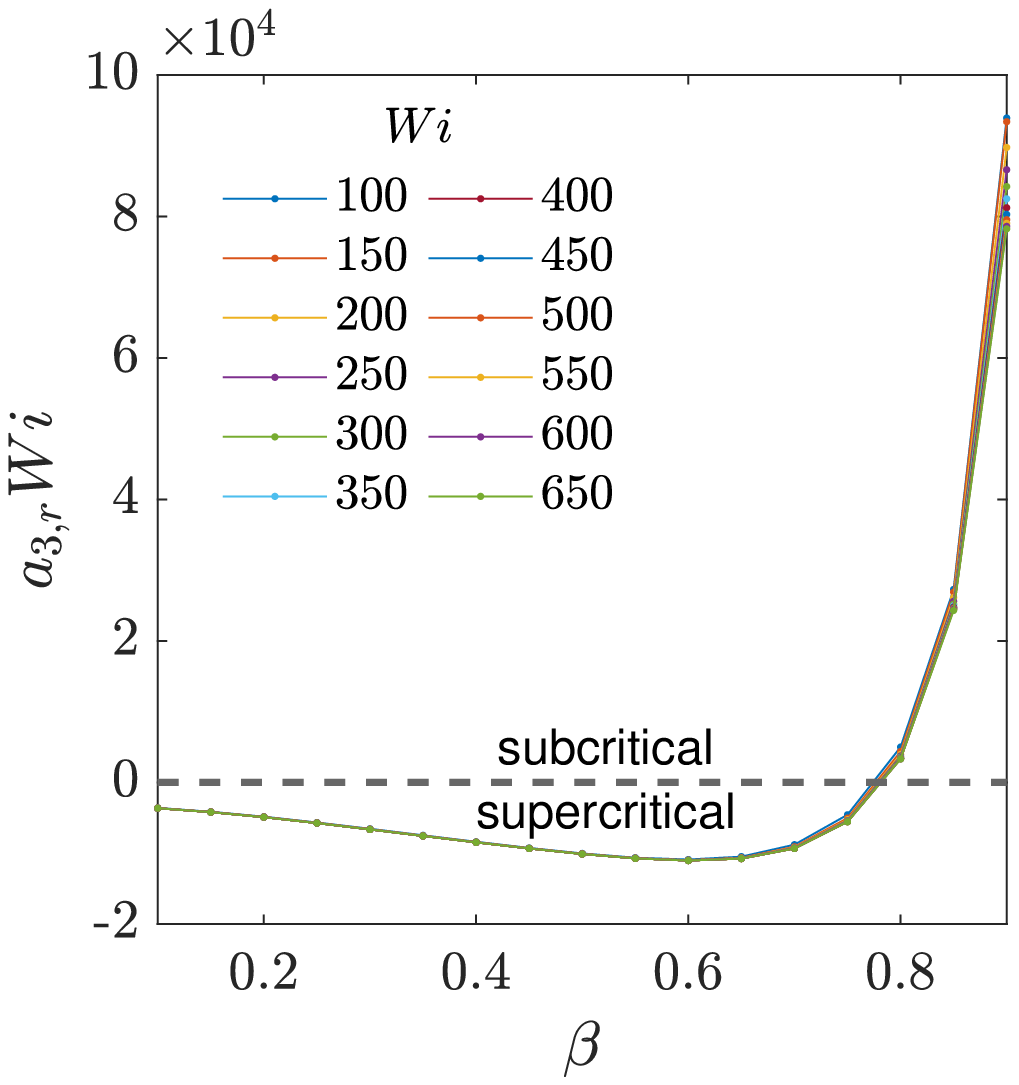}\put(-185,145){$(a)$}
	\includegraphics[width=0.50\textwidth,trim=30 5 30 0,clip]{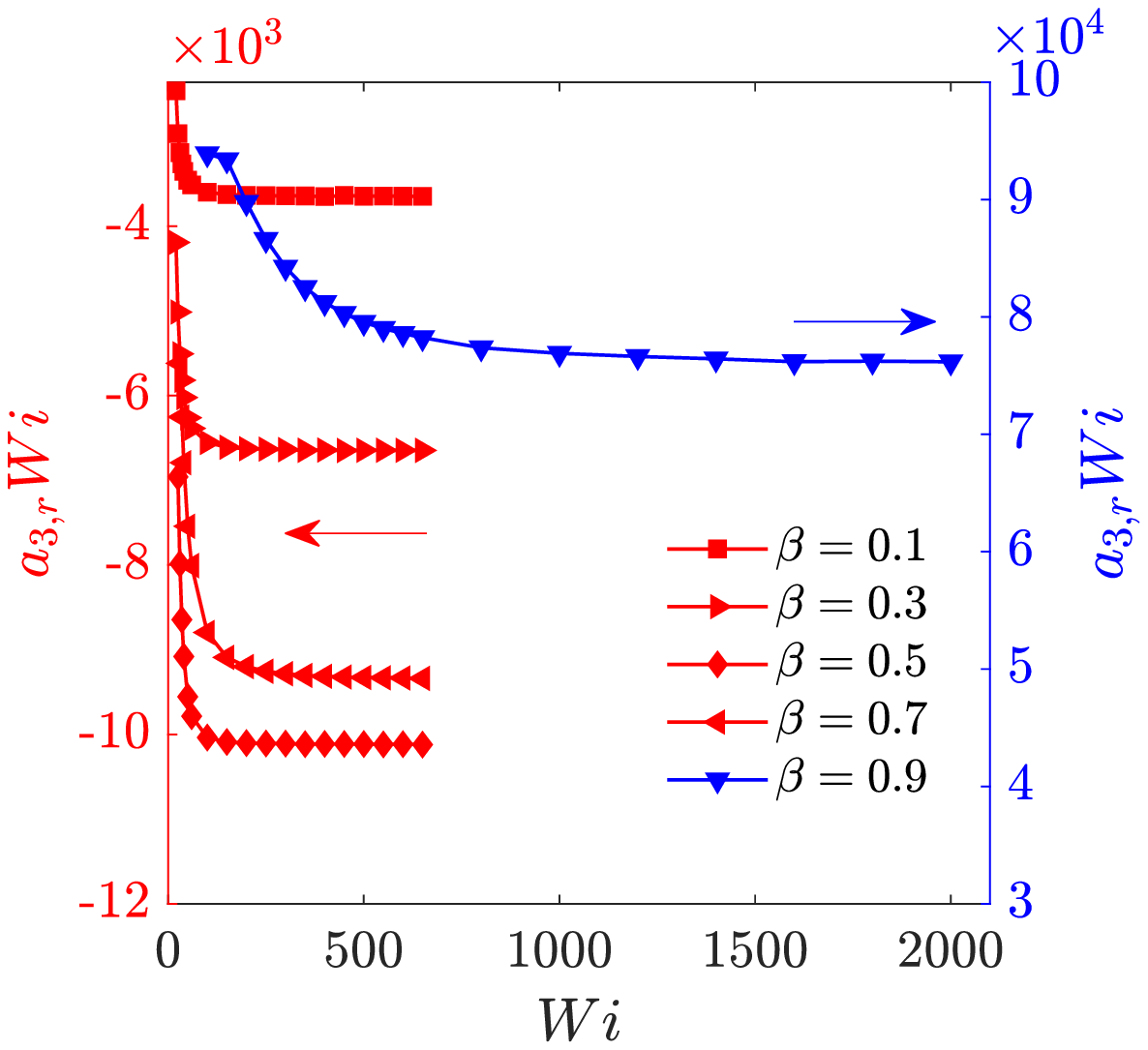}\put(-185,145){$(b)$}
	\caption{A scaling law of the Landau coefficient $a_3$: $(a)$ collapse of the $a_{3,r}Wi$-$\beta$ curves for different $Wi$; $(b)$ values of $a_{3,r}Wi$ as function of $Wi$ for some $\beta$. The horizontal dashed line in $(a)$ marks the bifurcation boundary.}
	\label{Fig:a3Wi_scaling}
\end{figure}

\Dongdong{In the end, we report a scaling law of the Landau coefficient $a_{3}$. } 
\Dongdong{In the results of linear stability analysis, \cite{Chaudhary2021Linear} has reported scalings of critical Reynolds number $Re_c \propto [E(1-\beta)]^{-3/2}$ and critical wavenumber $\alpha_c \propto [E(1-\beta)]^{-1/2}$ for $E(1-\beta) \ll 1$; they justified these scalings by means of a scaling analysis for the boundary layer near the pipe centreline $r=0$. For the present weakly nonlinear analysis, we have also observed a scaling of the Landau coefficient $a_3$, as shown in figure \ref{Fig:a3Wi_scaling}$(a)$. The real part $a_{3,r}$ scales with $1/Wi$ (with $Wi\in[100,650]$) at fixed values of $\beta$ and the scaling also holds for the imaginary part $a_{3,i}$ (not shown). At a first glance, the scaling at a large value of $\beta$ looks inferior; this is because the $Wi$ is not large enough for these $\beta$. Figure \ref{Fig:a3Wi_scaling}$(b)$ shows the variation of $a_{3,r}Wi$ as a function of $Wi$ at several viscosity ratios. When $\beta=0.1, 0.3, 0.5$ or $0.7$, the variation of $a_{3,r}Wi$ is large at low Weissenberg numbers $Wi<100$ and the value of $a_{3,r}Wi$ remains almost constant in the range of $100<Wi<650$. Within the same range of $Wi$ for $\beta=0.9$, however, the value of $a_{3,r}Wi$ changes with $Wi$ significantly and only asymptotically approaches a constant value for much higher $Wi$. These results suggest that the scaling result of $a_{3,r}Wi$ is a high-$Wi$ phenomenon (strong elasticity) and larger $\beta$ (lower polymer concentrations) entails higher $Wi$ to present the scaling law. Considering the scaling results in the linear regime found by \cite{Garg2018Viscoelastic} and \cite{Chaudhary2021Linear} and the scaling in the weakly nonlinear regime discovered in the current work, it would be very interesting to reveal the connection of these two scalings; however, due to space limit, this direction of work, and more possible scalings, will be pursued in a future work.}

\section{Conclusions} \label{Conclusions}

In this paper, we have performed a weakly nonlinear stability analysis of viscoelastic pipe flows of Oldroyd-B fluids based on a multiple-scale expansion method. Specifically, the disturbance is restricted to be axisymmetric, the same as that in \cite{Garg2018Viscoelastic} where the linear instability was reported for the first time. It is the existence of such linear instability that makes our weakly nonlinear analysis feasible and thus the convergence of multiple-scale expansion near the linear critical point can be guaranteed. This is a different situation compared to \cite{Meulenbroek2003Intrinsic,Meulenbroek2004Weakly} where neutral curves do not exist and they had to consider bifurcation from infinity. Resulting from the multiple-scale expansion of space, time, disturbance and control parameter, a complex Ginzburg-Landau equation governing the evolution of disturbance amplitude has been derived. The corresponding Landau coefficient has been evaluated based on two different formulations of the same problem, yielding results in good agreement. In addition, direct numerical simulations have been conducted to verify the bifurcation type  predicted by the present weakly nonlinear analysis. The key finding is that both supercritical and subcritical bifurcations exist in viscoelastic pipe flows and the viscosity ratio (or the polymer concentration) is the control parameter defining the bifurcation type.

The present weakly nonlinear analysis has been performed around the linear instability point. In our revisit of the linear eigenvalue problem, the results we have obtained (in terms of the eigenspectra and neutral curves) agree quantitatively well with those in the literature \citep{Garg2018Viscoelastic,Chaudhary2021Linear}. Moreover, we have found that the pipe Poiseuille flow of UCM fluids (with $\beta=0$) can be linearly unstable and such linear instability ceases to exist for $Wi<14.5$. The corresponding eigenfunction is not localized near the pipe axis, instead, it goes through the whole pipe cross-section. This result supplements those in \cite{Garg2018Viscoelastic} and \cite{Chaudhary2021Linear} where the authors find stable modes in a wide range of governing parameters for the UCM fluids. In addition, the corresponding weakly nonlinear results show that the bifurcation in UCM pipe Poiseuille flow in the inertial regime is supercritical. \Dongdong{It is noted that \cite{Meulenbroek2003Intrinsic} reported subcritical instability of pipe Poiseuille flow of UCM fluids in the inertialess limit; and its counterpart in channel is also subcritical \citep{Meulenbroek2004Weakly}}.

With increasing $\beta$ from the UCM limit, the above-mentioned supercritical bifurcation continues to exist at high polymer concentrations (small $\beta$) but the flow becomes subcritical for low polymer concentrations. In the experimental study of viscoelastic pipe Poiseuille flow by \cite{Samanta2013Elasto-inertial}, they also observed hysteresis in changing $Re$ at low polymer concentrations (signifying subcriticality), and the hysteresis disappeared at high polymer concentrations (implying supercriticality despite experimental uncertainties). These experimental results indicate that changing polymer concentration (or $\beta$) may affect the role the (lowest-order) nonlinearity plays in the flow bifurcation (being stabilising as in supercritical bifurcations or destabilising as in subcritical bifurcations). In terms of the bifurcation boundary defined by the viscosity ratio $\beta$, the present analysis predicts $\beta\approx 0.785$ (above which, the flow is subcritical and below which the flow is supercritical), close to the experimental value of $\beta \approx 0.855$ in \cite{Samanta2013Elasto-inertial}. The discrepancy may be attributed to several reasons, including that the experiments were not designed particularly for determining this bifurcation boundary in $\beta$ and that we have only considered Oldroyd-B model. Another experimental study by \cite{Chandra2020Early} on the flow transition in viscoelastic mircotube flows is also supportive of such a bifurcation boundary in terms of $\beta$. Our results can be viewed as a theoretical exploration of the supercritical and subcritical bifurcations in viscoelastic pipe flows from the perspective of the equations. \Dongdong{Lastly, a scaling law of $a_3$ with $Wi^{-1}$ is observed for values of $\beta\in[0.1,0.9]$ when $Wi$ is sufficiently large. It is conjectured that this scaling in the weakly nonlinear stability analysis is related to the scaling laws in the linear regime as found by \cite{Garg2018Viscoelastic} and \cite{Chaudhary2021Linear}. This newly discovered scaling law in the weakly nonlinear phase will be explained in a future work.}

As mentioned above, the current results are based on viscoelastic pipe flows of Oldroyd-B fluids. For more accurate quantitative comparisons to experiments, more realistic constitutive models deserve to be further analysed. For example, in \cite{Chaudhary2021Linear}, the authors had to resort to the FENE-P model to consider the shear-thinning effect in order to better match the scaling in experiments. The more realistic constitutive model may improve the prediction of bifurcation boundary we found in this work. \Dongdong{Besides, it may also be interesting to investigate the bifurcation boundary in other viscoelastic flows and possible scaling laws of the Landau coefficients there. In revising this paper, we found that \cite{Buza2021} submitted a weakly nonlinear stability analysis of viscoelastic channel flows. Lastly, one can also consider to perturb $Re$ or $Wi$ with fixed $E$ and $\beta$, as the elastic number can facilitate the interpretation of the results.} 

The authors report no conflict of interest.\\
\\
DW is supported by a PhD scholarship (No. 201906220200) from the China Scholarship Council and the NUS research scholarship. MZ acknowledges the financial support of Ministry of Education, Singapore (a Tier 2 grant with the WBS no. R-265-000-661-112).

\begin{acknowledgments}
\end{acknowledgments}

\begin{appendix}

\Dongdong{
\section{An illustration of bifurcation types}\label{Appendix_subsuper}

We present here the bifurcation types as shown in figure \ref{Fig:bifurcationtypes}. 
The solid curves represent stable solutions and the dashed curves unstable solutions. The grey dotted circle represents a neighbourhood of the linear critical condition, where our computation is carried out. Panel ($a$) shows a supercritical bifurcation (with $a_{3,r}<0$ in equation \ref{eq:GL-t2 scale}) and panel ($b$) a subcritical one (with $a_{3,r} > 0$). The arrows in the figures indicate that unstable solutions will eventually become stable (linearly or nonlinearly), but via different bifurcation routes. Routes 1 to 4 represent stable spiral, supercritical bifurcation, subcritical bifurcation and stable spiral, respectively. Panel ($c$) shows a subcritical bifurcation but may appear as a supercritical one in experiments (black dots). The red dotted line represents the disturbance level of the system. This case is particularly relevant for a flow which is very sensitive to the external disturbance (including the systematic one).

\begin{figure}
	\centering
	\includegraphics[width=0.80\textwidth]{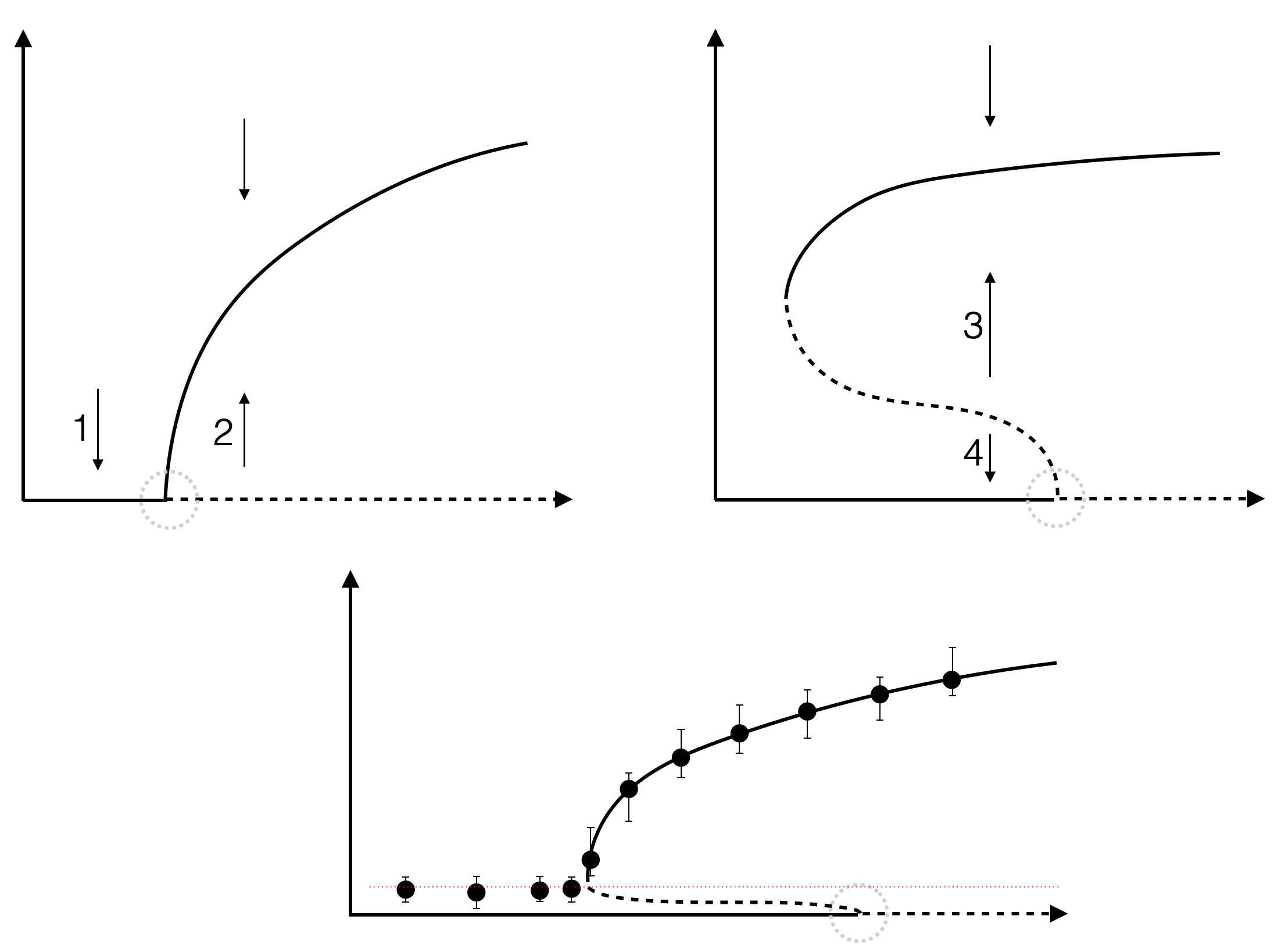}
	\put(-325,215){$(a)$}
	\put(-157,215){$(b)$}
	\put(-247,85){$(c)$}
	\put(-317,165){\rotatebox{0}{$|A|$}}\put(-234,98){$Re$}
	\put(-149,165){\rotatebox{0}{$|A|$}}\put(-74,98){$Re$}
	\put(-238,45){\rotatebox{0}{$|A|$}}\put(-146,-2){$Re$}
	\caption{Illustration of different bifurcation types ($y$-axis: disturbance amplitude $|A|$; $x$-axis: the control parameter $Re$) near the linear critical points (marked with grey dotted circles). Solid curves are stable solutions and dashed curves unstable solutions; arrows indicate the evolution direction of solutions. $(a)$ supercritical bifurcation ($a_{3,r}<0$); $(b)$ subcritical bifurcation ($a_{3,r}>0$); $(c)$ a subcritical bifurcation but may appear to be supercritical in experiments (black dots). The red dotted line indicates the systematic disturbance, which cannot be entirely removed.}
	\label{Fig:bifurcationtypes}
\end{figure}
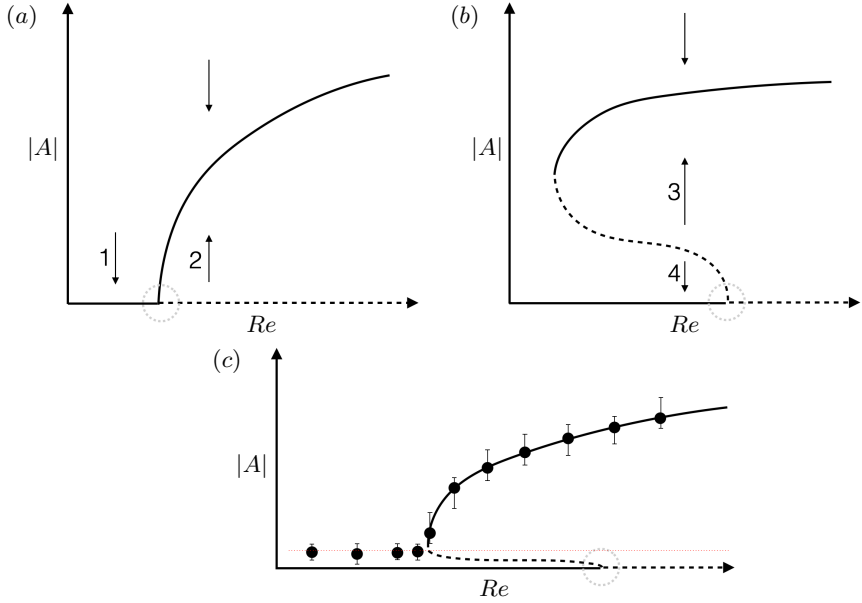

\section{Governing equations in the $\psi$-$g$ formulation}\label{Appendix_psig}

The short notations $F(g_{rr})$ and $F(g_{rz})$ in the governing equation of $\psi$ are
\begin{subequations}
\begin{flalign}
F(g_{rr})=\, &C_{rz} \frac{\partial^2 g_{rr}}{\partial r^2} + \left( 2 C'_{rz} + \frac{C_{rz}}{r} \right) \frac{\partial g_{rr}}{\partial r} + \left( C''_{rz} + \frac{C'_{rz}}{r} - \frac{C_{rz}}{r^2} \right) g_{rr} \notag \\ &+ \left(\frac{C_{rz}^2}{C_{rr}} - C_{rr}\right) \frac{\partial^2 g_{rr}}{\partial r \partial z}  + \left( \left(\frac{C_{rz}^2}{C_{rr}}\right)' - C'_{rr} - \frac{C_{rr}}{r}\right) \frac{\partial g_{rr}}{\partial z} - C_{rz}\frac{\partial^2 g_{rr}} {\partial z^2}, && \\
F(g_{rz})=\, &S \frac{\partial^2 g_{rz}}{\partial r^2} + \left( 2 S' + \frac{S}{r} \right) \frac{\partial g_{rz}}{\partial r} + \left( S'' + \frac{S'}{r} - \frac{S}{r^2} \right) g_{rz} \notag \\ &+ \frac{2S C_{rz}}{C_{rr}} \frac{\partial^2 g_{rz}}{\partial r \partial z} + \left(\frac{2S C_{rz}}{C_{rr}}\right)' \frac{\partial g_{rz}}{\partial z} - S\frac{\partial^2 g_{rz}} {\partial z^2}, && 
\end{flalign}
\end{subequations}
and the governing equations of $\boldsymbol{g}$ using the geometric decomposition are
\begin{subequations}
\begin{flalign}
\frac{\partial g_{rr}}{\partial t} = - U_z \frac{\partial g_{rr}}{\partial z} - \frac{ g_{rr}}{Wi} - 2 \frac{\partial^2 \psi}{\partial r \partial z} + \frac{C'_{rr}}{C_{rr}} \frac{\partial \psi}{\partial z} - \frac{2 C_{rz}}{C_{rr}} \frac{\partial^2 \psi}{\partial z^2} - \frac{N_{c_{rr}}}{C_{rr}}, &&
\end{flalign}
\begin{flalign}
\frac{\partial g_{rz}}{\partial t} =& - U_z \frac{\partial g_{rz}}{\partial z} - \frac{ g_{rz}}{Wi} + \frac{C_{rr}}{S} \frac{\partial^2 \psi}{\partial r^2} + \frac{C_{rr}}{ r S} \frac{\partial \psi}{\partial r} + \frac{2C_{rz}}{S} \frac{\partial^2 \psi}{\partial r \partial z} -\frac{C_{rr}}{r^2 S } \psi \notag \\
&+ \left( \frac{C_{rz}}{ r S} + \frac{C'_{rz}}{S} - \frac{C'_{rr} C_{rz}}{C_{rr} S} \right) \frac{\partial \psi}{\partial z} + \left( \frac{2 C_{rz}^2}{C_{rr} S} - \frac{ C_{zz}}{S} \right) \frac{\partial^2 \psi}{\partial z^2} \notag \\
&+ \frac{C_{rr} U'_z}{S} g_{rr} + \frac{C_{rz} N_{c_{rr}}}{C_{rr}S} - \frac{N_{c_{rz}}}{S}, &&
\end{flalign}
\begin{flalign}
\frac{\partial g_{\theta\theta}}{\partial t} = - U_z \frac{\partial g_{\theta\theta}}{\partial z} - \frac{ g_{\theta\theta}}{Wi}  + \left(\frac{C'_{\theta\theta}}{C_{\theta\theta}} - \frac{2}{r}\right) \frac{\partial \psi}{\partial z}  - \frac{N_{c_{\theta\theta}}}{C_{\theta\theta}}, &&
\end{flalign}
\begin{flalign}
\frac{\partial g_{zz}}{\partial t} =& - U_z \frac{\partial g_{zz}}{\partial z} - \frac{ g_{zz}}{Wi} + 2 \frac{\partial^2 \psi}{\partial r \partial z} + \left( \frac{2}{r} +  \frac{C'_{rr}C_{rz}^2 - 2 C_{rr}C_{rz}C'_{rz} + C_{rr}^2 C'_{zz}}{C_{rr} S^2} \right) \frac{\partial \psi}{\partial z} \notag \\
&+ \frac{2 (C_{rr}C_{rz}C_{zz} - C_{rz}^3)}{C_{rr}S^2} \frac{\partial^2 \psi}{\partial z^2} + \frac{2C_{rr}U'_{z}}{S}g_{rz} - \frac{C_{rz}^2 N_{c_{rr}}}{C_{rr} S^2} + \frac{2C_{rz} N_{c_{rz}}}{ S^2} - \frac{C_{rr} N_{c_{zz}}}{ S^2}. &&
\end{flalign}
\end{subequations}
}

\section{Operators in the $up$-$c$ formulation}\label{Appendix_operators}

\subsection{original operators}\label{Appendix_original_operators}

In the $up$-$c$ formulation,  the weight matrix $\boldsymbol{M}$, linear operator $\boldsymbol{L}$ and nonlinear operator $\boldsymbol{N}$ are given as
\begin{equation}
\boldsymbol{M} = diag(I,I,0,I,I,I,I), 
\end{equation}
\begin{equation}
\boldsymbol{L} = 
\begin{pmatrix}
L_{r,r} & 0 & L_{r,p} & L_{r,rr} & L_{r,rz} & L_{r,\theta\theta} & 0 \\
L_{z,r} & L_{z,z} &  L_{z,p}& 0 &  L_{z,rz} &0 &L_{z,zz} \\
L_{p,r}& L_{p,z} & 0& 0& 0 & 0 & 0 \\
L_{rr,r}&0 &0 & L_{rr,rr}& 0 & 0 & 0\\
L_{rz,r} & L_{rz,z} & 0 & L_{rz,rr} & L_{rz,rz} & 0 & 0\\
L_{\theta\theta,r} & 0 & 0 & 0 & 0 & L_{\theta\theta,\theta\theta} & 0 \\
L_{zz,r} & L_{zz,z} & 0 & 0 & L_{zz,rz} & 0 & L_{zz,zz} \\
\end{pmatrix},
\end{equation}
where the non-zero elements are
\begin{align}
&L_{r,r} = -U_{z} \frac{\partial }{\partial z} + \frac{\beta}{Re} \left( \nabla^{2} - \frac{1}{r^{2}} \right),
& &L_{r,p}=- \frac{\partial}{\partial r}, \notag \\
&L_{r,rr}=\frac{1-\beta}{ReWi} \left( \frac{\partial }{\partial r} + \frac{1}{r}  \right),
& &L_{r,rz}=\frac{1-\beta}{ReWi} \frac{\partial }{\partial z}, \notag \\
&L_{r,\theta\theta}=\frac{1-\beta}{ReWi} \left(- \frac{1}{r}  \right),
& &L_{z,r}= -U_{z}', \notag\\
&L_{z,z}=- U_{z} \frac{\partial }{\partial z}  + \frac{\beta}{Re} \nabla^{2},
& &L_{z,p}= - \frac{\partial}{\partial z}, \notag\\
&L_{z,rz}=\frac{1-\beta}{ReWi} \left( \frac{\partial }{\partial r} + \frac{1}{r} \right),
& &L_{z,zz}=\frac{1-\beta}{ReWi} \frac{\partial }{\partial z}, \notag\\
&L_{p,r}=\frac{\partial }{\partial r} + \frac{1}{r},
& &L_{p,z}=\frac{\partial }{\partial z}, \\
&L_{rr,r}= - C_{rr}'  + 2 C_{rr}\frac{\partial }{\partial r} + 2 C_{rz}\frac{\partial }{\partial z},
& &L_{rr,rr}=- U_{z} \frac{\partial }{\partial z}  - \frac{1}{Wi}, \notag\\
&L_{rz,r}=-  C_{rz}' + C_{rz}\frac{\partial }{\partial r} + C_{zz}\frac{\partial }{\partial z},
& &L_{rz,z}= C_{rr}\frac{\partial }{\partial r} + C_{rz}\frac{\partial }{\partial z}, \notag\\
&L_{rz,rr}= U_{z}',
& &L_{rz,rz}=- U_{z} \frac{\partial }{\partial z} - \frac{1}{Wi}, \notag\\
&L_{\theta\theta,r}=- C_{\theta \theta}' - 2 C_{\theta \theta}\frac{1}{r},
& &L_{\theta\theta,\theta\theta}=- U_{z} \frac{\partial }{\partial z} - \frac{1}{Wi}, \notag\\
&L_{zz,r}= - C_{zz}',
& &L_{zz,z}= 2 C_{rz}\frac{\partial }{\partial r}+ 2 C_{zz}\frac{\partial }{\partial z}, \notag \\
&L_{zz,rz}= 2 U_{z}',
& &L_{zz,zz}=- U_{z} \frac{\partial}{\partial z} - \frac{1}{Wi}, \notag
\end{align}
\begin{equation}
\boldsymbol{N} =- (N_{u_{r}},N_{u_{z}},0,N_{c_{rr}},N_{c_{rz}},N_{c_{\theta \theta}},N_{c_{zz}})^T.
\end{equation}

\subsection{Multiple-scale expansion of operators}\label{Appendix_expansion_operators}
In the $up$-$c$ formulation,  the weight matrix $\boldsymbol{M}$ is a constant matrix and thus
\begin{equation}
\boldsymbol{M}_0 = \boldsymbol{M}, \ \ \ \ \ \boldsymbol{M}_1 = \boldsymbol{M}_2 = \boldsymbol{0}, \ \ \ \ \ \text{so} \, \boldsymbol{M}_1^\circ = \boldsymbol{M}_2^\circ = \boldsymbol{0}.
\end{equation}
The subscale operators for $\boldsymbol{L} = \boldsymbol{L}_{0} + \epsilon \boldsymbol{L}_{1} + \epsilon^{2} \boldsymbol{L}_{2} + O(\epsilon^{3})$ are described as follows. $\boldsymbol{L}_{0}$ can be obtained by simply replacing $\frac{\partial}{\partial z}$ with $\frac{\partial}{\partial z_0}$, and also replacing $Re$ with $Re_c$ in $\boldsymbol{L}$. $\boldsymbol{L}_{1}$ is not so straightforward and is given as
\begin{equation}
\boldsymbol{L}_1 = 
\begin{pmatrix}
L_{r,r1} & 0 & 0 & 0 & L_{r,rz1} & 0 & 0 \\
0 & L_{z,z1} &  L_{z,p1}& 0 &  0 &0 &L_{z,zz1} \\
0& L_{p,z1} & 0& 0& 0 & 0 & 0 \\
L_{rr,r1}&0 &0 & L_{rr,rr1}& 0 & 0 & 0\\
L_{rz,r1} & L_{rz,z1} & 0 & 0 & L_{rz,rz1} & 0 & 0\\
0 & 0 & 0 & 0 & 0 & L_{\theta\theta,\theta\theta 1} & 0 \\
0 & L_{zz,z1} & 0 & 0 & 0 & 0 & L_{zz,zz1} \\
\end{pmatrix} = \boldsymbol{L}_1^\circ \frac{\partial}{\partial z_1},
\end{equation}
where the non-zero elements are
\begin{align}
&L_{r,r1} = -U_{z} \frac{\partial }{\partial z_1} + \frac{\beta}{Re_c} \left( 2 \frac{\partial}{\partial z_0} \frac{\partial}{\partial z_1} \right),
& &L_{r,rz1}=\frac{1-\beta}{Re_c Wi} \frac{\partial }{\partial z_1}, \\
&L_{z,z1}=- U_{z} \frac{\partial }{\partial z_1}  + \frac{\beta}{Re_c} \left(2 \frac{\partial}{\partial z_0} \frac{\partial}{\partial z_1} \right),
& &L_{z,p1}= - \frac{\partial}{\partial z_1},
& &L_{z,zz1}=\frac{1-\beta}{Re_c Wi} \frac{\partial }{\partial z_1}, \notag\\
&L_{p,z1}=\frac{\partial }{\partial z_1},
& &L_{rr,r1}=2 C_{rz}\frac{\partial }{\partial z_1},
& &L_{rr,rr1}=- U_{z} \frac{\partial }{\partial z_1}, \notag\\
&L_{rz,r1}=C_{zz}\frac{\partial }{\partial z_1},
& &L_{rz,z1}=  C_{rz}\frac{\partial }{\partial z_1},
& &L_{rz,rz}=- U_{z} \frac{\partial }{\partial z_1}, \notag\\
&L_{\theta\theta,\theta\theta}=- U_{z} \frac{\partial }{\partial z_1},
& &L_{zz,z1}= 2 C_{zz}\frac{\partial }{\partial z_1},
& &L_{zz,zz1}=- U_{z} \frac{\partial}{\partial z_1}. \notag
\end{align}
and $\boldsymbol{L}_1^\circ$ is constructed by moving $\frac{\partial}{\partial z_1}$ out of $\boldsymbol{L}_1$. $\boldsymbol{L}_{2}$ is given as
\begin{equation}
\boldsymbol{L}_2 = 
\begin{pmatrix}
L_{r,r2} & 0 & 0 & L_{r,rr2} & L_{r,rz2} &L_{r,\theta\theta 2} & 0 \\
0 & L_{z,z2} & 0& 0 &  L_{z,rz2} &0 &L_{z,zz2} \\
0& 0 & 0& 0& 0 & 0 & 0 \\
0&0 &0 & 0& 0 & 0 & 0\\
0 & 0 & 0 & 0 & 0 & 0 & 0\\
0 & 0 & 0 & 0 & 0 & 0 & 0 \\
0 & 0 & 0 & 0 & 0 & 0 & 0 \\
\end{pmatrix} = \boldsymbol{L}_{2Re} + \boldsymbol{L}_{2}^\circ \frac{\partial^2 }{\partial z_1^2},
\end{equation}
where the non-zero elements are
\begin{align}
&L_{r,r2} = -\frac{\beta}{Re_c^2} \left( \frac{\partial^{2}}{\partial r^{2}} + \frac{1}{r}\frac{\partial}{\partial r} + \frac{\partial^{2}}{\partial z_0^{2}} - \frac{1}{r^2}\right) + \frac{\beta}{Re_c} \frac{\partial^{2}}{\partial z_1^{2}}, \\
&L_{r,rr2} = -\frac{1-\beta}{Re_c^2 Wi} \left( \frac{\partial}{\partial r} + \frac{1}{r} \right), \ \ \ \ 
L_{r,rz2} = -\frac{1-\beta}{Re_c^2 Wi} \frac{\partial}{\partial z_0}, \ \ \ \ 
L_{r,\theta\theta 2} = \frac{1-\beta}{Re_c^2 Wi} \left( \frac{1}{r} \right), \notag\\
&L_{z,z2} = -\frac{\beta}{Re_c^2} \left( \frac{\partial^{2}}{\partial r^{2}} + \frac{1}{r}\frac{\partial}{\partial r} + \frac{\partial^{2}}{\partial z_0^{2}} \right) + \frac{\beta}{Re_c} \frac{\partial^{2}}{\partial z_1^{2}}, \notag\\
&L_{r,rz2} = -\frac{1-\beta}{Re_c^2 Wi} \left( \frac{\partial}{\partial r} + \frac{1}{r} \right), \ \ \ \ 
L_{r,zz2} = -\frac{1-\beta}{Re_c^2 Wi} \frac{\partial}{\partial z_0}, \notag
\end{align}
and $\boldsymbol{L}_{2}^\circ$ is constructed with the terms having $\frac{\partial^2 }{\partial z_1^2}$; and $\boldsymbol{L}_{2Re}$ consists of the remaining terms. In the nonlinear term $\boldsymbol{N}= \epsilon^{2} \boldsymbol{N}_{2} + \epsilon^{3} \boldsymbol{N}_{3} + O(\epsilon^{4})$, the subscales are
\begin{subeqnarray}
\boldsymbol{N}_{2} &=&- (N_{u_{r}2},N_{u_{z}2},0,N_{c_{rr}2},N_{c_{rz}2},N_{c_{\theta \theta}2},N_{c_{zz}2})^T,\\
\boldsymbol{N}_{3} &=&- (N_{u_{r}3},N_{u_{z}3},0,N_{c_{rr}3},N_{c_{rz}3},N_{c_{\theta \theta}3},N_{c_{zz}3})^T.
\end{subeqnarray}
where the non-zero elements are given as

\begin{align}
&N_{u_{r}2} = u_{r,1} \frac{\partial u_{r,1}}{\partial r} + u_{z,1} \frac{\partial u_{r,1}}{\partial z_0}, N_{u_{z}2} = u_{r,1} \frac{\partial u_{z,1}}{\partial r} + u_{z,1} \frac{\partial u_{z,1}}{\partial z_0}, \notag\\
&N_{c_{rr}2} = u_{r,1} \frac{\partial c_{rr,1}}{\partial r} + u_{z,1} \frac{\partial c_{rr,1}}{\partial z_0} - 2 c_{rr,1} \frac{\partial u_{r,1}}{\partial r} - 2 c_{rz,1} \frac{\partial u_{r,1}}{\partial z_0}, \\
&N_{c_{rz}2} = u_{r,1} \frac{\partial c_{rz,1}}{\partial r} + u_{z,1} \frac{\partial c_{rz,1}}{\partial z_0} - c_{rr,1} \frac{\partial u_{z,1}}{\partial r} - c_{rz,1} \frac{\partial u_{z,1}}{\partial z_0} - c_{rz,1} \frac{\partial u_{r,1}}{\partial r} - c_{zz,1} \frac{\partial u_{r,1}}{\partial z_0}, \notag\\
&N_{c_{\theta \theta}2} = u_{r,1} \frac{\partial c_{\theta \theta,1}}{\partial r} + u_{z,1} \frac{\partial c_{\theta \theta,1}}{\partial z_0} - \frac{2 u_{r,1}}{r} c_{\theta \theta,1},\notag\\
&N_{c_{zz}2} = u_{r,1} \frac{\partial c_{zz,1}}{\partial r} + u_{z,1} \frac{\partial c_{zz,1}}{\partial z_0} - 2 c_{rz,1} \frac{\partial u_{z,1}}{\partial r} - 2 c_{zz,1} \frac{\partial u_{z,1}}{\partial z_0},\notag \\
\text{and}\notag\\
&N_{u_{r}3} = u_{r,1} \frac{\partial u_{r,2}}{\partial r} + u_{r,2} \frac{\partial u_{r,1}}{\partial r} + u_{z,1} \frac{\partial u_{r,2}}{\partial z_0} + u_{z,1} \frac{\partial u_{r,1}}{\partial z_1} + u_{z,2} \frac{\partial u_{r,1}}{\partial z_0},\notag\\
&N_{u_{z}3} = u_{r,1} \frac{\partial u_{z,2}}{\partial r} + u_{r,2} \frac{\partial u_{z,1}}{\partial r} + u_{z,1} \frac{\partial u_{z,2}}{\partial z_0} + u_{z,1} \frac{\partial u_{z,1}}{\partial z_1} + u_{z,2} \frac{\partial u_{z,1}}{\partial z_0}, \notag\\
&N_{c_{rr}3} = u_{r,1} \frac{\partial c_{rr,2}}{\partial r} +u_{r,2} \frac{\partial c_{rr,1}}{\partial r} + u_{z,1} \frac{\partial c_{rr,2}}{\partial z_0} + u_{z,1} \frac{\partial c_{rr,1}}{\partial z_1} + u_{z,2} \frac{\partial c_{rr,1}}{\partial z_0} \notag\\
& \ \ \ \ \ \ \  - 2 c_{rr,1} \frac{\partial u_{r,2}}{\partial r}  - 2 c_{rr,2} \frac{\partial u_{r,1}}{\partial r} - 2 c_{rz,1} \frac{\partial u_{r,2}}{\partial z_0} - 2 c_{rz,1} \frac{\partial u_{r,1}}{\partial z_1} - 2 c_{rz,2} \frac{\partial u_{r,1}}{\partial z_0}, \notag\\
&N_{c_{rz}3} =  u_{r,1} \frac{\partial c_{rz,2}}{\partial r} +u_{r,2} \frac{\partial c_{rz,1}}{\partial r} + u_{z,1} \frac{\partial c_{rz,2}}{\partial z_0} + u_{z,1} \frac{\partial c_{rz,1}}{\partial z_1} + u_{z,2} \frac{\partial c_{rz,1}}{\partial z_0} \notag\\
& \ \ \ \ \ \ \ - c_{rr,1} \frac{\partial u_{z,2}}{\partial r} - c_{rr,2} \frac{\partial u_{z,1}}{\partial r} - c_{rz,1} \frac{\partial u_{z,2}}{\partial z_0} - c_{rz,1} \frac{\partial u_{z,1}}{\partial z_1} - c_{rz,2} \frac{\partial u_{z,1}}{\partial z_0} \notag\\
& \ \ \ \ \ \ \  - c_{rz,1} \frac{\partial u_{r,2}}{\partial r} - c_{rz,2} \frac{\partial u_{r,1}}{\partial r} - c_{zz,1} \frac{\partial u_{r,2}}{\partial z_0} - c_{zz,1} \frac{\partial u_{r,1}}{\partial z_1} - c_{zz,2} \frac{\partial u_{r,1}}{\partial z_0}, \\
&N_{c_{\theta \theta}3} = u_{r,1} \frac{\partial c_{\theta \theta,2}}{\partial r} +u_{r,2} \frac{\partial c_{\theta \theta,1}}{\partial r}+ u_{z,1} \frac{\partial c_{\theta \theta,2}}{\partial z_0} + u_{z,1} \frac{\partial c_{\theta \theta,1}}{\partial z_1} + u_{z,2} \frac{\partial c_{\theta \theta,1}}{\partial z_0} \notag\\
& \ \ \ \ \ \ \ - \frac{2 }{r} c_{\theta \theta,1} u_{r,2} - \frac{2 }{r} c_{\theta \theta,2} u_{r,1},\notag\\
&N_{c_{zz}3} = u_{r,1} \frac{\partial c_{zz,2}}{\partial r} +u_{r,2} \frac{\partial c_{zz,1}}{\partial r} + u_{z,1} \frac{\partial c_{zz,2}}{\partial z_0} + u_{z,1} \frac{\partial c_{zz,1}}{\partial z_1} + u_{z,2} \frac{\partial c_{zz,1}}{\partial z_0} \notag\\
& \ \ \ \ \ \ \ - 2 c_{rz,1} \frac{\partial u_{z,2}}{\partial r}  - 2 c_{rz,2} \frac{\partial u_{z,1}}{\partial r} - 2 c_{zz,1} \frac{\partial u_{z,2}}{\partial z_0} - 2 c_{zz,1} \frac{\partial u_{z,1}}{\partial z_1} - 2 c_{zz,2} \frac{\partial u_{z,1}}{\partial z_0},\notag
\end{align}
and $\boldsymbol{N}_{3}^{\circ}$ in equation \eqref{eq:Landau-coefficients} can be obtained by singling out the terms of the linear wave $e^{i\alpha z_0 + \mu t_0}$ and then moving the $A$-related coefficient out of the original expression, i.e. in form of $|A|^2 A \boldsymbol{N}_{3}^{\circ} e^{i\alpha z_0 + \mu t_0}$.

\subsection{Adjoint operators}\label{Appendix_adjoint_operators}

For the $up$-$c$ formulation, we take the coefficient matrix in the inner product \eqref{eq:inner-product-definition} to be the identity matrix $\boldsymbol{W}=\boldsymbol{I}$ and the adjoint operators can be then derived from equation \eqref{eq:adjoint}. As a result, $\boldsymbol{M}_0=\boldsymbol{M}_0^\dagger$ is self-adjoint and $\boldsymbol{L}_0^\dagger$ is given as
\begin{equation}
\boldsymbol{L}_0^\dagger = 
\begin{pmatrix}
L_{r,r0}^\dagger &L_{r,z0}^\dagger & L_{r,p0}^\dagger & L_{r,rr0}^\dagger & L_{r,rz0}^\dagger & L_{r,\theta\theta 0}^\dagger & L_{r,zz0}^\dagger \\
0 & L_{z,z0}^\dagger &  L_{z,p0}^\dagger& 0 &  L_{z,rz0}^\dagger &0 &L_{z,zz0}^\dagger \\
L_{p,r0}^\dagger& L_{p,z0}^\dagger & 0& 0& 0 & 0 & 0 \\
L_{rr,r0}^\dagger&0 &0 & L_{rr,rr0}^\dagger&  L_{rr,rz0}^\dagger & 0 & 0\\
L_{rz,r0}^\dagger & L_{rz,z0}^\dagger & 0 & 0 & L_{rz,rz0}^\dagger & 0 & L_{rz,zz0}^\dagger\\
L_{\theta\theta,r0}^\dagger & 0 & 0 & 0 & 0 & L_{\theta\theta,\theta\theta 0}^\dagger & 0 \\
0 & L_{zz,z0}^\dagger & 0 & 0 & 0 & 0 & L_{zz,zz0}^\dagger \\
\end{pmatrix},
\end{equation}
where the non-zero elements are
\begin{align}
&L_{r,r0}^\dagger=U_{z} \frac{\partial }{\partial z_0} + \frac{\beta}{Re_c} \left( \frac{\partial^{2}}{\partial r^{2}} + \frac{1}{r}\frac{\partial}{\partial r} + \frac{\partial^{2}}{\partial z_0^{2}} - \frac{1}{r^2} \right),
& &L_{r,z0}^\dagger=-U_{z}', \notag\\
&L_{r,p0}^\dagger=-\frac{\partial }{\partial r},
& &L_{r,rr0}^\dagger= -2\frac{\partial }{\partial r} -  \frac{2}{r}- 2 C_{rz}\frac{\partial }{\partial z_0},\notag\\
&L_{r,rz0}^\dagger=- 2 C_{rz}' - C_{rz}\frac{\partial }{\partial r} - \frac{C_{rz}}{ r} - C_{zz}\frac{\partial }{\partial z_0},
& &L_{r,\theta\theta 0}^\dagger= \frac{2}{r}, \ \
L_{r,zz0}^\dagger=- C_{zz}', \notag\\
&L_{z,z0}^\dagger= U_{z} \frac{\partial }{\partial z_0}  + \frac{\beta}{Re_c} \left( \frac{\partial^{2}}{\partial r^{2}} + \frac{1}{r}\frac{\partial}{\partial r} + \frac{\partial^{2}}{\partial z_0^{2}} \right),
& &L_{z,p0}^\dagger=-\frac{\partial }{\partial z_0}, \notag\\
&L_{z,rz0}^\dagger=-\frac{\partial }{\partial r}-\frac{1}{r} - C_{rz}\frac{\partial }{\partial z_0}, & \\
&L_{z,zz0}^\dagger= -2 C_{rz}\frac{\partial }{\partial r} -2 C_{rz}'- \frac{2 C_{rz}}{ r} - 2 C_{zz}\frac{\partial }{\partial z_0},
& &L_{p,r0}^\dagger=\frac{\partial}{\partial r} + \frac{1}{ r}, \notag\\
&L_{p,z0}^\dagger= \frac{\partial}{\partial z_0},
& &L_{rr,r0}^\dagger=\frac{1-\beta}{Re_c Wi} \left( -\frac{\partial }{\partial r} \right), \notag\\
&L_{rr,rr0}^\dagger= U_{z} \frac{\partial }{\partial z_0}  - \frac{1}{Wi},
& &L_{rr,rz0}^\dagger= U_{z}', \notag\\
&L_{rz,r0}^\dagger=\frac{1-\beta}{Re_c Wi}  \left( -\frac{\partial }{\partial z_0} \right),
& &L_{rz,z0}^\dagger=\frac{1-\beta}{Re_c Wi} \left( -\frac{\partial }{\partial r} \right), \notag\\
&L_{rz,rz0}^\dagger= U_{z} \frac{\partial }{\partial z_0} - \frac{1}{Wi},
& &L_{rz,zz0}^\dagger= 2 U_{z}', \notag\\
&L_{\theta\theta,r0}^\dagger=\frac{1-\beta}{Re_c Wi} \left(- \frac{1}{r}  \right),
& &L_{\theta\theta,\theta\theta 0}^\dagger=U_{z} \frac{\partial }{\partial z_0} - \frac{1}{Wi}, \notag\\
&L_{zz,z0}^\dagger=\frac{1-\beta}{Re_c Wi} \left( -\frac{\partial }{\partial z_0} \right),
& &L_{zz,zz0}^\dagger=U_{z} \frac{\partial }{\partial z_0} - \frac{1}{Wi}. \notag
\end{align}
The boundary conditions for the adjoint problem are the same as those for the direct problem.

\section{Validation of the Landau coefficients using direct numerical simulations}\label{Appendix_DNS_validation}

\subsection{Numerical methods for the DNS}
We have conducted direct numerical simulation (DNS) to evaluate the Landau coefficient and validate our weakly nonlinear results. To simulate the pressure-driven viscoelastic pipe flow, an additional pressure gradient term $4(1+\alpha_f)/Re \, \hat{\boldsymbol{e}}_z$ is added to the right hand side of the Navier-Stokes equation \eqref{eq:continuity-NS} where $\alpha_f$ is a fluctuating coefficient adjusted at every time step to maintain a constant flux in the flow. The governing equation of the conformation tensor remains the same as in equation \eqref{eq:conformation-tensor}. This equation system is solved by a numerical solver developed based on an open-source code \emph{openpipeflow} \citep{Willis2017Openpipeflow}. Fourier-Chebyshev pseudo-spectral method and a high order finite-difference method is adopted in the homogeneous (azimuthal and axial) directions and in the radial direction respectively. The no-slip boundary condition for velocity and the incompressibility condition for the flow field are enforced by a pressure-Poisson equation formulation and an influence-matrix technique is applied to enforce continuity. A second-order predictor-corrector scheme based on the Crank-Nicolson method is employed for the time-stepping. All the DNS are performed with an initial condition constructed with the eigenfunction obtained from the corresponding linear eigenvalue problem. The radial resolution is set by $N_r$ being the number of nodes in the radial direction and the time step size $dt$ can be fixed or adaptive subject to a CFL number of 0.25. \Dongdong{A more detailed description of this code has been reported in \cite{Sun2021}.}

\subsection{Convergence check of the DNS results}

\begin{figure}
	\centering
	\includegraphics[width=0.45\textwidth,trim= 30 0 65 0,clip]{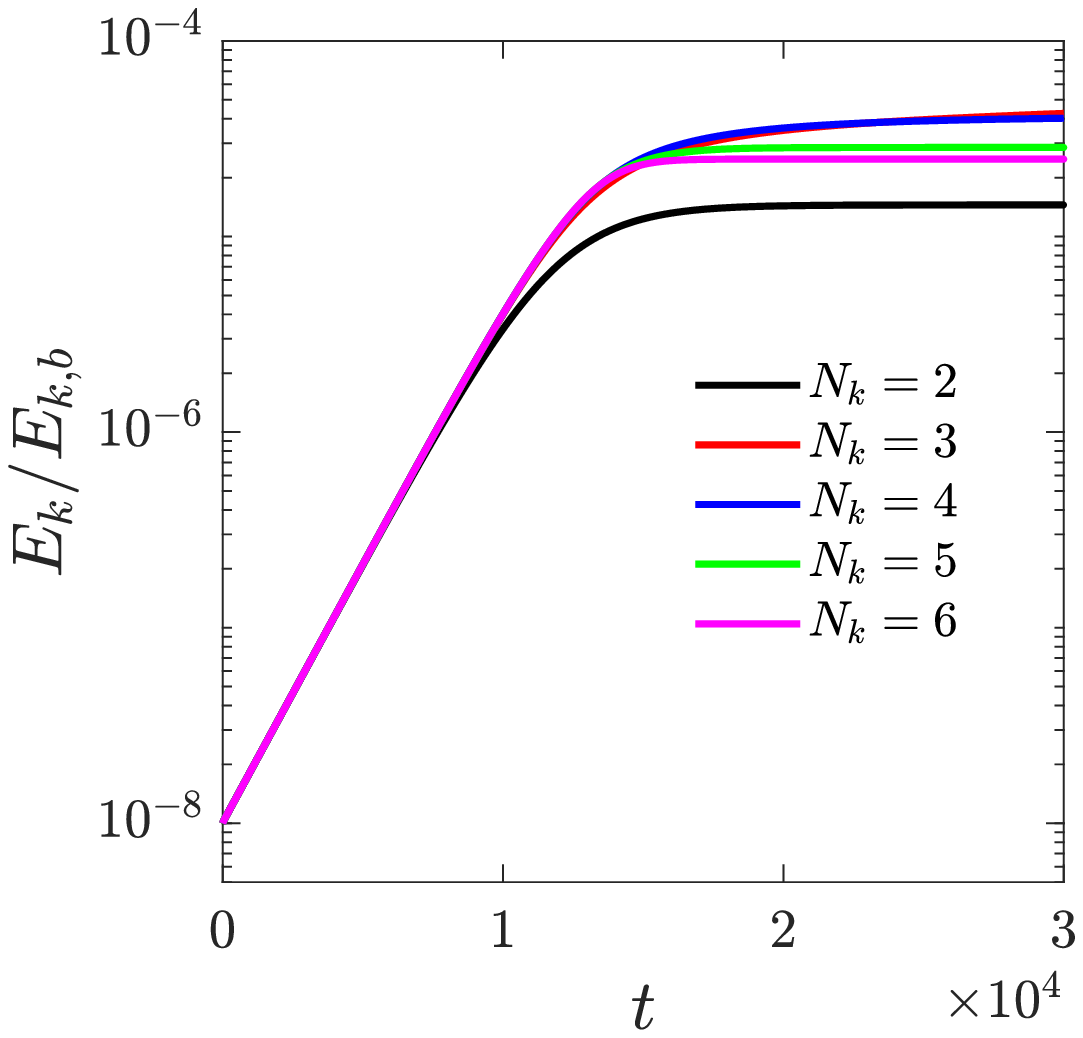}
	\put(-173,150){$(a)$}
	\includegraphics[width=0.45\textwidth,trim= 30 0 65 0,clip]{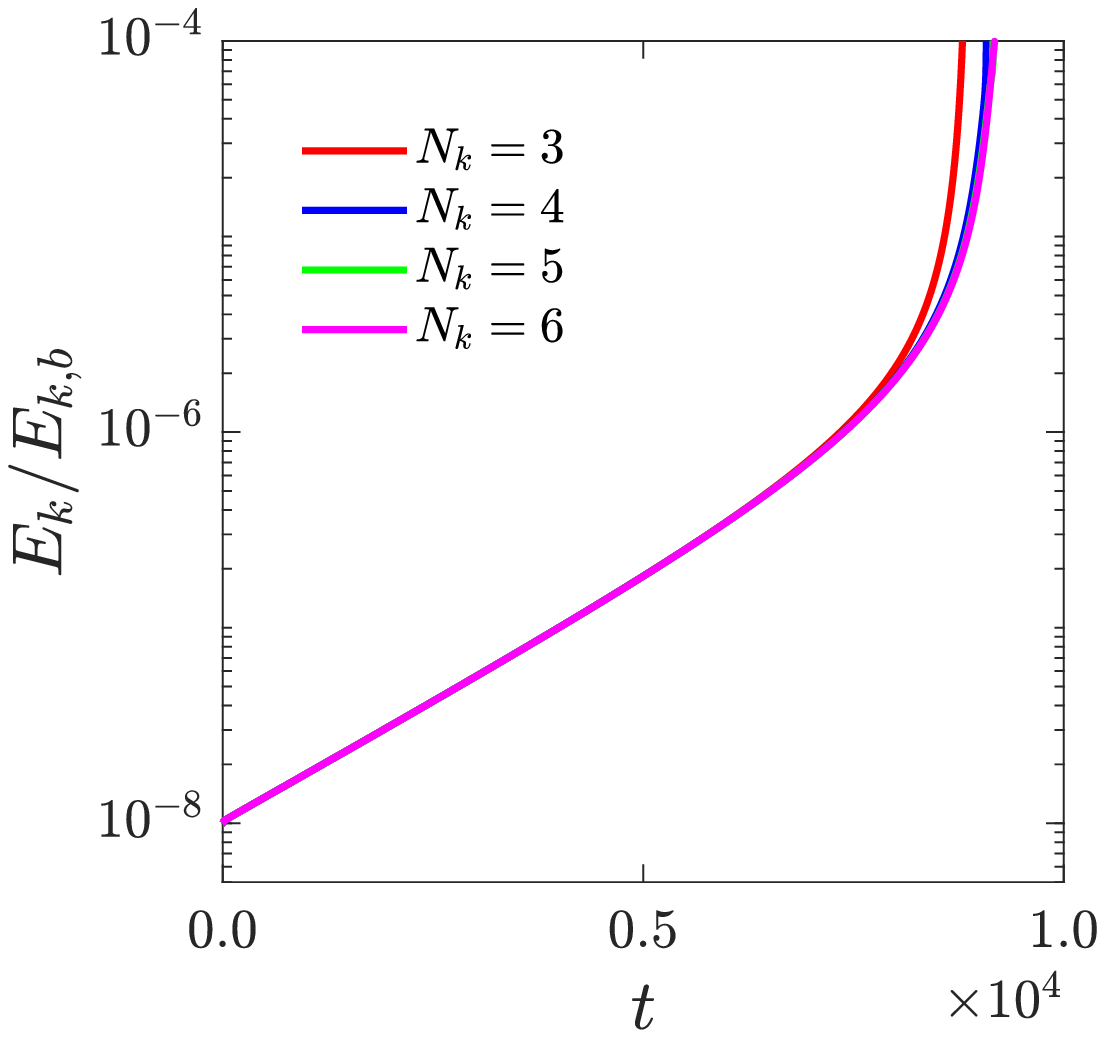}
	\put(-173,150){$(b)$}
	\caption{Evolution of disturbance kinetic energy (normalized by the kinetic energy of the corresponding Newtonian base flow $E_{k,b}$) at: $(a)$ $Re=280$, $\alpha_{c}=0.515525$, $Wi=65$ and $\beta=0.65$; $(b)$ $Re=90$, $\alpha_{c}=1.550722$, $Wi=65$ and $\beta=0.90$. These results are obtained from DNS and $N_k$ is number of Fourier modes in the axial direction. Generally, $N_{k}\ge 4$ suffices to capture the weak nonlinearity for the evaluation of the Landau coefficients.}
	\label{Fig:kinetic_energy_DNS}
\end{figure}

To provide an additional validation, we use the DNS method to investigate the evolution of a small disturbance, which starts from the linear phase, undergoes a weakly nonlinear phase and then saturates. As there is no diffusion term in the original governing equation of conformation tensor, sharp gradient can cause numerical instability. A widely adopted strategy to solve this problem is to add small amount of global artificial diffusion to the equation (see \cite{Sureshkumar1995} and \cite{Lopez2019Dynamics}). However, we note that even a quite small amount of artificial diffusion can alter the linear and weakly nonlinear phases, and thus we choose to not add any artificial diffusion. This strategy is feasible because only few modes need to be resolved for the evaluation of the Landau coefficient $a_{3,r}$ (because the multiple-scale expansion is based on several linear and nonlinear modes and we do not need to resolve all the nonlinearities). In this case, increased spacial resolution in the radial direction and smaller timestep size are needed to be applied: we find that the resolution of $N_k=2$ (modes in streamwise direction), $N_r=256$ and $dt=0.001$ suffices to accurately resolve the linear phase at $Re=800$, $\alpha=1$, $Wi=30$ and $\beta=0.65$ (in this case, the growth rate in the linear phase evaluated using DNS is $-0.01420$, which agrees well with the linear analysis prediction value $-0.01425$; and the relative error is about 0.35\%). The number of Fourier modes in the axial direction $N_k$ has also been examined as shown in figure \ref{Fig:kinetic_energy_DNS}. We found that inclusion of more modes indeed makes the computation more difficult. However, only the initial derivation from the linear phase is necessary to evaluate the Landau coefficient $a_{3,r}$. The result indicates that the essential nonlinearities can be faithfully resolved with $N_{k}\ge 4$.

\subsection{Comparison of the DNS results with weakly nonlinear predictions}
As we are mainly interested in the global amplification of disturbance amplitudes, we will consider the following simplified Landau model
\begin{equation} \label{eq: Landau_amplitude}
\frac{d |A|}{d t} = c_{1} |A| + c_{3} |A|^{3},
\end{equation}
where $|A|$ is the amplitude of the perturbation, $c_{1}$ is the linear growth rate of the amplitude and $c_{3}$ determines the bifurcation type. The original Landau equation is in a series form and the truncation up to the third order is consistent with our theoretical derivation. In order to compare quantitatively with our weakly nonlinear analysis results, the global amplitude $|A|$ in DNS is defined based on the total energy of the disturbances contained in the mode \Dongdong{($k=1,n=0$, i.e. the first axial Fourier mode in axisymmetric flows) } as
\begin{equation}\label{eq:total_energy}
|A| = \sqrt{\frac{1}{2} \int_{0}^{1} \left( \left(|\tilde{u}_r|^2 + |\tilde{u}_z|^2 \right)  +  \frac{1-\beta}{Re Wi}  \left( |\tilde{g}_{rr}|^{2}+2|\tilde{g}_{rz}|^{2}+|\tilde{g}_{\theta\theta }|^{2}+|\tilde{g}_{zz}|^2\right) \right) \, rdr}.
\end{equation}

\begin{figure}
	\centering
	\includegraphics[width=0.33\textwidth,trim= 40 0 65 0,clip]{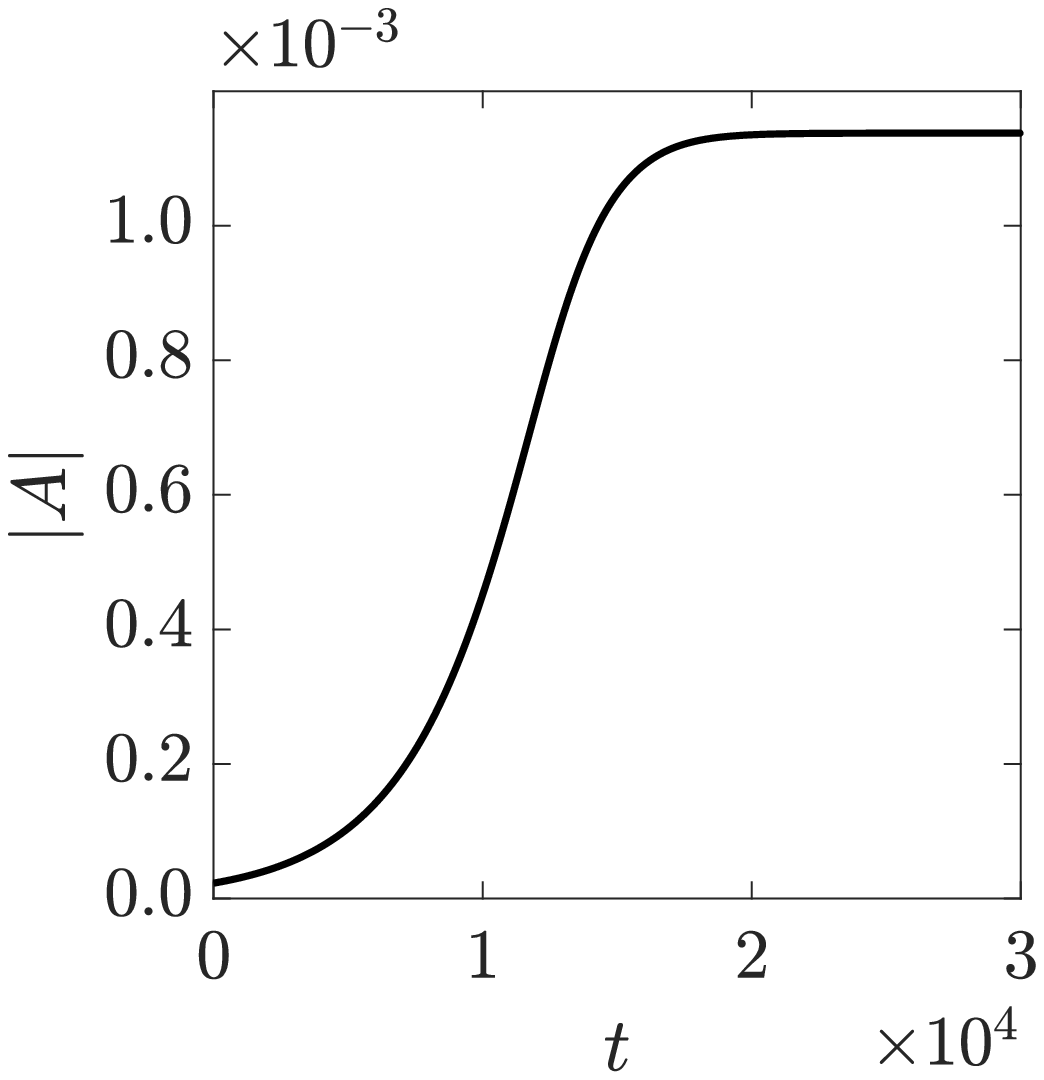}
	\put(-129,110){$(a)$}
	\includegraphics[width=0.33\textwidth,trim= 40 0 65 0,clip]{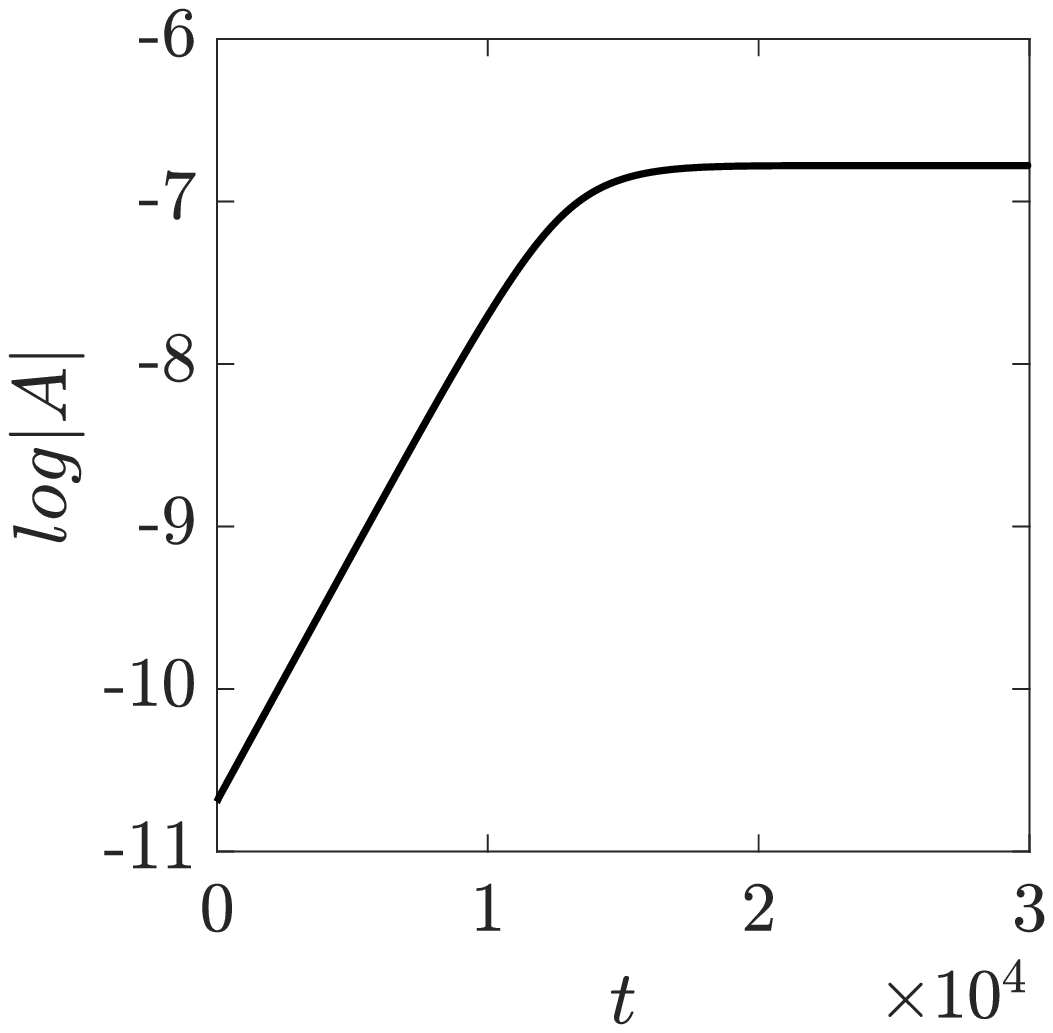}
	\put(-129,110){$(b)$}
	\includegraphics[width=0.33\textwidth,trim=40 0 65 0,clip]{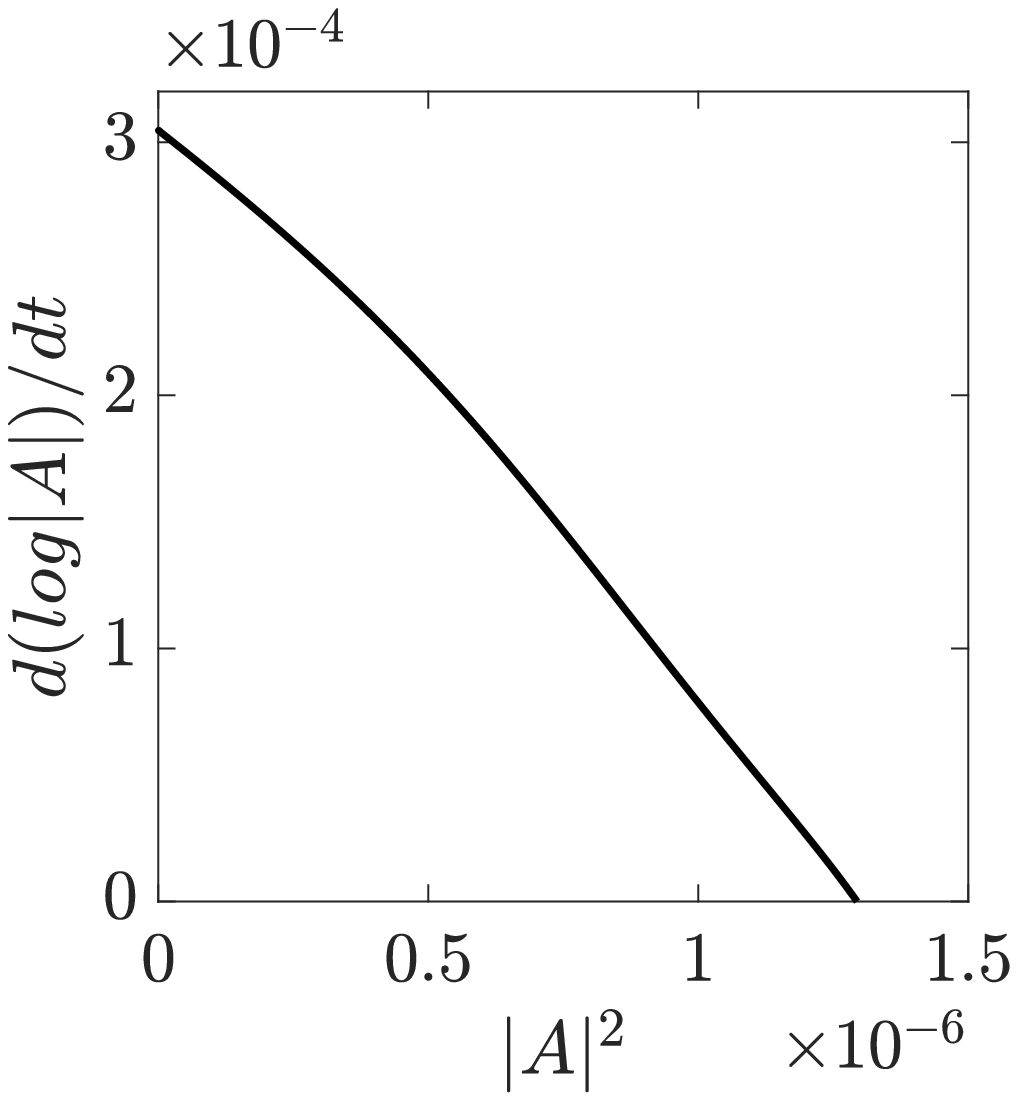}
	\put(-123,110){$(c)$}
	\caption{Evolution and post-processing of disturbance amplitude $|A|$ at $Re=280$, $\alpha_{c}=0.515525$, $Wi=65$ and $\beta=0.65$ obtained with $N_r=400$ and $N_k=5$ from the DNS: $(a)$ original signal; $(b)$ log plot of $(a)$ demonstrating the linear growth and nonlinear saturation; $(c)$ time derivative of $log|A|$ versus $|A|^{2}$ with the vertical intercept giving the linear growth rate $c_{1} \approx 3.05 \times 10^{-4}$ and the gradient near this intercept point giving the Landau coefficient $c_{3} \approx -168$.}
	\label{Fig:Re280_a3}
\end{figure}

Based on the convergence check of the kinetic energy evolution curve, $N_k=5$ is applied in DNS for the evaluation of Landau coefficient. The evolution of $|A|$ as defined in equation \eqref{eq:total_energy} is recorded as shown in figure \ref{Fig:Re280_a3}$(a)$ where $|A|$ grows and finally saturates. From the log plot shown in figure \ref{Fig:Re280_a3}$(b)$, an apparent linear growth stage can be observed followed by a nonlinear saturation phase. The Landau coefficient $c_{1}$ and $c_{3}$ can be calculated by recasting equation \eqref{eq: Landau_amplitude} in the form of $\frac{d (log|A|)}{d t} = c_{1} + c_{3} |A|^{2},$ and plotting $d (log|A|)/dt$ versus $|A|^{2}$ so that the vertical intercept point gives an estimation of $c_{1}$ and the gradient near this point is an approximation of $c_{3}$ \citep{Thompson2001Kinematics}. Such a plot is shown in figure \ref{Fig:Re280_a3}$(c)$ and the Landau coefficients can be determined as $c_{1} \approx 3.05 \times 10^{-4}$ and $c_{3} \approx -168$. It should be noted that the Ginzburg-Landau equation \eqref{eq:GL-t2 scale} derived from the weakly nonlinear analysis is in a $t_{2}$ timescale, while the simplified Landau equation \eqref{eq: Landau_amplitude} used for the DNS data is in a $t$ timescale, and the coefficients are related by $c_{1}=a_{1}(Re-Re_{c})$ and $c_{3}=a_{3,r}$. For the present case with $Re=280$ and $Re_{c}=265.572335$, $a_{1}(Re-Re_{c}) \approx 3.13\times 10^{-4}$, which is close to $c_{1}$ with a relative error of about $2.6\%$; \Dongdong{$a_{3,r} \approx -154$ is in agreement with $c_{3}$ with a relative error of about $9.1\%$ (see table \ref{Tab:convergence-pipe-upc} for the value of $a_{1}$ and also $a_{3,r}$)}. The sources of the error can be: (1) few $N_k$ modes; (2) relative distance to the linear criticality condition. A further DNS test at $Re=270$ being closer to the critical value $Re_{c}=265.572335$ with other parameter unchanged, gives $c_{3} \approx -164$, which is also closer to the theoretical prediction $a_{3,r} \approx -154$ with a smaller relative error of $6.5\%$. \Dongdong{If $Re$ is too close to the critical value, the exponential-growth phase will take a much long time}. In addition to the supercritical cases, the weakly nonlinear analysis of a subcritical case at $Re_c=84.644523$, $\alpha_{c}=1.550722$, $Wi=65$, $\beta=0.90$ predicts $a_{3,r} \approx 1179$, which also agrees with the DNS results $c_{3} \approx 1241$ obtained at $Re=90$ with a relative error of $5.3\%$. For this subcritical case, although the DNS finally blows up as shown in figure \ref{Fig:kinetic_energy_DNS}$(b)$, the weak nonlinearity relating to the initial derivation can still be captured, suggesting that the simplification by using few axial modes in DNS is reasonable. The DNS results double-check our theoretical prediction and verify the bifurcation types in different conditions.

\section{Linear instability in the UCM limit}\label{app_UCM}

\begin{figure}
	\centering
	\includegraphics[width=0.48\textwidth,trim= 0 0 0 0,clip]{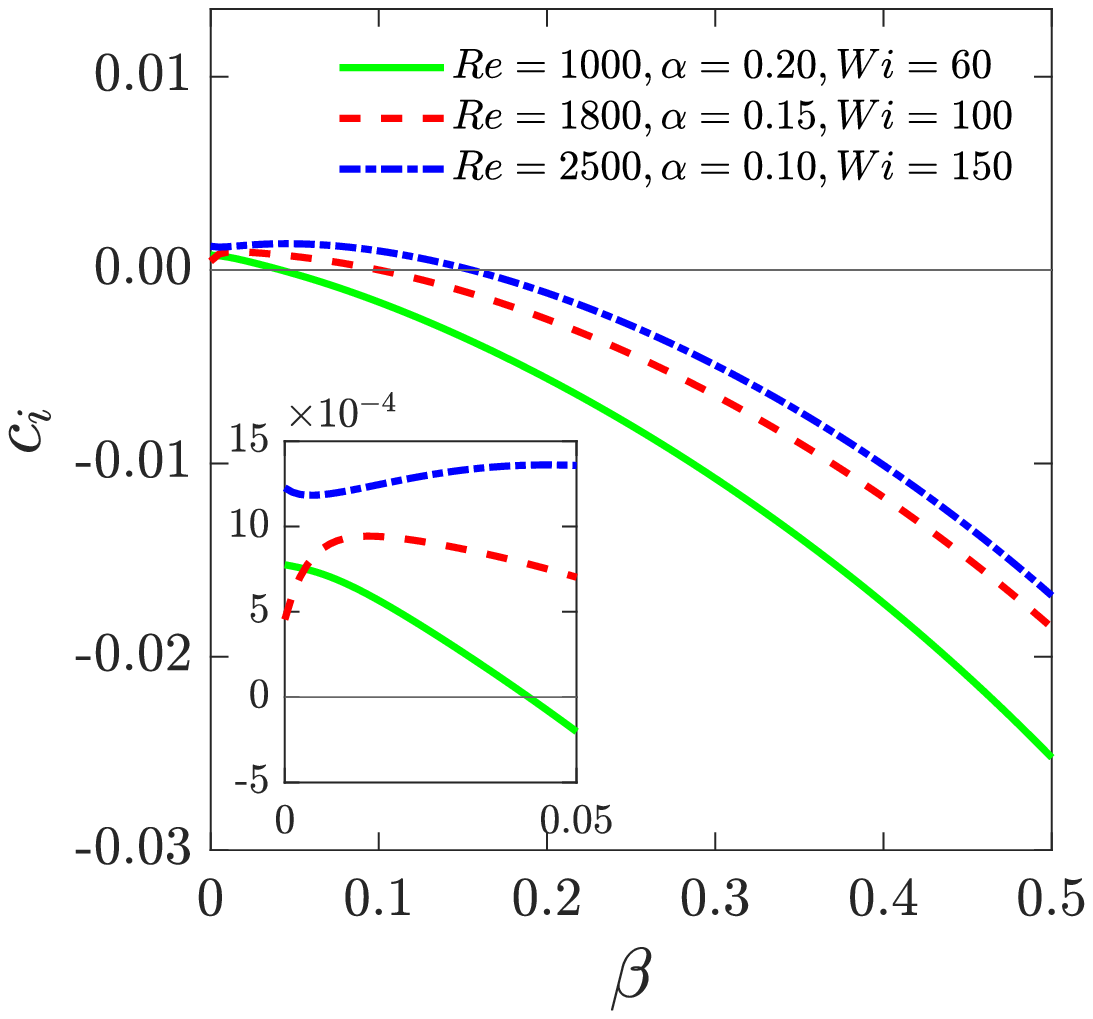}
	\put(-183,160){$(a)$}
	\includegraphics[width=0.48\textwidth,trim= 0 0 0 0,clip]{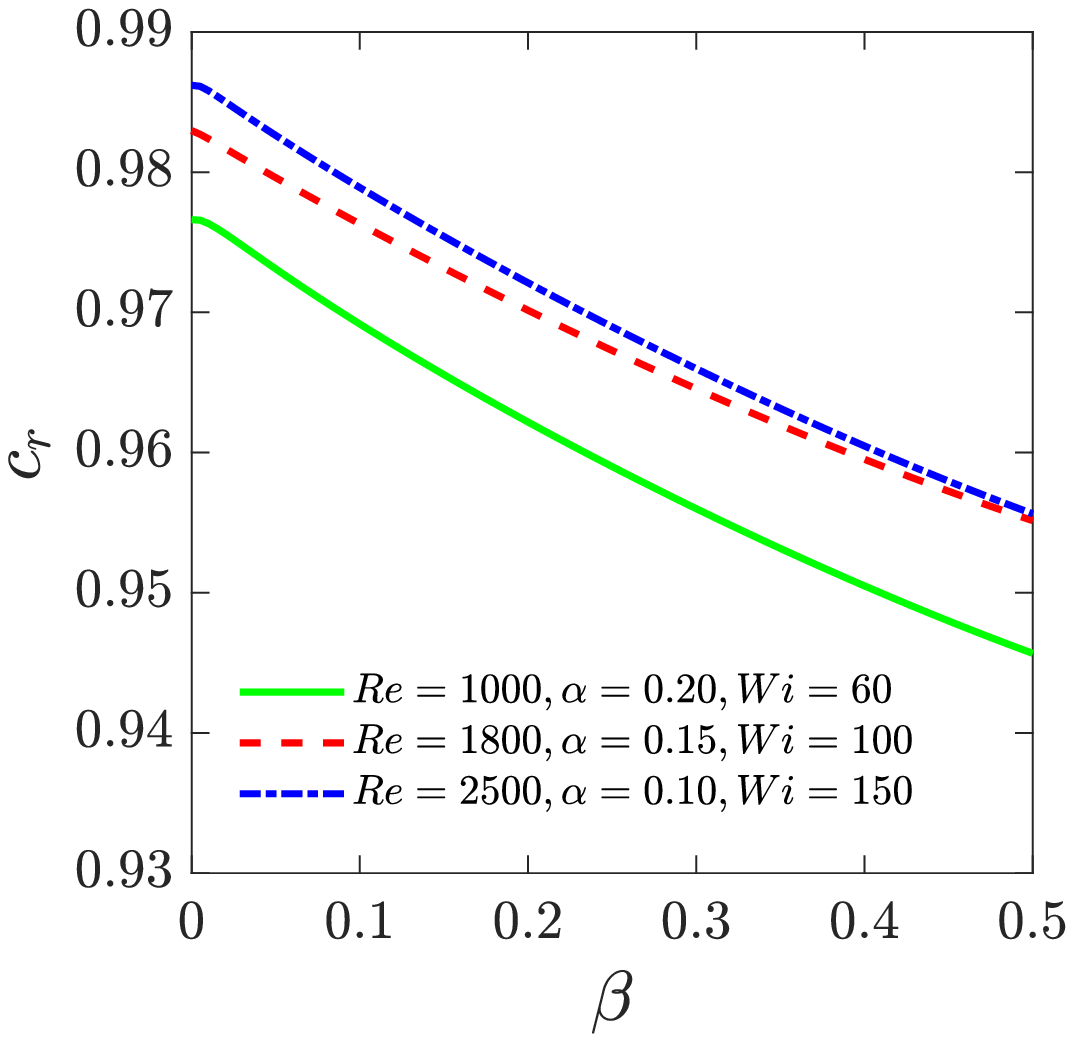}
	\put(-183,160){$(b)$}
	\caption{Continuous variation of $c=\omega/\alpha$ with viscosity ratio $\beta$ at three sets of parameters: $(a)$ imaginary part $c_i$; $(b)$ real part $c_r$. Linear instability exists ($c_i>0$) in the range of $\beta$ above the horizontal black line in $(a)$.}
	\label{Fig:c_beta}
\end{figure}

\begin{figure}
	\centering
	\includegraphics[width=0.33\textwidth,trim=40 5 20 0,clip]{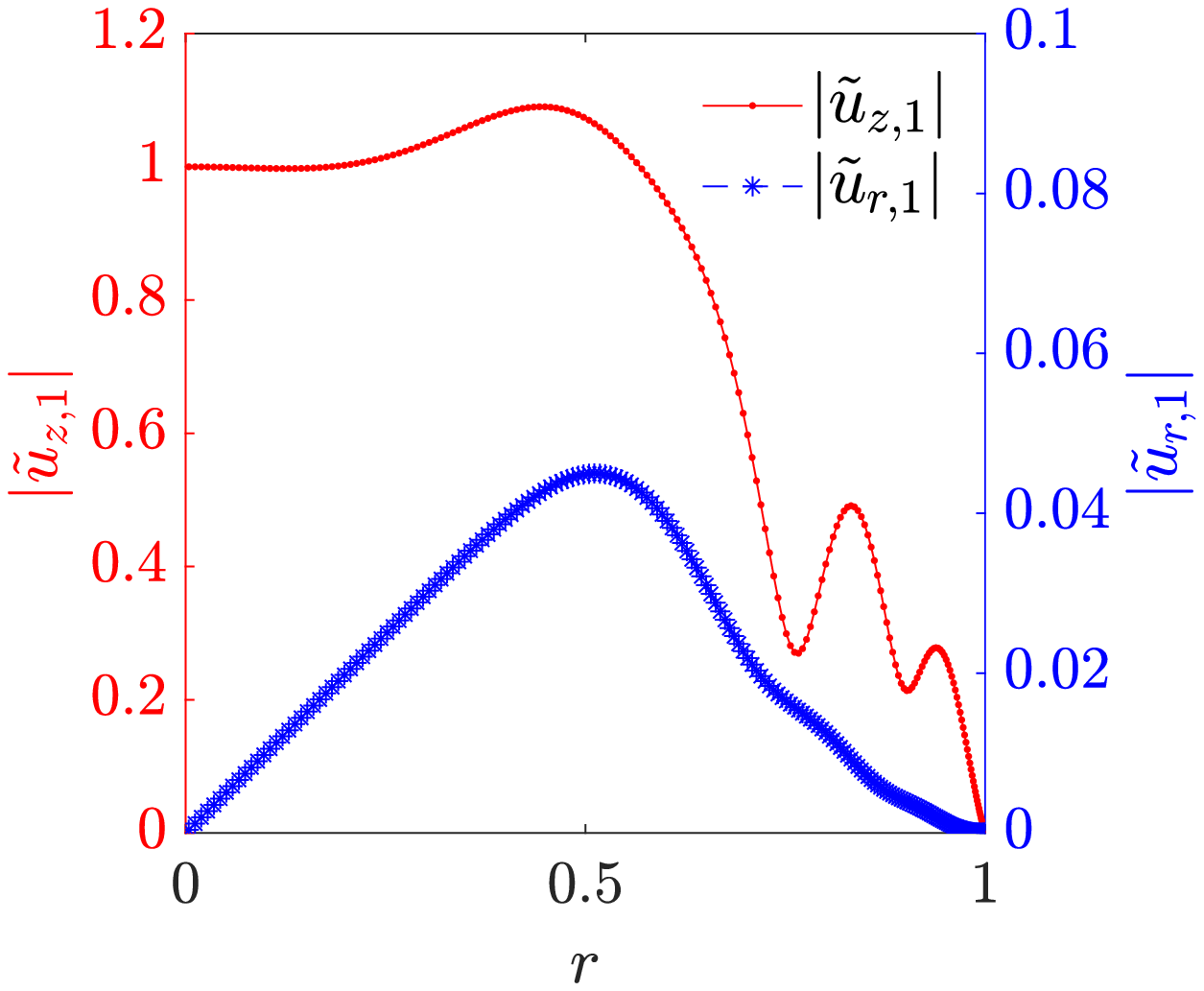}\put(-128,93){$(a)$}
	\includegraphics[width=0.33\textwidth,trim=30 5 30 0,clip]{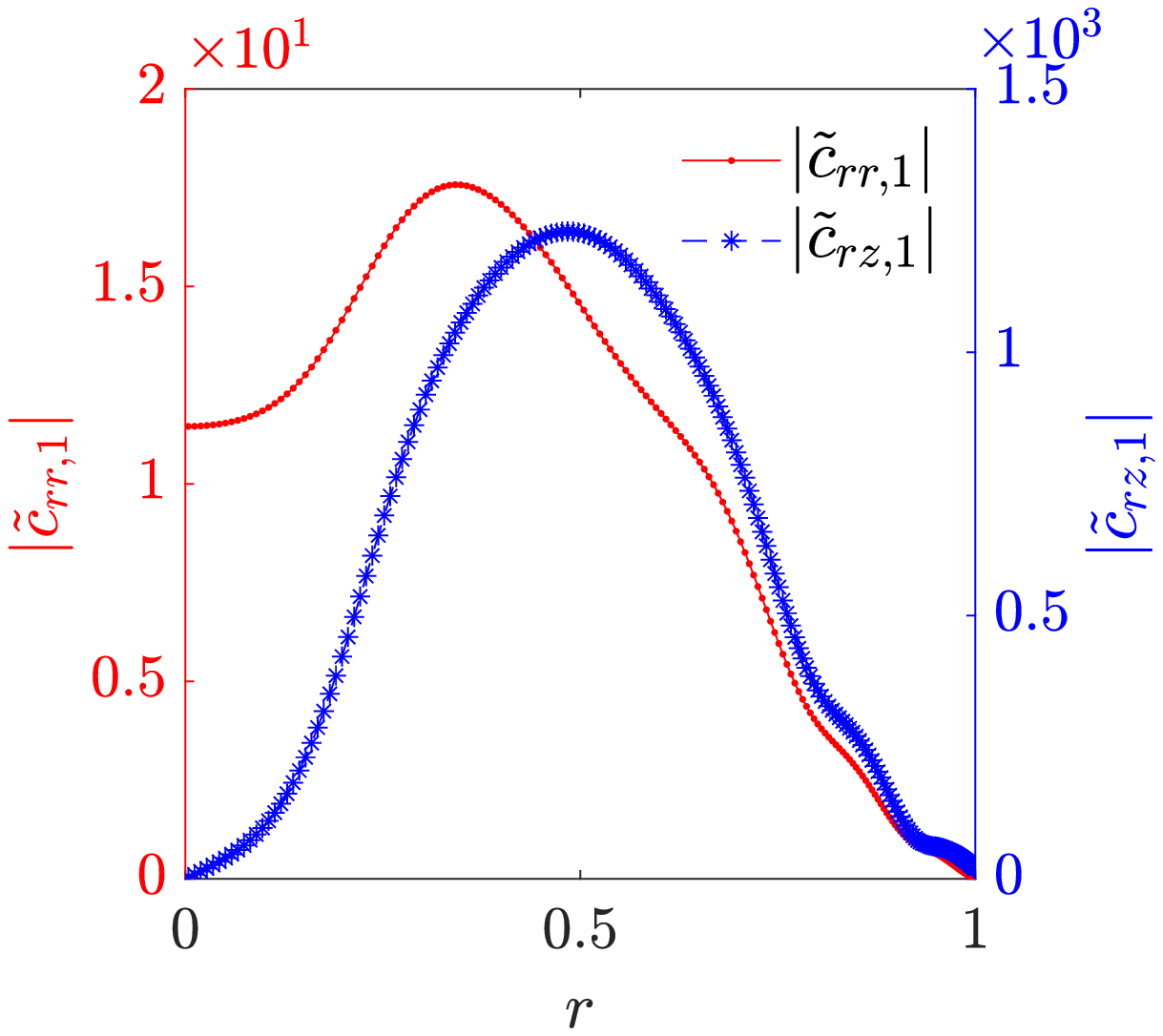}\put(-122,93){$(b)$}
	\includegraphics[width=0.33\textwidth,trim=30 5 30 0,clip]{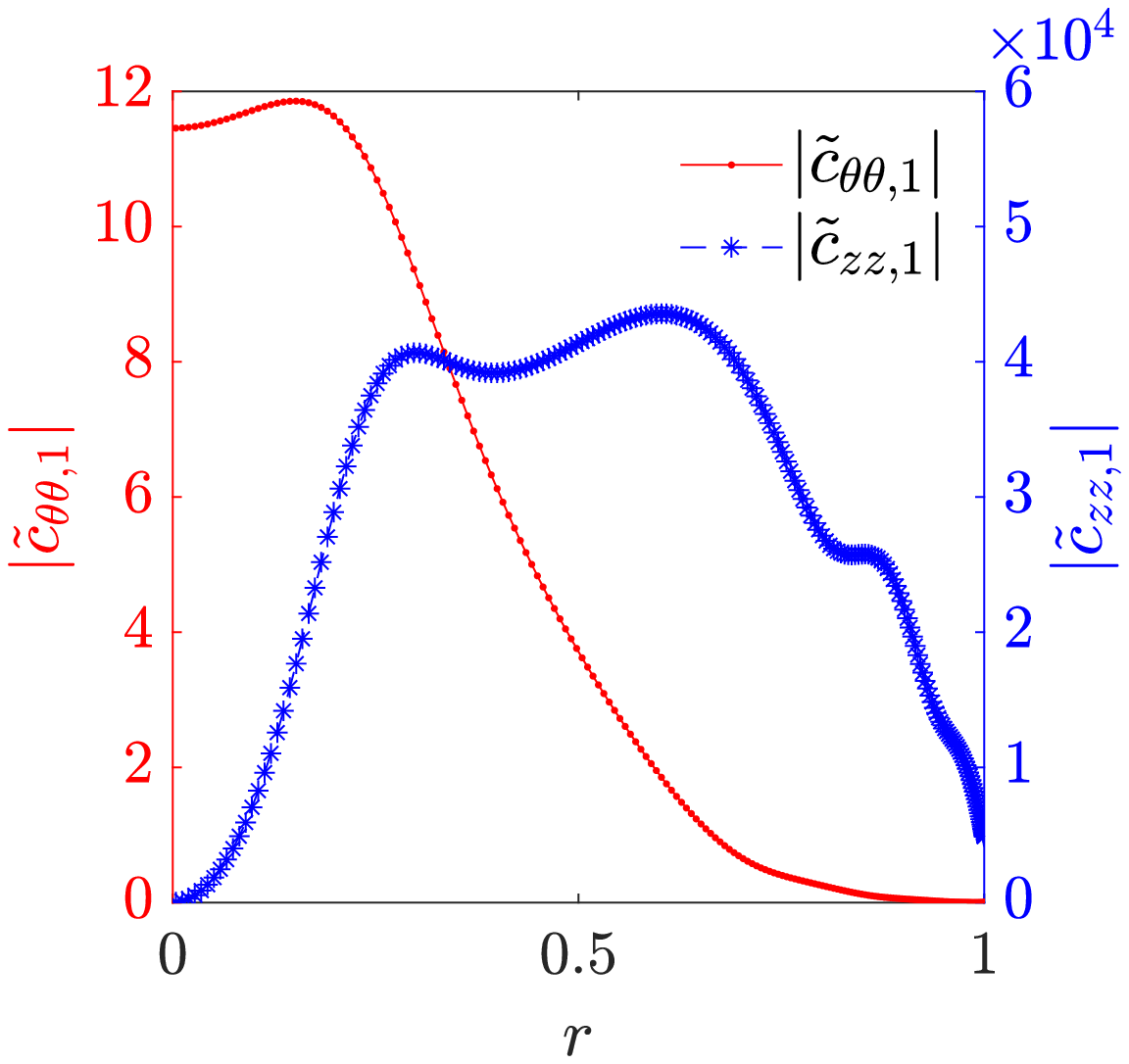}\put(-122,93){$(c)$}	
	\caption{Eigenfunctions corresponding with the unstable mode shown in figure \ref{Fig:UCM_eigens}$(a)$ at Reynolds number $Re=1000$, wavenumber $\alpha=0.2$ and Weissenberg number $Wi=60$: $(a)$ velocity components $\tilde{u}_{z,1}$ and $\tilde{u}_{r,1}$; $(b)$ conformation tensor components $\tilde{c}_{rr,1}$ and $\tilde{c}_{rz,1}$; $(c)$ conformation tensor components $\tilde{c}_{\theta\theta,1}$ and $\tilde{c}_{zz,1}$.}
	\label{Fig:UCM_eigenfunctions}
\end{figure}

In this appendix, we provide more results on the viscoelastic pipe flows of UCM fluids. The eigenspectra as shown in figure \ref{Fig:UCM_eigens} have a typical structure consisting of high-frequency Gorodtsov-Leonov (HFGL) modes \citep{Gorodtsov1967Linear}, a continuous spectrum in a balloon-like shape due to finite spatial resolutions \citep{Chaudhary2019Elasto-inertial}, and some discrete modes. Such a structure agrees well with that described in \cite{Chaudhary2021Linear} for the eigenspectrum of pipe Poiseuille flows with UCM fluids. Theoretically, the HFGL modes should distribute around a horizontal line $\omega_i/\alpha = - 1/(2\alpha Wi)$, which predicts $\omega_i=-0.008$ in good agreement with that shown in panel figure \ref{Fig:UCM_eigens}$(a)$ where $Re=1000$, $\alpha=0.2$ and $Wi=60$, and $\omega_i=-0.034$ in panel $(b)$ where $Re=260$, $\alpha=0.94$ and $Wi=14.5$. In addition, the continuous spectrum should locate around another horizontal line $\omega_i/\alpha = - 1/(\alpha Wi)$; and this theoretical prediction also agrees with the obtained continuous spectra where $\omega_i=-0.017$ for panel $(a)$ and $\omega_i=-0.069$ for panel $(b)$. In addition to the convergence of the eigenspectra, the complex wave speed $c=\omega/\alpha$ is shown to be continuous with decreasing $\beta$ even to zero as plotted in figure \ref{Fig:c_beta}. For the three sets of parameters explored, linear instability can be observed at small $\beta$ and it continues to exist when $\beta=0$ which is the UCM limit. Similar to the linear instability found by \cite{Garg2018Viscoelastic}, this mode is also a centre mode. The eigenfunctions of the unstable mode presented in figure \ref{Fig:UCM_eigens}$(a)$ are plotted in figure \ref{Fig:UCM_eigenfunctions}. The eigenfunctions are not localized near pipe axis $r=0$. Instead, they go through the whole cross-section of pipe and seem to be somehow like the eigenfunctions of wall modes which have large variations near the wall (see the variation of $\tilde{u}_{z,1}$ and $\tilde{c}_{zz,1}$).

\begin{figure}
	\centering
	\includegraphics[width=0.45\textwidth,trim= 20 0 40 0,clip]{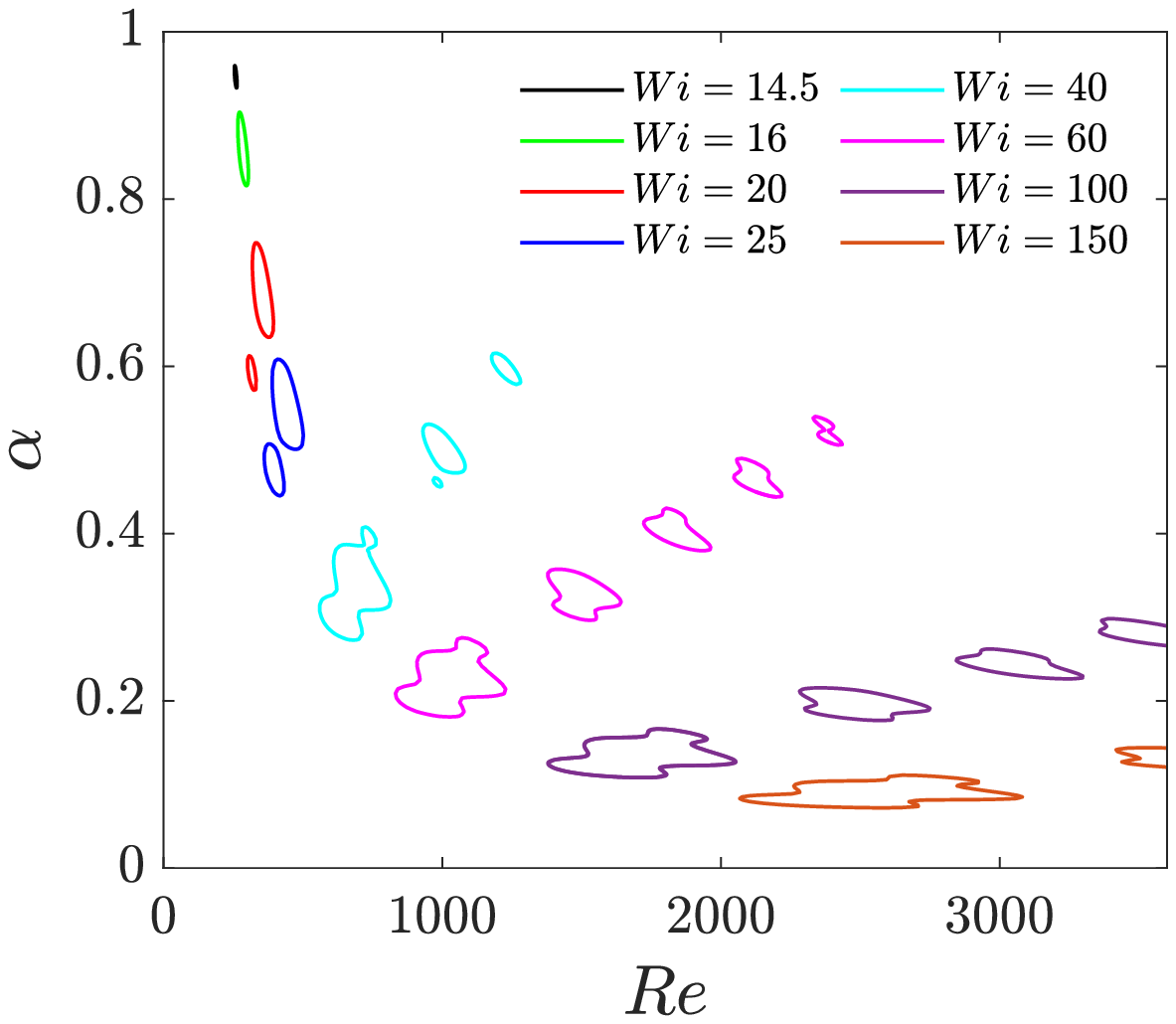}
	\put(-173,134){$(a)$}
	\includegraphics[width=0.47\textwidth,trim= 10 0 30 0,clip]{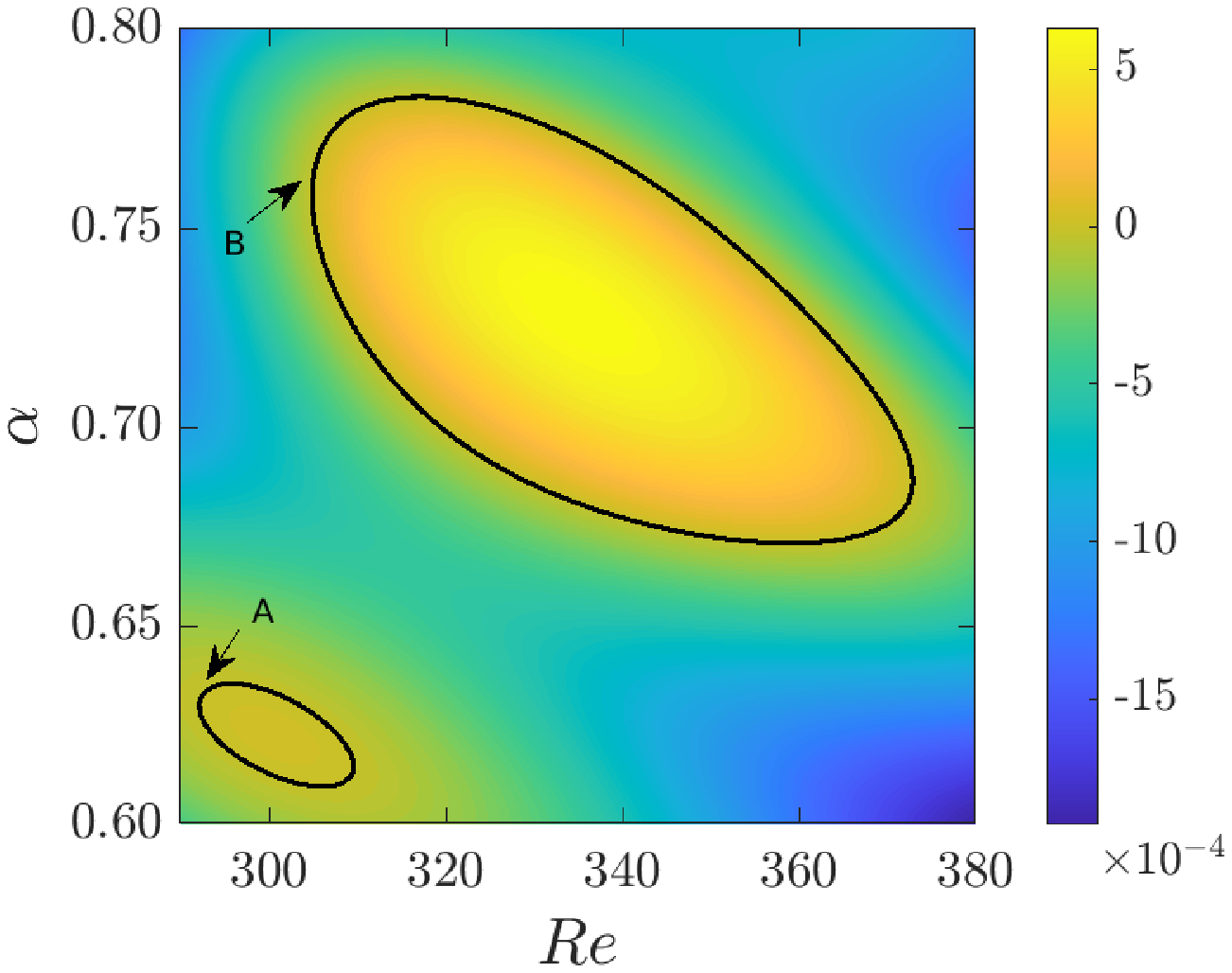}
	\put(-183,134){$(b)$}
	\caption{Neutral curves for pipe flow of UCM fluids: $(a)$ at varying Weissenberg numbers $Wi$; $(b)$ at $Wi=19$ with the colorbar indicating the linear growth rate $\mu_r$. The two arrows in $(b)$ point at the two critical points.}
	\label{Fig:NC-UCM-Wi}
\end{figure}

To examine the variation of instability with changing $Wi$, several neutral curves in the $(\alpha,Re)$ plane are plotted in figure \ref{Fig:NC-UCM-Wi}. As expected, these neutral curves are also in form of closed loops like those with finite viscosity ratios (see figure \ref{Fig:linear_Wi65beta65}$(a)$), but in more complex shapes with multiple loops. At higher $Wi$, the linear instability is limited in a very narrow range of low wavenumbers, covering a broad range of $Re$; and the loop moves towards small $Re$ and high wavenumber as $Wi$ decreases. Notably, when $Wi$ is decreased, the loops may split and disappear. Such complex behaviour causes the discontinuity of the critical parameters and Landau coefficients at $Wi=19$ as described in Section \ref{results}. Finally, at about $Wi=14.5$, the loop shrinks to almost a very small area (at $Re\approx 255$) and then the linear instability completely disappears (at least in our search of the linear instability for the current parameters). This type of behaviour of neutral curves (being closed, disconnected and may split) has been reported in other complex fluid systems, for instance, in stably stratified plane Poiseuille flows \citep{Fujimura1997Degenerate}, convection in a multicomponent fluid layer \citep{Terrones1989Onset}, and the parallel shear flow between two inclined plates with different temperatures \citep{Chen1989Stability}. In the present case, we can see that polymer elasticity (characterized by $Wi$) plays an important role in stabilizing UCM pipe flows, in contrast to the destabilizing effect in UCM channel flows \citep{Porteous1972Linear,Chaudhary2019Elasto-inertial}. The lower bound of linear instability (reached at about $Wi=14.5$) aforementioned shows that the UCM pipe flow is linearly stable at low $Re$, in analogous to the absence of linear instability at low $Re$ in UCM channel flows reported in some studies in early works \citep{Ho1977Stability,Lee1986Stability}. 

\end{appendix}

\bibliographystyle{jfm}
\bibliography{references.bib}

\end{document}